\documentclass{article}
\usepackage[utf8]{inputenc}
\usepackage{mathtools}
\usepackage[table]{xcolor}
\usepackage{amssymb,graphicx}
\usepackage{epstopdf}
\usepackage{amsmath,amsfonts}
\usepackage{epsfig}
\usepackage{xcolor}
\usepackage[T1, T2A]{fontenc}
\usepackage[english]{babel}
\usepackage{float}
\usepackage{soul}
\usepackage{graphicx}
\usepackage{subcaption}
\usepackage{geometry}
\usepackage{hyperref}
\usepackage{psfrag}
\usepackage{slashed}
\usepackage{bm}
\usepackage{siunitx}
\usepackage{soul}
\begin{document}
\begin{flushright}
INR-TH-2021-008
\end{flushright}
\vspace{10pt}
\begin{center}
  {\LARGE \bf Nonsingular cosmological models\\ with strong gravity in the past} \\
\vspace{20pt}
%\medskip
Y.~Ageeva$^{a,b,c,}$\footnote[1]{{\bf email:}
    ageeva@inr.ac.ru}, P. Petrov$^{a,}$\footnote[2]{{\bf email:}
    petrov@inr.ac.ru},
  V. Rubakov$^{a,b,}$\footnote[3]{{\bf email:} rubakov@inr.ac.ru}\\
\vspace{15pt}
  $^a$\textit{
Institute for Nuclear Research of
         the Russian Academy of Sciences,\\  60th October Anniversary
  Prospect, 7a, 117312 Moscow, Russia}\\
\vspace{5pt}
$^b$\textit{Department of Particle Physics and Cosmology,
  Physics Faculty, M.V.~Lomonosov
  Moscow State University, \\Leninskie Gory 1-2,  119991 Moscow, Russia
  }\\
\vspace{5pt}
$^c$\textit{
  Institute for Theoretical and Mathematical Physics,
  M.V.~Lomonosov Moscow State University,\\ Leninskie Gory 1,
119991 Moscow,
Russia
}
    \end{center}
    \vspace{5pt}
\begin{abstract}
%
%  {\bf ABSTRACT NOT EDITED}
  In scalar-tensor
    Horndeski theories, nonsingular cosmological models --- bounce and
  genesis --- are problematic because of potential ghost and/or
  gradient instabilities.  One
  way to get around this  obstacle
  is to send the 
  effective Planck mass to zero in the asymptotic past
  (``strong gravity in the past''). One may suspect
    that this feature is
  a signal of a strong coupling  problem
  at early times.
   However, the classical treatment
   of the cosmological  background
   is legitimate, provided that the strong
  coupling energy scale remains  at
    all times much higher than the scale associated with
  the classical evolution. We construct
  various models of this sort, namely (i) bouncing Universe which proceeds
  through inflationary epoch to kination (expansion 
  within general relativity, driven
  by massless scalar field); (ii) bouncing Universe with
  kination stage immediately after bounce; (iii) combination of genesis
  and bounce, with the Universe starting from flat space-time, then contracting
  and bouncing to the expansion epoch; (iv) ``standard'' genesis evading the
  strong coupling problem in the past. All these models are stable, and
  perturbations about the backgrounds are not superluminal.  
\end{abstract}

\section{Introduction}
\label{sec:intro}
Nonsingular cosmological models --- bouncing cosmology
and genesis from Minkowski space
--- are
of continuous interest as
alternatives to or completions of 
inflation. Provided that the spatial curvature is negligible,
a prerequisite for the construction of these models
is  the stable violation of
the null energy condition (and, more generally, null convergence condition).
It is known since
2010~\cite{Creminelli:2010ba,Deffayet:2010qz,Kobayashi:2010cm}
that the latter feature can exist in Horndeski
theories~\cite{Horndeski:1974wa} (for  reviews see, e.g.,
%, Fairlie:1991qe, Luty:2003vm, Nicolis:2004qq, Nicolis:2008in, Deffayet:2010qz, Kobayashi:2010cm, Padilla:2012dx,
Refs. \cite{Rubakov:2014jja,Kobayashi:2019hrl}).
These are scalar-tensor modifications of gravity, with the Lagrangians
containing second derivatives of both the metric and
scalar field, and yet with the
second-order equations of motion. Indeed, within Horndeski theories,
numerous explicit examples of
stable early
genesis~\cite{Creminelli:2010ba,Creminelli:2012my,Hinterbichler:2012fr,Elder:2013gya,Pirtskhalava:2014esa,Nishi:2015pta,Kobayashi:2015gga}
and bouncing~\cite{Qiu:2011cy,Easson:2011zy,Cai:2012va,Osipov:2013ssa,Qiu:2013eoa,Cai:2012va,Koehn:2013upa,Battarra:2014tga,Qiu:2015nha,Ijjas:2016tpn}
stages were constructed.

However, within Horndeski theory, these cosmologies typically
suffer from either  singularities or
gradient and/or ghost instabilities at some earlier or later
%\marginpar{\bf new refs}
stage (possibly well after the initial genesis epoch and, likewise,
well before or well after the bounce).
In earlier papers, this property was observed
explicitly in most
cases where the evolution
was followed by sufficiently distant past and
future~\cite{Pirtskhalava:2014esa,Kobayashi:2015gga,Cai:2012va,Koehn:2013upa,Battarra:2014tga,Qiu:2015nha,Ijjas:2016tpn}
(see Ref.~\cite{Dobre:2017pnt}
for the discussion of other problematic properties of
the model of Ref.~\cite{Ijjas:2016tpn}). Later on, the problem has
been formulated as a no-go theorem~\cite{Libanov:2016kfc,Kobayashi:2016xpl}.
Namely, in the unitary gauge and in the spatially flat Friedmann-Lemaître-Robertson-Walker
  background $ds^2 = dt^2 - a^2(t) \delta_{ij} dx^i dx^j$,
the quadratic
actions for tensor (transverse traceless) perturbation $h_{ij}$
and scalar perturbation $\zeta$ have the forms
\begin{subequations}
  \label{jul25-21-1}
\begin{align}
 \mathcal{ S}_{hh} & =\int dt d^3x \frac{ a^3}{8}\left[
        \mathcal{ G}_T
        \dot h_{ij}^2
        -\frac{\mathcal{ F}_T}{a^2}
        h_{ij,k} h_{ij,k} \right] \; ,
 \\
   \mathcal{ S}_{ss} &=\int dt d^3x a^3\left[
        \mathcal{ G}_S
        \dot\zeta^2
        -\frac{\mathcal{ F}_S}{a^2}
        \zeta_{,i}\zeta_{,i}
        \right] \; .
\end{align}
\end{subequations}
The theorem states that if in a Horndeski theory
the background is nonsingular during
the entire evolution
$-\infty < t < +\infty$, the coefficient $  \mathcal{ G}_T$ is
strictly positive at all times, and
the following two integrals are divergent at lower and upper
limits, respectively,
\begin{subequations}
  \begin{align}
\label{no-go_integral}
    \int_{-\infty}^{t} a(t) (\mathcal{ F}_T +\mathcal{ F}_S) dt &= \infty \; ,
    \\
   \int_t^{+\infty} a(t) (\mathcal{ F}_T +\mathcal{ F}_S) dt &=
  \infty \; ,
\end{align}
  \end{subequations}
then $\mathcal{ F}_S < 0$ and/or  $\mathcal{ F}_T < 0$
in some time interval, i.e.,
there exists either ghost or gradient instability (or both).
Adding extra scalar fields, conventional or Galileon, does  not
improve the situation~\cite{Kolevatov:2016ppi,Akama:2017jsa}.

As a digression, we emphasize that like most
cosmology model builders, we stick to
the study of
{\it homogeneous and isotropic} backgrounds and their stability 
against {\it linearized} perturbations. Like most others, we are confident
that
the standard (3+1) decomposition with algebraic gauge conditions
(unitary gauge in our case) is adequate {\it for this particular purpose}:
the wave equations derived from the actions \eqref{jul25-21-1}
are manifestly strongly hyperbolic for positive $\mathcal{ G}_T$,
$\mathcal{ F}_T$, $ \mathcal{ G}_S$,  $\mathcal{ F}_S$ and manifestly
elliptic for negative sound speeds squared (but we tend to agree with
Ref.~\cite{Ijjas:2018cdm} that the algebraic gauges, including unitary,
may not be convenient for analyzing the evolution at a fully nonlinear level).
We leave aside the issue of
stability at the nonlinear level and, even more so, the issue of
well posedness
of general backgrounds in Horndeski theories; the latter issues are discussed,
e.g., in
Refs.~\cite{Ijjas:2018cdm,Papallo:2017qvl,Allwright:2018rut,Kovacs:2019jqj}.
In this regard, positivity of  $\mathcal{ G}_T$,
$\mathcal{ F}_T$, $ \mathcal{ G}_S$, and  $\mathcal{ F}_S$ is necessary,
albeit possibly not a sufficient condition for a healthy cosmological model.

One way to deal with instabilities implied by the no-go theorem
is to arrange for (or merely
declare) a sufficiently low energy scale of the UV completion and
make sure that the unstable modes (with energies below this scale)
do not have enough time to
develop~\cite{Pirtskhalava:2014esa,Creminelli:2006xe,Koehn:2015vvy,deRham:2017aoj}.
Another is to
get around 
these
instabilities altogether
\cite{Cai:2016thi,Creminelli:2016zwa,Cai:2017dyi,Kolevatov:2017voe,Mironov:2019qjt}
by  making use of beyond
Horndeski~\cite{Zumalacarregui:2013pma, Gleyzes:2014dya} or more general
degenerate higher-order scalar-tensor
theories~\cite{Langlois:2015cwa,Langlois:2018dxi}, which, however, have
problems with superluminality~\cite{Mironov:2020pqh}.
In this paper we follow yet 
another route
\cite{Kobayashi:2016xpl}, namely, we
stick to the Horndeski theory and
ensure that the coefficients
 $\mathcal{ G}_T$,
$\mathcal{ F}_T$, $ \mathcal{ G}_S$, and  $\mathcal{ F}_S$
 (``effective
  Planck masses'' squared)
in the
quadratic actions for perturbations \eqref{jul25-21-1}
sufficiently rapidly decay as
one goes backwards in time to $t \to -\infty$, so that the
integral in the left-hand side of Eq.~\eqref{no-go_integral}
is actually convergent. In this way we relax the assumption
of the no-go theorem and construct Horndeski models without
gradient or ghost instabilities (at the linearized level).
A peculiarity of
this case is that the
gravitational and scalar interactions are strong at early times,
which, among other things, signalizes  a potentially strong coupling
problem.\footnote{There exists even more radical proposal
  that the effective Planck masses squared
  $\mathcal{ G}_T$ and $\mathcal{ F}_T$ vanish
  at some finite time $t_0$, i.e.,
  $\mathcal{ G}_T,  \mathcal{ F}_T
  \propto (t-t_0)^2$~\cite{Ijjas:2016vtq}. It remains to be seen
  whether or not models of this sort are tractable within classical field
  theory.}  
For brevity, we refer to this property as ``strong gravity in the past.''

In the latter class of models, the fact that the
effective Planck masses
tend to zero as $t\to -\infty$ does not necessarily mean that the
classical field theory treatment of the (homogeneous and isotropic)
cosmological evolution is not
legitimate at early times~\cite{Ageeva:2018lko,Ageeva:2020gti}.
Indeed, the classical analysis {\it is} valid, provided that
the quantum strong coupling
    energy scale $E_{strong}$
    stays well above
    the energy scale of the classical evolution $E_{class}$
    (for power-law evolution, the latter is $E_{class} \sim |t|^{-1}$
    as $t\to -\infty$).
    This issue was considered in
    Refs.~\cite{Ageeva:2018lko,Ageeva:2020gti,Ageeva:2020buc}
    in the framework of the class of models suggested in
    Ref.~\cite{Kobayashi:2016xpl}.
    Using the  dimensional analysis, it was shown, order by order in
   perturbation theory, that
    there actually exists a region in the parameter
    space where the classical
    field theory treatment is legitimate.\footnote{We note in passing that
another model with vector field and power-law background solution
was constructed in Ref.~\cite{Petrov:2020vlq}; it
describes  stable early genesis which is
legitimately treated within classical field theory.}
     It is worth noting, though, that the parameters of the
     concrete model
    given in Ref.~\cite{Kobayashi:2016xpl} do not belong to this region.
    Interestingly, the result of the all-order analysis
of Ref.~\cite{Ageeva:2020buc}
coincides with the result of Ref.~\cite{Ageeva:2020gti} obtained by
studying
the cubic order only: both lead to the same
constraints on the parameters.\footnote{A would-be
  caveat in the analysis of
  Ref.~\cite{Ageeva:2020buc} is that it did not give explicit comparison
  of strong coupling energy scales emerging at different orders
  of perturbation theory. The subtlety
  is that when
  the strong coupling scale inferred from a higher order term
  is below that
  coming from lower order  ones,
  the naive estimate for the strong
  coupling scale may break down~\cite{Bellazzini:2020cot}. The
   dominance of the cubic order
   shows that this is not the case in models we
  consider.}
 
 Explicit Horndeski models with strong gravity in the past
have not been constructed so far. It is the purpose of this paper
to fill this gap: we
%In this paper we
introduce several Horndeski cosmologies
of this sort, which are stable at all times;
%which admit solutions with
%strong gravity in the past;
we emphasize that we always work
in the Jordan frame. We ensure that these models are
free of the strong coupling problem, i.e.,
$E_{class} \ll E_{strong}$  at all
 times, even though $E_{class} \to 0$ as $t \to -\infty$.
  Our cosmologies are complete in the sense that at late times
 the Universe expands in a 
standard way: at large positive $t$, the
models
turn into general relativity with a conventional massless
scalar field that drives the expansion.
This is kination epoch which is assumed to end up with reheating
through, say, one of the mechanisms of
Refs.~\cite{ArmendarizPicon:1999rj,BazrafshanMoghaddam:2016tdk}. 
The least straightforward part of our
construction is to ensure the (linear)
stability
of the solutions during the entire evolution. 
We also make sure that
the speed of the perturbations about our backgrounds
does not exceed the speed of light.
So, our cosmologies are exotic
but healthy (modulo possible pathologies at nonlinear level).

The first model, elaborated in greater detail, 
is the bouncing Universe. 
In the asymptotic past the Universe contracts with
the power-law behavior of the scale factor, then the contraction terminates
and expansion begins (bounce). Depending on the choice of the
Lagrangian, the expansion epoch may or may not pass through the
inflationary stage; we give examples of both scenarios.
As described above, we follow the evolution up to the kination epoch.

Another
  model is a combination of genesis and bounce: the Universe
starts from the flat space-time, then contracts, passes through
the bounce and then evolves in the same way as in the first example.
For completeness, we design yet another ``standard'' genesis
model (in which the Universe expands from the beginning),
which satisfies the condition~\cite{Ageeva:2020buc}
of the absence of
strong coupling
and thus 
improves on the model
of Ref.~\cite{Kobayashi:2016xpl}.

This
  paper is organized as follows. We introduce our subclass of
  models from the Horndeski
class
in Sec.~\ref{sec:model}, where we also discuss general properties of
these models.
Bouncing Universes are constructed in Sec.~\ref{sec:bounce}.
Models with genesis are presented in Sec.~\ref{sec:genesismodels}.
We conclude in Sec.~\ref{sec:summary}. In Appendix~\ref{app:strong_coupl}
we derive the condition of
the absence of strong coupling at early times in the models of
Sec.~\ref{sec:bounce}, while in Appendix~\ref{app:variable_u} we give
details of our numerical treatment of the models of
Sec.~\ref{sec:genesismodels}.

\section{Generalities}
\label{sec:model}

In this paper we 
consider
 a subclass of the Horndeski theories. The general form of the Lagrangian
for this
subclass is
    \begin{eqnarray}
    \cal L&=&G_2(\phi, X)-G_3(\phi, X)\Box \phi+ G_4(\phi,X)R + G_{4X}\big[(\Box \phi)^2 - (\nabla_{\mu}\nabla_{\nu}\phi)^2\big],
    \label{Hor_L}\\
        X &=& -\frac{1}{2}g^{\mu\nu}\partial_{\mu}\phi\partial_{\nu}\phi,
    \nonumber%    \label{Hor_L}
    \end{eqnarray}
where 
$\Box \phi = g^{\mu\nu} \nabla_\mu \nabla_\nu \phi$ and $(\nabla_{\mu}\nabla_{\nu}\phi)^2 = \nabla_{\mu}\nabla_{\nu}\phi \nabla^{\mu}\nabla^{\nu}\phi$, and  $R$ is the Ricci scalar. The metric
signature is $(-,+,+,+)$. Unlike 
the general Horndeski theory,
the Lagrangian  \eqref{Hor_L} involves three arbitrary functions
$G_{2,3,4}$ rather than four. We
  recall that we always work
in the Jordan frame.

It is convenient for our purposes to work in the Arnowitt-Deser-Misner (ADM) formalism. The ADM form
of the metric is
 \begin{equation*}
     ds^2=-N^2 dt^2 +  
        \gamma_{ij}\left( dx^i+N^i dt\right)\left(dx^j+N^j dt\right) ,
    \end{equation*}
 where  $\gamma_{ij}$ is the three-dimensional metric,
 $N$ is the lapse function and  $N_i=\gamma_{ij}N^j$ 
    is the shift  function.
 In ADM terms, the action for the Horndeski
theory subclass \eqref{Hor_L} has the form 
    \begin{equation}
    \label{adm_action}
        \mathcal{S} = \int{\sqrt{-g}d^4x \mathcal{L}},
    \end{equation}
    with~\cite{Kobayashi:2019hrl}
%    \marginpar{\bf deleted $B_4$}
\begin{align}
\label{adm_lagr}
        \mathcal{L} =  A_2 (t, N) + A_3 (t, N) K 
        +  A_4 (t, N)
        (K^2 - K_{ij}^2) + B_4 (t, N) R^{(3)} \text{,}
    \end{align}
where
\begin{equation*}
    A_4(t,N) = - B_4(t,N) - N\frac{\partial B_4(t,N)}{\partial N},
\end{equation*}
         and
$^{(3)} R_{ij}$ is the Ricci tensor made of $\gamma_{ij}$,
   $\sqrt{-g} = N\sqrt{\gamma}$,
    $K= \gamma^{ij}K_{ij}$, $^{(3)} R = \gamma^{ij} \phantom{0}^{(3)} R_{ij}$ and
    \begin{align*}
      K_{ij} &\equiv\frac{1}{2N}\big(\dot\gamma_{ij} -\,^{(3)}\nabla_{i}N_{j}-\;^{(3)}\nabla_{j}N_{i}\big) \; .
      %,  \\
      %  ^{(3)}R_{ij}  &\equiv \partial_{k}\;^{(3)}\Gamma^{k}_{ij}-\partial_{i}\;^{(3)}\Gamma^{k}_{kj}+
      %  \;^{(3)}\Gamma^{k}_{lk}\;^{(3)}\Gamma^{l}_{ij}-\;^{(3)}\Gamma^{k}_{li}\;^{(3)}\Gamma^{l}_{jk}, \\
       % \sqrt{-g} &= N\sqrt{\gamma} \text{,} \;\;\; \gamma = \text{det}(^{(3)}\gamma_{ij}),
    \end{align*}

 %   \vspace{1cm}

 %   {\bf TO BE CLARIFIED}
    
    The relationship between the two  formalisms is established by
    choosing the equal-time slices as slices of constant $\phi$ and
    defining the 
    time coordinate
    in such a way that $\phi (t)$ is a prescribed monotonous
    function, $\dot{\phi} > 0$ (as an example,
    it is convenient
    to choose at large positive times 
    $\mbox{e}^{\phi} = t$).  This gives
 %   \marginpar{\bf Correct in this paragraph?}
    \begin{equation*}
%\label{N}
     N^{-1} = \frac{\sqrt{2X}}{\dot{\phi}(t)} \; .
\end{equation*}
    Then one has~\cite{Gleyzes:2014dya, Gleyzes:2013ooa, Fasiello:2014aqa}
  %  \marginpar{\bf refs needed}
    \begin{align}
        G_2 =& A_2 - 2XF_{\phi} \text{,} \label{ADM-trans2}\\
        G_3 =& - 2XF_X - F \text{,} \label{ADM-trans3}\\
        G_4 =& B_4 \text{,} \label{ADM-trans4}
    \end{align}
where
    \begin{equation}
    \label{F}
        F_X = - \frac{A_3}{\left(2X\right)^{3/2}} - \frac{B_{4\phi}}{X} \text{.}
    \end{equation}
    It is worth noting that general relativity (GR) description of gravity
      is restored for $B_4 = -A_4 = M_P^2/2 = \mbox{const}$, where $M_P$
      is the reduced Planck mass, which we set equal to 1 in what follows.
      Note also that
    the transition from the ``covariant'' formulation \eqref{Hor_L}
    to ADM action \eqref{adm_action} is not unique: it depends on the choice
    of the function $\phi(t)$. Thus, one can impose additional constraints on
    the functions $A_2$, $A_3$, $B_4$,
    or, in other words, on the background
    solution. We will use this freedom in Sec.~\ref{sec:early_times}.
%{\bf END TO BE CLARIFIED}
    
    %    \vspace{1cm}

    The equations of motion for homogeneous, isotropic, and spatially
    flat background
    are obtained by setting $N=N(t)$, $\gamma_{ij} = a^2(t) \delta_{ij}$
    and 
    varying the 
    action \eqref{adm_action} with the respect to $N(t)$ and $a(t)$.
    They read~\cite{Kobayashi:2011nu}
    %\begin{align*}
    %    \mathcal{E} &=0,\\
    %    \mathcal{P} &=0,
    %\end{align*}
%where
    \begin{subequations}
    \label{eoms}
    \begin{align}
         &  (NA_2)_N + 3NA_{3N}H + 6N^2(N^{-1}A_4)_N H^2 = 0,\\
      &  A_2
        -6A_4H^2-\frac{1}{N}\frac{d}{dt}\left( A_3+4A_4H \right) = 0 \; ,
    \end{align}
    \end{subequations}
    where $H=\dot{a}/(aN)$ is the Hubble parameter.
    To perform the stability analysis, one writes
        \begin{align*}
        N &=N_0(t) (1+\alpha),\nonumber\\
        N_{i} &=\partial_{i}\beta +  N^T_i,
        %\; \text{where} \; \partial_i N^{Ti}=0,
        \nonumber\\
        \gamma_{ij} &=a^{2}(t) \Big(\text{e}^{2\zeta}(\text{e}^{h})_{ij} + \partial_i\partial_j Y 
        + \partial_i W^T_j + \partial_j W^T_i\Big) \; ,
        \end{align*}
        where
        $a(t)$ and $N_0(t)$ are background solutions,
        $\partial_i N^{Ti}=0$ and
         \begin{align*}
        (\text{e}^h)_{ij} &=\delta_{ij}+h_{ij}+\frac{1}{2}h_{ik}h_{kj}+\frac{1}{6}h_{ik}h_{kl}h_{lj}+
        \cdots, \quad h_{ii}=0, \quad
        \partial_{i}h_{ij}=0 \; .
    \end{align*}
    %Through this paper, we will denote the background value of the lapse 
    %function by $N$ where this notation does not lead to any confusion. 
        Note that the ADM formulation automatically implies
        the unitary gauge, $\delta \phi =0$.
        The residual gauge freedom is fixed
        by setting $Y = 0$ and $W^T_i = 0$, so  that the
    spatial part of the metric reads
    \begin{align*}
        \gamma_{ij} &=a^{2}\text{e}^{2\zeta}(\text{e}^{h})_{ij} \; .
    \end{align*}
    Variables
      $\alpha$, $\beta$, and $N^T_i$ enter the action without temporal
    derivatives;
    the dynamical degrees of freedom are $\zeta$
    and transverse traceless $h_{ij}$, i.e., scalar and tensor perturbations.

    In what
      follows we omit subscript 0 in the notation for the background
    lapse function. Then the  quadratic
    action for tensor perturbations reads~\cite{Kobayashi:2015gga}
    \begin{eqnarray}
        \mathcal{ S}_{hh}=\int dt d^3x \frac{N a^3}{8}\left[
        \mathcal{ G}_T
        \frac{\dot h_{ij}^2}{N^2}
        -\frac{\mathcal{ F}_T}{a^2}
         h_{ij,k} h_{ij,k}
        \right]
        \label{tensor2},
    \end{eqnarray}
where 
    \begin{align*}
     %\label{stability_func}
         \mathcal{ G}_T &=  -2A_4,\\
         \mathcal{ F}_T &= 2B_4.
    \end{align*}
Likewise,  the  quadratic action for
scalar perturbation $\zeta$ is~\cite{Kobayashi:2015gga}
    %\marginpar{\bf ref. needed}
\begin{eqnarray}
        \mathcal{ S}_{ss}=\int dt d^3x N a^3\left[
        \mathcal{ G}_S
        \frac{\dot\zeta^2}{N^2}
        -\frac{\mathcal{ F}_S}{a^2}
        \zeta_{,i}\zeta_{,i}
        \right]
        \label{scalar2},
\end{eqnarray}
 where
    \begin{subequations}
    \label{eq:Fs_Gs_form}
    \begin{eqnarray}
        \mathcal{ F}_S&=&\frac{1}{a N}\frac{d}{d t}\left(\frac{a}{\Theta}\mathcal{ G}_T^2\right)
        -\mathcal{ F}_T, 
        %\label{eq:Fs_form}
        \\
        \mathcal{ G}_S&=&\frac{\Sigma }{\Theta^2}\mathcal{ G}_T^2+3\mathcal{ G}_T %\label{eq:Gs_form}
    \end{eqnarray}
    \end{subequations}
 with
     %\label{stability_func2}
    \begin{align*}
      \Sigma&=
      N A_{2N}+\frac{1}{2}N^2A_{2NN}+
      \frac{3}{2}N^2A_{3NN}H+3\big(2A_4-2NA_{4N}+N^2A_{4NN}\big)H^2, 
        \\
        \Theta&=2H\Big(\frac{NA_{3N}}{4H}-A_4 + NA_{4N}\Big).
    \end{align*}
%and in these formulas above one should put  $N = N_0$ after differentiation with respect to $N$. 
To avoid ghost and gradient
    instabilities, one requires that
    \begin{align}
        \mathcal{ F}_S, \mathcal{ G}_S, \mathcal{ F}_T, \mathcal{ G}_T>0.
        \label{stability_conditions}
    \end{align}
We also require that the speed of perturbations  does 
not exceed the  speed of light,
\begin{subequations}
\label{velocities}
\begin{align}
    c_T^2 &= \frac{\mathcal{F}_T}{\mathcal{G}_T} \leq 1, \\
    c_S^2 &= \frac{\mathcal{F}_S}{\mathcal{G}_S}\leq 1 \; .
\end{align}
\end{subequations}
It has been
argued that the latter conditions are  necessary for  
the existence of the UV  completion 
\cite{Adams:2006sv,deRham:2013hsa}.

It is worth noting that under rescaling
  of the Lagrangian functions
\begin{equation}
  A_2 (t, N) \to \lambda^2 A_2 (\lambda t, N)\;, \quad \quad
  A_3 (t, N) \to \lambda A_3 (\lambda t, N)\;, \quad \quad
   B_4 (t, N) \to  B_4 (\lambda t, N)
\label{apr2-21-10}
\end{equation}
with constant $\lambda$, one has
$A_4 (t, N) \to  A_4 (\lambda t, N)$, and
solutions to equations of motion \eqref{eoms} scale as
    \begin{equation}
    H (t) \to \lambda H (\lambda t)\;, \quad \quad
    N(t) \to N (\lambda t) \; ,
    \label{apr2-21-11}
    \end{equation}
    while the coefficients of the quadratic action transform as
    \begin{equation}
      {\cal G}_T (t) \to  {\cal G}_T (\lambda t) \; ,
      \quad \dots \; , \quad    {\cal F}_S (t) \to  {\cal F}_S (\lambda t) \; .
       \label{apr2-21-12}
    \end{equation}
    In particular, stability and subluminality
    conditions
      remain intact under
      the transformation \eqref{apr2-21-10}. The
        scaling property of
      \eqref{apr2-21-10},  \eqref{apr2-21-11}, and \eqref{apr2-21-12}
      implies, in particular, that the overall
      time scale of evolution can be chosen at one's will, so that at
      epochs described by GR it is safely longer
      than the Planck time.

%\vspace{1cm}

%{\bf NOT EDITED FROM HERE TO END OF PAPER}

\section{Bouncing Universes}
\label{sec:bounce}

\subsection{\textit{Ansatz}}
\label{sec:ansatz-bounce}
Our purpose in this paper is to design the functions
$A_2$, $A_3$, $A_4$, and $B_4$
in such a way that the model admits a cosmological solution of
interest.
%which is stable and subluminal at all times and is free of the
%strong coupling problem at early stage.
To this end, we do not need to work
in complete generality. To construct bouncing cosmologies in this Section,
we make use of the following \textit{Ansatz}:
%for the
%functions defining the model:
%\marginpar{\bf set $f_0=1$}
\begin{subequations}
\label{A_ansatz}
	\begin{align}
	&A_2 =  \frac{1}{2} f^{-2\mu -2} \cdot a_2 (t,N) \text{,} \label{A2_ansatz} \\
	  &A_3 =  \frac{1}{2} f^{-2\mu -1} \cdot a_3 (t,N) \text{,}
          \label{A3_ansatz}\\
	&A_4   = -B_4 = -\frac{1}{2}  f^{-2\mu} \text{,}
	\end{align}
\end{subequations}
where $\mu>0$ is a time-independent
parameter which may be different for different
cosmologies,\footnote{Note 
that our parameter $\mu$ was denoted by $\alpha$ in 
Refs.\cite{Kobayashi:2016xpl,Ageeva:2020gti}.} $f = f(t)$ is a 
positive function of time which is extracted as a prefactor in
\eqref{A2_ansatz} and \eqref{A3_ansatz} for convenience. 
%in order to obtain stable bounce without strong coupling regime at early times.
Functions $a_2$, $a_3$ are given by
\begin{subequations}
  \label{apr22-21-1}
\begin{align}
&a_2(t,N) = \Big(\frac{x(t)}{N^2} + \frac{v(t)}{N^4}\Big),\\
&a_3(t,N) = \frac{y(t)}{N^3}.
\end{align}
\end{subequations}
Thus, our \textit{Ansatz}  generalizes
Ref.~\cite{Kobayashi:2016xpl} and involves
four arbitrary functions of time $f(t)$, 
$x(t)$, $v(t)$, and $y(t)$. The construction of concrete cosmological models
boils down to the design of these functions.

Given the \textit{Ansatz}~\eqref{A_ansatz}, the background equations of
motion~\eqref{eoms} are
\begin{subequations}
\label{eoms_all_substitute}
\begin{align}
    &\left(-\frac{x(t)}{N^2} - \frac{3 v(t)}{N^4}\right) - \frac{9 y(t)\cdot f\cdot H}{N^3} 
    + 6 f^2\cdot H^2  = 0, \label{eom_1} \\
    &\left(\frac{x(t)}{N^2} + \frac{v(t)}{N^4}\right) + 6 f^2 \cdot H^2 
    + \frac{(2 \mu + 1)  \cdot \dot{f}}{N}\left( \frac{y(t)}{N^3} - 4 f \cdot H\right) 
    - 
    \frac{f}{N}\frac{d}{dt}\left( \frac{y(t)}{N^3} - 4 f\cdot H\right) = 0,
    \label{eom_2}
\end{align}
\end{subequations}
while the functions entering 
\eqref{eq:Fs_Gs_form} are given by
\begin{align*}
  \Theta &=  f^{-2\mu-1} \left( f \cdot H
    - \frac{3 y(t)}{4 N^3}\right), \\
    \Sigma &=  \frac{f^{-2\mu-2}}{2 N^4} \cdot \Big(6 v(t) + 18 y(t) \cdot f\cdot H \cdot N 
    + x(t) \cdot N^2 - 6 f^2 \cdot H^2 \cdot N^4 \Big)\; .
\end{align*}
Thus
\begin{subequations}
\label{stability_all_subtitute}
\begin{align}
    \mathcal{F}_T &= \mathcal{G}_T=  f^{-2 \mu}, \\ 
    %\mathcal{G}_T &=   f^{-2 \mu},\\
    \mathcal{ F}_S&=
     f^{-2 \mu} \cdot\left( \frac{f\cdot H}{f\cdot H 
     - \frac{3  y}{4 N^3}} - 1 \right) 
     + \frac{1}{N}\frac{d}{d t}\left(\frac{ f^{-2 \mu+1}}{f\cdot H 
     - \frac{3 y }{4 N^3}}\right),\\
    \mathcal{ G}_S&=  f^{-2 \mu} \left(\frac{6 v 
    + 18 y \cdot f\cdot H \cdot N
    + x \cdot N^2 - 6 f^2 \cdot H^2 \cdot N^4}{ 2
        N^4\cdot\Big(f\cdot H 
    - \frac{3  y}{4  N^3}\Big)^2} + 3\right). \label{G_s_full}
\end{align}
\end{subequations}
We always choose the functions $f(t), \dots y(t)$ in such a way that
inequalities \eqref{stability_conditions} and \eqref{velocities} are satisfied.

%\section{Bouncing Universes}

\subsection{Bounce followed by inflation}
\label{sec:bounce_to_inflation}
In this Section, we construct a linearly stable bounce solution which evolves through
the following
stages:
\begin{itemize}
\item \textit{contraction}, with
  power-law behavior of the
  scale factor
    \item \textit{bounce} %where Hubble parameter starts to grow, then changes its sign;
    \item \textit{inflation}, with the Hubble parameter
      %and lapse 
      %function tend to
 almost constant in time
      %constant value;
\item \textit{kination},   with the Horndeski field
  reduced to a massless scalar field.
%  driven by  \textcolor{magenta}{a}
%  free massless scalar field. 
\end{itemize}
Unlike at contraction and bounce,
gravity at inflation and kination is described by conventional
GR, and the expansion
is driven by the scalar field.

To build the model, we make use of the following approach. We
design the functions $f(t)$, $x(t)$, $v(t)$, and $y(t)$ in
\eqref{A_ansatz} and  \eqref{apr22-21-1}
for the contraction, inflation,
and kination epochs separately and then construct smooth
interpolations between these epochs. One of these interpolating
stages involves
bounce, and we have to figure out the conditions for its realization. 
Needless to say, we have to ensure stability and absence of
superluminality throughout the whole evolution.
Clearly, the construction involves a lot of guesswork, some of
which is sketched in what follows. The existence of a consistent
solution is ultimately proven by a numerical example.

\subsubsection{Early times: the Universe contracts}
\label{sec:early_times}
We begin with the earliest epoch, i.e., large negative times.
We require the power-law contraction with constant lapse function,
%\textcolor{magenta}{behavior} of scalar factor. 
%As well, the Hubble parameter and the background of the lapse function should b%e as follows:
\begin{align}
    H = -\frac{\chi}{(-t)}, \quad N = 1, \quad  \chi >0, \quad t\to-\infty.
    \label{hubble_bounce}
\end{align}
Here we set $N=1$
by making
use of the ambiguity of transition from covariant to ADM
formalism pointed out in Sec.~\ref{sec:model} after
  Eq.~\eqref{F}.
As we discussed there,
this is equivalent to imposing a constraint on the functions
$A_2, \dots, A_4$, or, in other words, on $f(t), \dots,y(t)$.
We will encounter this constraint in due course, see Eq.~\eqref{mar3-21-2}.

%Without loss of generality, we set $N = 1$. Indeed, this value can be achieved 
%by the redefinition of time.
%Thus, requiring the contraction \textcolor{magenta}{behavior}, we immediately find the first 
%restriction on the model parameters, which comes from 
%\begin{equation}
%\label{restriction_1}
%    \chi>0.
%\end{equation}
The desired behavior~\eqref{hubble_bounce}
is
achieved by choosing
%\marginpar{\bf replaced subscript $o$ by $0$}
%in the same way as it was done in  \cite{Kobayashi:2016xpl}. That is why, we 
%choose the Lagrangian functions asymptotic \textcolor{magenta}{behavior} as follows
\begin{subequations}
\label{x_y_v_early}
\begin{equation}
\label{f_past}
    f =  -ct, \quad c>0,
\end{equation}
\begin{equation}
    x(t) = x_0, \quad 
    v(t)  = v_0, \quad
    y(t) = y_0,
\end{equation}
\end{subequations}
where $c$, $x_0$, $v_0$, and $y_0$ are constant parameters.
%, which numerical values 
%we give later in a concrete example.
Our next purpose is to find the complete set of constraints
on these parameters.
There are several sources of these constraints.

(i) We have to ensure that the background equations \eqref{eoms_all_substitute}
are satisfied (with $N=1$). Making use of
\eqref{hubble_bounce} and \eqref{x_y_v_early} one finds that
the background equations reduce to algebraic equations 
\begin{subequations}
  \label{mar3-21-1}
\begin{align}
         x_0 + 3v_0 - 9 y_0\cdot c \cdot\chi - 6 c^2\cdot \chi^2&= 0, \\
         x_0 + v_0 + 6 c^{2}\cdot \chi^2 - (2\mu + 1)(y_0 + 4\chi c)c &= 0.
\end{align}
\end{subequations}
    This set of equations determines the Hubble coefficient $\chi$ and also
    constrains the values of $c$, $x_0$, $v_0$, and $y_0$. The latter
    constraint is precisely the one that ensures $N=1$.
    For an appropriate root\footnote{The second root is inconsistent with
      the all-time stability of the set up.}
    of \eqref{mar3-21-1}, the constraint can be written as follows:
  %  \marginpar{\bf why this solution, not the other?}
    %\begin{subequations}
    %\label{vo_chi} 
 \begin{align}
  v_0 &= \frac{1}{192}\Big[ 48 c^2 \cdot (2\mu+1)^2 - 96 x_0 + 120 c\cdot y_0\cdot (2\mu+1) - 81 y_0^2 
  \nonumber \\
    &-\big(4c\cdot (2\mu+1) + 9y_0\big)\cdot \sqrt{3}\cdot \sqrt{48 c^2 \cdot(2\mu + 1)^2 - 64 x_0 + 27y_0^2 +
      24c \cdot y_0\cdot (2\mu +1)} \Big].
  \label{mar3-21-2}
\end{align}
%
% \begin{align}
 % v_0 &= \frac{1}{192}\Big[ 48 c^2 \cdot (2\mu+1)^2 - 96 x_0
  %  \nonumber \\
   % &- 4 c \cdot (2\mu +1) \cdot \big(-30 y_0 + \sqrt{3}\cdot \sqrt{48 c^2 \cdot (2\mu + 1)^2 
    %  - 64x_0 + 27 y_0^2 + 24 c \cdot y_0\cdot (2\mu +1)}\big)
    %\nonumber \\
    %&- 9 y_0 \cdot \big(9 y_0 
    %+ \sqrt{3}\sqrt{48 c^2 \cdot(2\mu + 1)^2 - 64 x_0 + 27y_0^2 +
    %  24c \cdot y_0\cdot (2\mu +1) }\big) \Big],
%\end{align}
 %\marginpar{\bf can this constraint be simpler written?}
 Then the Hubble coefficient is given by
\begin{align}
  \chi &= \frac{2  x_0 + 4  v_0
    - y_0 \cdot c  \cdot (2\mu+1)}{9 y_0\cdot c + 4  c^2 \cdot (2\mu +1)}.
  \label{mar3-21-3}
\end{align}
So, the first set of constraints on the parameters defining  the model
at early times
is that $v_0$ is not arbitrary but is
given by Eq.~\eqref{mar3-21-2}, it must be real
(argument of square root must be positive), and
the Hubble parameter given by \eqref{mar3-21-3} must be positive,
\begin{equation*}
%\label{restriction_1}
    \chi>0.
\end{equation*}

%\end{subequations}

%Actually, the setup \eqref{hubble_bounce} together with Friedman equations \eqr%ef{eoms} 
%lead to the following equation of state at early times: 
%\begin{equation*}
%    p = \Big(\frac{2}{3 \chi} - 1\Big) \rho.
%\end{equation*}
%Let us mark, that the contracting Universe can become inhomogeneous and anisotr%opic 
%due to the Belinsky–Lifshits–Khalatnikov phenomenon \cite{Lifshitz:1963ps}, \ci%te{Belinsky:1970ew}, 
%\cite{Belinskii:1972sg}. A way to avoid this problem is to assume that the domi%nant matter 
%(galileon, in our case) at the contracting stage has super-stiff equation of st%ate 
%\cite{Erickson:2003zm}. So, our second restriction on model parameters is $p>\r%ho$, or:
%\begin{equation}
%\label{restriction_2}
%    \chi < 1/3.
%\end{equation}

(ii) The second set of constraints comes from the 
stability 
requirement \eqref{stability_conditions} and the absence of superluminal
propagation
\eqref{velocities}.  We make use of \eqref{stability_all_subtitute} and write
%Again using background \eqref{hubble_bounce}, 
%asymptotics of Lagrangian functions \eqref{x_y_v_early}, expressions \eqref{eq:%Fs_Gs_form},  
%\eqref{stability_func} for $\mathcal{ F}_T$, $\mathcal{ F}_S$ etc, we obtain:
\begin{subequations}
\label{restriction_3}
\begin{align}
    \mathcal{ F}_T = \mathcal{ G}_T =   (-c\cdot t)^{-2\mu} ,
    \label{asy_Ft}
\end{align}
\begin{align}
   \mathcal{ F}_S =  (-c\cdot t)^{-2\mu} \cdot \frac{4c\cdot (1-2\mu )
   - 3y_0}{4 c\cdot \chi + 3y_0} ,
   \label{asy_FS}
\end{align}
\begin{align}
   \mathcal{ G}_S =  (-c\cdot t)^{-2\mu} \cdot \frac{48c^2\cdot\chi\cdot (2\mu + 1) - 16x_0 
   + 12 c\cdot y_0\cdot(2\mu + 3\chi +1)+27y_0^2}{(4 c\cdot \chi + 3y_0)^2} ,
\end{align}
\begin{align}
   c_S^2 =  \frac{(4 c \cdot \chi + 3y_0)\cdot(4c\cdot(1-2\mu)-3y_0)}{48c^2\cdot\chi\cdot(2\mu+1)-16x_0 
   +12c\cdot y_0\cdot(2\mu+3\chi+1)+27y_0^2} \; .
\end{align}
\end{subequations}
Thus, the constraints $ \mathcal{ F}_T,  \mathcal{ G}_T > 0$ are
satisfied automatically    and $c_T\equiv 1$,
while the constraints
$\mathcal{ F}_S,  \mathcal{ G}_S > 0$, $c_S^2 \leq 1$
are nontrivial (but time  independent).

(iii) One more constraint comes from the desire to get around the
no-go theorem of Refs.\cite{Libanov:2016kfc,Kobayashi:2016xpl}.
A necessary condition for having a stable bouncing solution in the
Horndeski theory with GR asymptotics as $t\to +\infty$,
is \cite{Kobayashi:2016xpl}
\begin{equation*}
%\label{no-go_integral}
       \int_{-\infty}^{t} a(t) (\mathcal{ F}_T +\mathcal{ F}_S) dt < \infty \; ,
\end{equation*}
i.e., this integral
must be convergent in
the lower limit of integration. At large negative times we have
\begin{align*}
    a &\propto (-t)^{\chi},\\
    \mathcal{ F}_T,   \mathcal{ F}_S &\propto(-t)^{-2\mu} \; ,
    %,\\
    %\mathcal{ F}_S&\propto(-t)^{-2\mu}.
\end{align*}
so the convergence of the integral requires
\begin{align}
\label{restriction_4}
    %\mu > \frac{1}{2}(1+\chi).
\chi < 2\mu -1 \; .
\end{align}

(iv) Yet another constraint is obtained by requiring that despite the fact
that   $\mathcal{ F}_T$,   $\mathcal{ F}_S$,   $\mathcal{G}_T$, and 
$\mathcal{ G}_S$ (effective Planck masses squared) tend to zero
as $t\to -\infty$, the background evolution
can be described classically at early times.
We consider this issue in  Appendix \ref{app:strong_coupl} along the lines
of Ref.~\cite{Ageeva:2020buc}. The outcome is simple:
the classical treatment of early time evolution is legitimate provided
that 
\begin{equation}
\label{restriction_5a}
    \mu<1 \; .
\end{equation}
Note that this constraint together with \eqref{restriction_4}
implies that
\begin{equation}
\chi <1 \; ,
\label{mar3-21-5}
  \end{equation}
i.e., the contraction velocity $|\dot{a}|$ increases.
%(in the time-reversed expanding Universe  this would mean
%decelerated expansion).

(v) Finally, there is a constraint that has to do with
the Belinsky--Khalatnikov--Lifshitz phenomenon
\cite{Lifshitz:1963ps, Belinsky:1970ew,Belinskii:1972sg}.
In our framework this phenomenon manifests itself in the behavior of
superhorizon tensor (and also scalar) modes in the
contracting Universe\footnote{The action \eqref{tensor2} for
  tensor perturbations (with $\mathcal{ F}_T= \mathcal{G}_T$) is
  the same as the action for tensor modes in GR in the
  background metric with $a_{eff} = a  \mathcal{G}_T^{1/2}$,
  $N_{eff} =  \mathcal{G}_T^{1/2}$. The combination
  $|a_{eff} H_{eff} = \dot{a}_{eff}/N_{eff}|$ behaves as
  $(-t)^{\chi-1}$,
  i.e., in view of \eqref{mar3-21-5}  it tends to zero as
  $t \to -\infty$ and grows as the Universe contracts.
  Thus, a mode of  sufficiently small
  conformal momentum $k$ is subhorizon at early times,
   $k/a_{eff} \gg H_{eff}$,
  and becomes superhorizon at later times. Note that
  in view of \eqref{restriction_4}
    and \eqref{restriction_5a}, the effective
    Universe  with scale factor $a_{eff} \propto (-t)^{\chi - \mu}$ is
    expanding, rather than
    contracting.  The same properties are characteristic
    of scalar perturbations as well.}: in the BKL case, one of the two
solutions for
a superhorizon mode of given conformal momentum grows as $t$ increases
and diverges in the formal limit $t \to 0$ (while  another solution
stays constant in time). This means that the Universe becomes strongly
anisotropic and inhomogeneous at late times, which is undesirable
(see, e.g., Ref.~\cite{Erickson:2003zm} for discussion). 
To avoid BKL, one makes sure that
the time-dependent superhorizon solution decays, instead of growing,
as $t$ increases  towards zero.
In our framework, the equation of motion for
superhorizon perturbation 
  is obtained from \eqref{tensor2} with
  spatial derivatives neglected,
  \[
  \frac{1}{a^3  \mathcal{G}_T} \frac{d}{dt} \left(a^3  \mathcal{G}_T
  \dot{h}_{ij} \right) = 0 \; .
  \]
  One of its solutions is constant in time, while another is
  \[
  h_{ij} \propto \int~dt~ \frac{1}{a^3  \mathcal{G}_T} \propto
  (-t)^{2\mu -3\chi+1} \, .
  \]
  It decays as $t$ increases towards zero for
  \begin{equation*}
    2\mu +1 >  3\chi \;.
  \end{equation*}
  %This is our last constraint
  %on the parameters characterizing  the early epoch.  
This constraint ensures also that the BKL phenomenon is absent for
scalar perturbations. Given that $\mu <1$, it is weaker than
\eqref{restriction_4}.

\begin{figure}[htb!]
\centering
\begin{minipage}{0.3\textwidth}
  \centering
\includegraphics[width=1.0\textwidth]{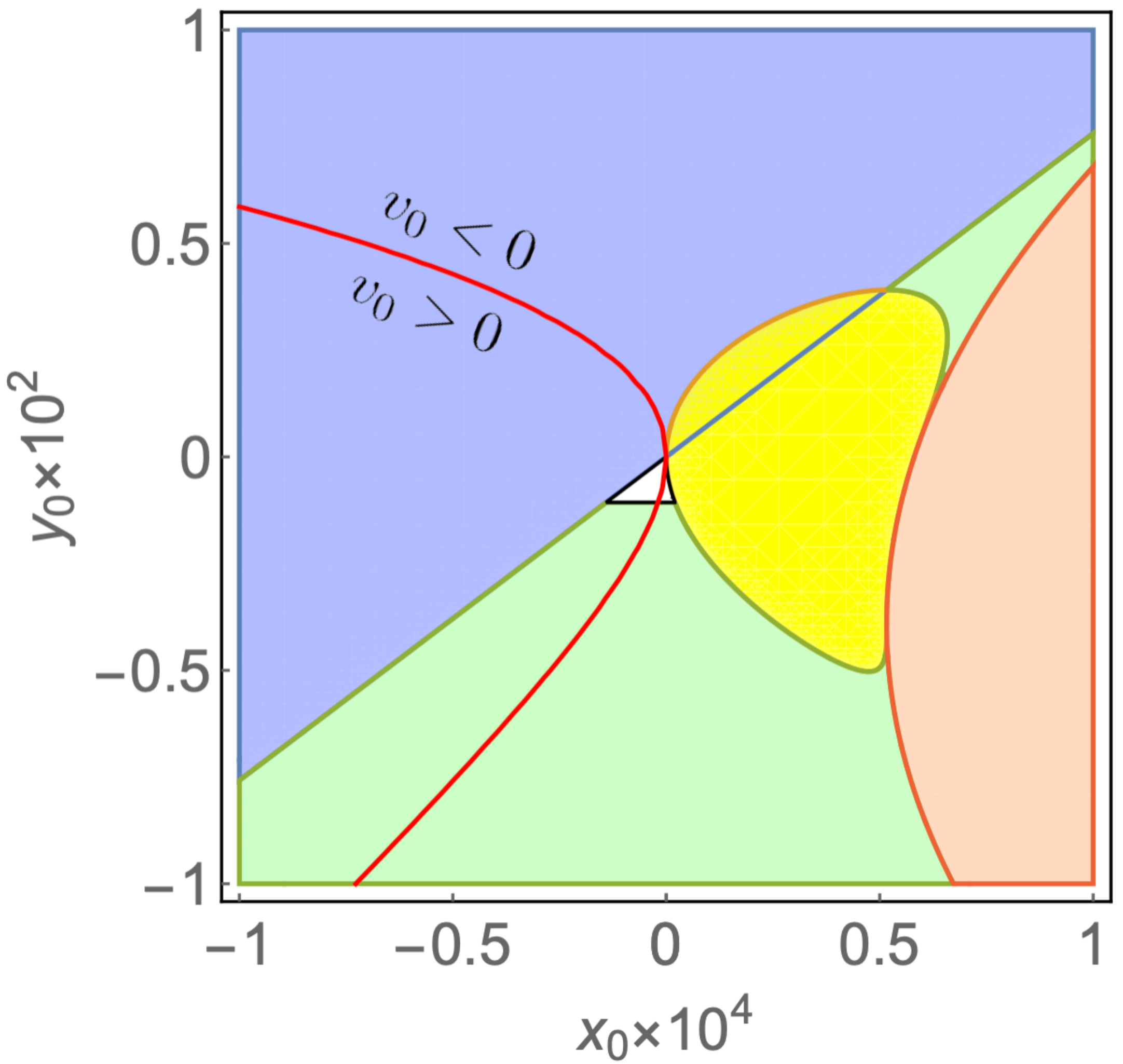}
\subcaption[first caption.]{}\label{fig:1a}
\end{minipage}%
\;\;\;\;
\begin{minipage}{0.3\textwidth}
  \centering
\includegraphics[width=1.0\textwidth]{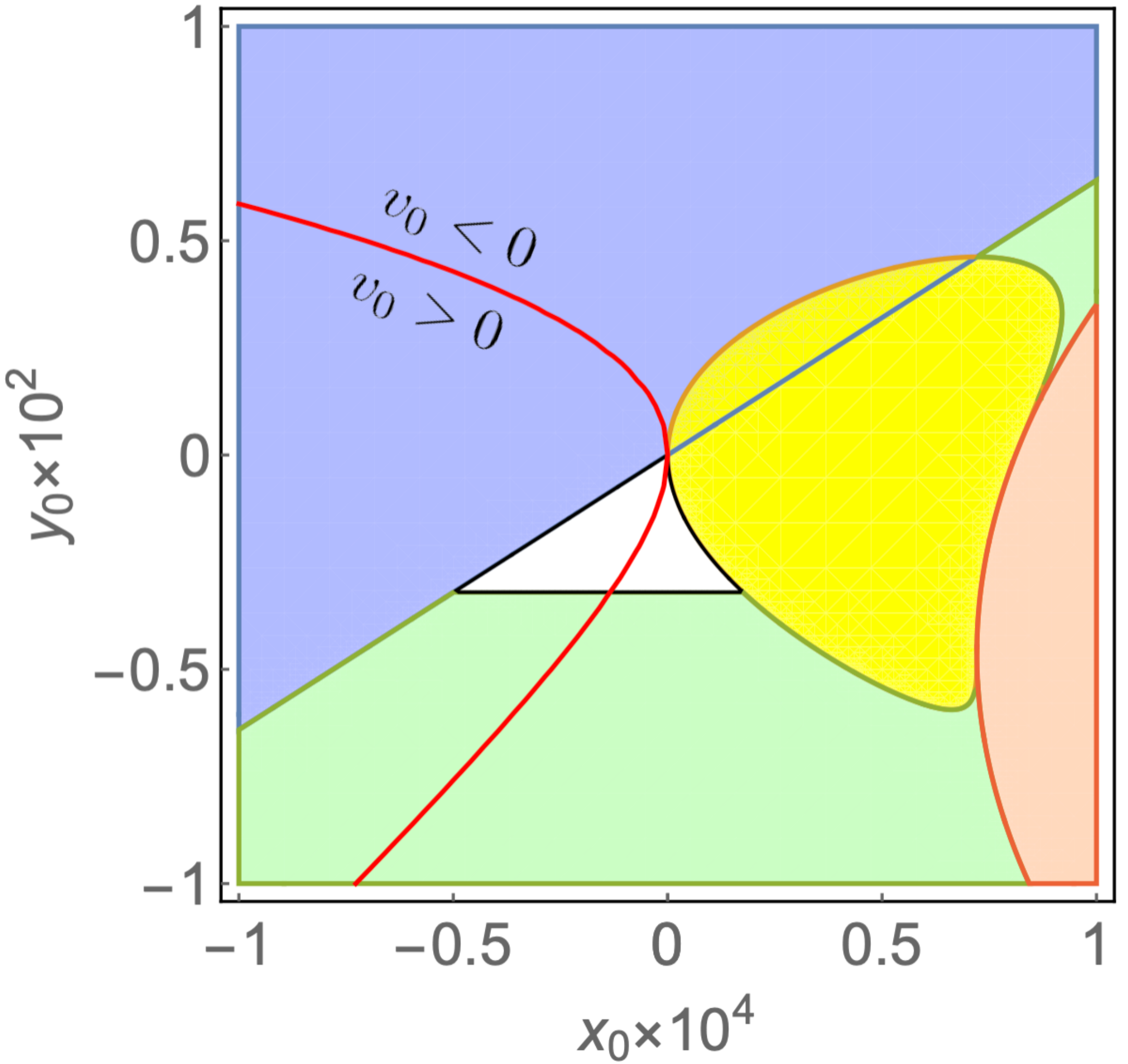}
\subcaption[second caption.]{}\label{fig:1b}
\end{minipage}%
\;\;\;\;
\begin{minipage}{0.3\textwidth}
  \centering
\includegraphics[width=1.0\textwidth]{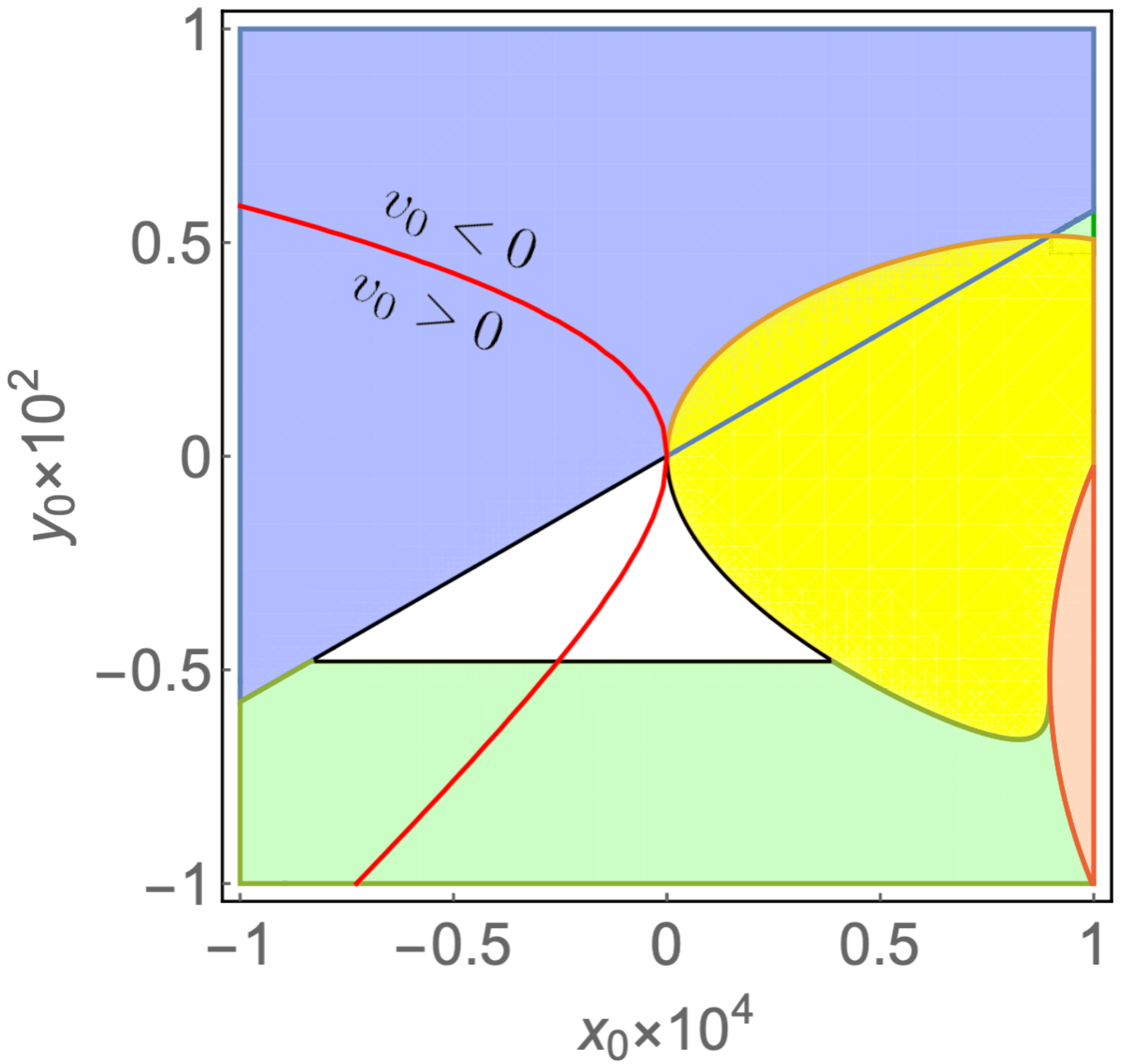}
\subcaption[third caption.]{}\label{fig:1c}
\end{minipage}

\caption{Space of parameters $x_0$ and $y_0$
  determinig the early time
  asymptotics \eqref{x_y_v_early} of the Lagrangian \eqref{A_ansatz}.
  Blue, yellow, green, and  pink patches are  regions
    forbidden by the constraints  
 $\chi>0$, $\mathcal{G}_S>0$,  $\mathcal{F}_S>0$  
    and constraint on $v_{0}$ (positivity of
    the argument of square root), respectively.
    The constraint
        coming from $c_S\leq 1$  is relevant
        as well, but it would not be visible in these figures;
        we show this constraint in Fig.~\ref{fig:region_resolution}.
        %do not draw it here for convenience and clarity of figures. } 
  Other conditions are weaker and not shown. 
 The red line corresponds to $v_0 = 0$. The regions to the right and left
 of this 
 line have $v_0 < 0$ and  $v_0 > 0$, respectively. 
 The white black-framed area shows the allowed range of  parameters $x_0$ 
 and $y_0$, where all  constraints
 of this Section are satisfied. 
 We set $\mu = 0.6$ in the left panel (Fig.~\ref{fig:1a}), 
 $\mu = 0.8$ in the central panel (Fig.~\ref{fig:1b}) and $\mu = 0.95$ 
 in the right panel (Fig.~\ref{fig:1c}); $c = 4 \cdot 10^{-3}$ everywhere.} \label{fig:region}
\end{figure}

Thus, the parameter $\mu$ in the
Lagrangian and the Hubble coefficient $\chi$
must belong to the intervals
[constraints  (i), (iii), and (iv)]
\begin{align*}
  \frac{1}{2} &< \mu < 1
  \\
  0 &< \chi < 2\mu -1 \; .
\end{align*}
To find the allowed range
  of parameters entering the Lagrangian,
we note that in accordance with Eqs.~\eqref{apr2-21-10},
\eqref{apr2-21-11}, and  \eqref{apr2-21-12}, 
%the Ansatz \eqref{A_ansatz} is invariant, modulo
%overall change ${\cal L} \to \lambda^{-2\mu} {\cal L}$, under
%rescaling $f \to \lambda f$, $a_2 \to \lambda^2 a_2$, $a_3 \to \lambda a_3$,
%i.e., $x \to \lambda^2 x$, $v \to \lambda^2 v$, $y \to \lambda y$, with
%time-independent $\lambda$.
 both equations of motion for background and
constraints coming from the absence of instability and superluminality
are invariant under rescaling $c \to \lambda c$,
 $x_0 \to \lambda^2 x_0$, $v_0 \to \lambda^2 v_0$, $y_0 \to \lambda y_0$.
This is explicit in \eqref{mar3-21-2}, \eqref{mar3-21-3} and
\eqref{restriction_3}.
%\textcolor{red}{
%  Furthermore,
The parameter
$v_0$ is given by Eq.~\eqref{mar3-21-2} in terms
of other parameters. So, it is sufficient to determine the
allowed range of $x_0$, $y_0$ for a given value of $\mu$ and
one value of $c$.
With the 
reduced Planck mass set equal to 1,
$c$ is roughly the
inverse characteristic time scale in Planck units,
so
  it should be small.\footnote{We could equally well
  set $c=1$ and after performing the whole analysis make use of
  the scaling properties \eqref{apr2-21-10},
  \eqref{apr2-21-11}, and  \eqref{apr2-21-12} to  obtain
  a model with
  the time scale of evolution much longer than the Planck time.
  It is worth keeping
    this in mind, but we take more intuitively transparent
   approach here.}
In our numerical example below we set
$c = 4 \cdot 10^{-3}$,
and here we stick to this choice. The allowed
ranges of
$x_0$ and $y_0$ for several  values of $\mu$ are  shown in Fig.~\ref{fig:region}.
%\textcolor{blue}{
In fact, the allowed range is not empty
 in the entire interval
   $1/2 < \mu < 1$;
%    \marginpar{\bf Check and insert correct number}
this is illustrated in  Fig.~\ref{fig:region}.

\begin{figure}[htb!]
\centering
\includegraphics[width=0.36\textwidth]{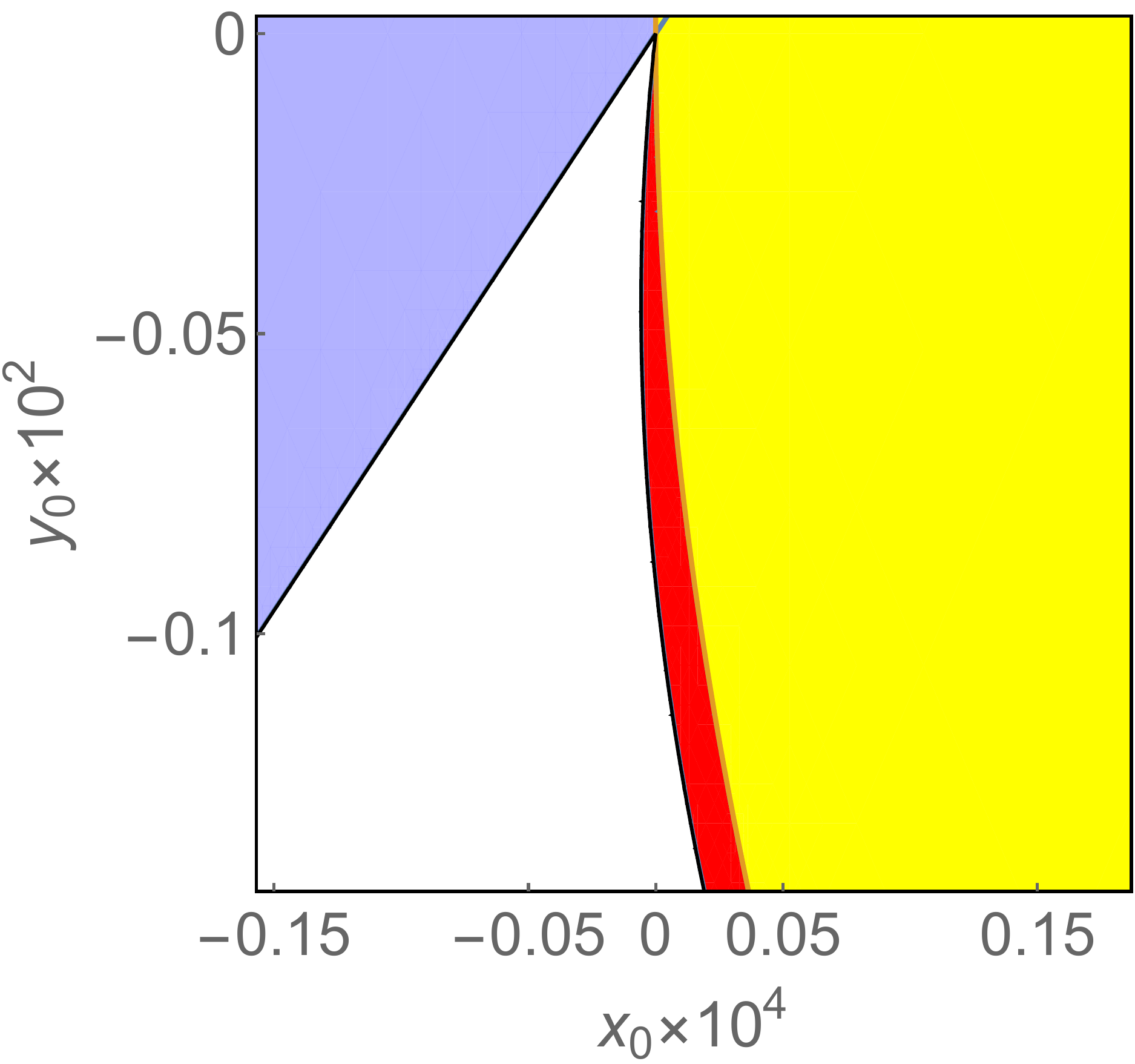}
\caption{ Part  of Fig.~\ref{fig:1b}
    with the constraint $c_S \leq 1$ added.
    Notice the  scales of the axes.
    Blue
      and yellow patches are again regions forbidden by the constraints
      $\chi>0$ and $\mathcal{G}_S>0$, respectively.
    The red patch is the new region 
    forbidden
    by the constraint $c_S \leq 1$:
    %\textcolor{violet}{Notice the  scales} of the axes.
    %\textcolor{violet}{Blue
    %  and yellow patches are again regions forbidden} by the constraints
    %$\chi>0$ and $\mathcal{G}_S>0$, respectively.
     the
   allowed (white)
    region is somewhat squeezed by the constraint $c_S \leq 1$.
  The constraint  $\mathcal{F}_S>0$ is not shown.}
%Here we again show the space of parameters $x_0$ and $y_0$
%  determining the early time
%  asymptotics \eqref{x_y_v_early} of the Lagrangian \eqref{A_ansatz}.
% Yellow black-framed area shows the allowed range of  parameters $x_0$ 
% and $y_0$, where all conditions of this section are satisfied. 
% We set 
% $\mu = 0.8$ and $c = 4 \cdot 10^{-3}$ here.}
\label{fig:region_resolution}
\end{figure}
It is
useful to note
that the asymptotics of $v(t)$ and $x(t)$
may be chosen in such a way that
\begin{equation}
\label{restriction_6}
    v_0>0, \quad x_0<0 \; .
\end{equation}
This is possible for all allowed values of $\mu$.
%  \begin{equation*}
%  \frac{1}{2} < \mu < 1.
%   \end{equation*}
 In what follows we consider this case  only.
%\marginpar{Check and insert correct number}

%\marginpar{Fig. to be added}

%{\bf FIGURE. Plane $(x_0, y_0)$ \textcolor{red}{(a) for
   % small $\mu$=''0.55'',
  %where allowed region is small; (b) $\mu=0.8$; (c) $\mu=0.95$}  
  %with regions allowed by (with $c=0.004$)

  %(1) positivity of square root in \eqref{mar3-21-2};

  %(2) constraint \eqref{restriction_1}

  %(3) constraint ${\cal G}_S > 0$

  %(4) constraint ${\cal F}_S > 0$

  %(5) constraint $c_S^2 \leq 1$

  %(6) constraint \eqref{restriction_4}

  %in different colors with overlap region in black. \textcolor{blue}{Also:
  %  show line $v_0 = 0$
  %and indicate where %$v_0 >0$.}
  %}

\subsubsection{Inflation after bounce}
\label{sec:infl-bounce}
As outlined in the beginning of this Section, our next step is to
describe the inflationary stage, and then discuss the transition from contraction
to inflation through bounce. Models of inflation in the  Horndeski theory
have been proposed in Refs.~\cite{ Kobayashi:2010cm,Kobayashi:2011nu,Pirtskhalava:2014esa,Kobayashi:2015gga,Hirano:2016gmv,Tahara:2020fmn}; 
here we employ the
construction similar to 
%Now, we want the smooth transition between contraction phase to the inflation. 
%This transition can be done by using the same idea, which was presented in
Ref.~\cite{Kobayashi:2016xpl}. Namely, exact exponential expansion occurs
when the
functions in the Lagrangian take constant values,
\begin{subequations}
\label{lagr_func_bounce_to_infl}
\begin{align}
    f &= 1, \label{f_future} \\
     x = x_1, \quad
    v &= v_1, \quad
     y = y_1,
\end{align}
\end{subequations}
where the choice \eqref{f_future}
is made to restore GR already at inflation (recall that $G_4 = -A_4 = f/2$).
 With the \textit{Ansatz}
\eqref{lagr_func_bounce_to_infl}, equations of motion
\eqref{eoms_all_substitute} read
%One can substitute \eqref{lagr_func_bounce_to_infl} into \eqref{eoms_all_substitute} 
%and it leads to the following equations of motion:
\begin{subequations}
\label{eoms_all_substitute_bounce_to_infl}
\begin{align}
  &\left(-\frac{x_1}{N^2} - \frac{3 v_1}{N^4}\right) -
  \frac{9 y_1\cdot H}{N^3} 
    + 6 H^2  = 0, \\
    &\left(\frac{x_1}{N^2} + \frac{v_1}{N^4}\right) + 6 H^2 = 0\; , 
    %- 
    %\frac{1}{N}\frac{d}{dt}\left( \frac{y_1}{N^3} - 4 H\right) = 0,
\end{align}
\end{subequations}
and we denote the (time-independent) solution to these equations by
$H = H_1$ and $N=N_1$. We require $H_1 > 0$, $N_1 >0$. 
%are constants and the solution of the above EoM. 
%It means that at these times the inflation takes place.  However, it is 
%not guaranteed at all, that the requirement \eqref{lagr_func_bounce_to_infl} le%ads to 
%the bounce at some moment of time, 
%since there may be not exist the solution, which connects the demanded asymptot%ics  
%\eqref{hubble_bounce} and ($H_1,N_1$). So, for each set of our parameters 
%one should check the existence of such a solution numerically.

Now, it is convenient to consider $H_1$, $N_1$, and $y_1$
as independent parameters, and express  $x_1$ and $v_1$ through
these parameters  using
\eqref{eoms_all_substitute_bounce_to_infl}:
\begin{subequations}
\label{x1_v1}
\begin{align}
    x_1 &= -\frac{3(8 H^2_1 \cdot N_1^3 - 3 H_1\cdot y_1)}{2 N_1}, \\
    v_1 &= \frac{3}{2}\big(4 H_1^2\cdot N_1^4 - 3  H_1 \cdot N_1 \cdot y_1\big)
\end{align}
\end{subequations}
Let us turn to the requirements of
%Our first constraint comes from the fact that we want to obtain the 
%inflation in the future, i.e.
%\begin{equation}
%\label{H1_N1}
%    H_1 > 0, \quad N_1 > 0.
%\end{equation}
%The second requirement is the necessity of
background stability 
and subluminal propagation of perturbations. Using
\eqref{stability_all_subtitute} 
and \eqref{x1_v1}, we arrive at the constraints
\begin{align*}
     \frac{3y_1}{4 H_1 \cdot N^3_1 - 3 y_1}  >0, 
\end{align*}
\begin{align*}
     64 H_1^2\cdot N_1^6 - 36 y_1\cdot H_1 \cdot N^3_1 + 9y_1^2 >0, 
\end{align*}
\begin{align*}
     \frac{y_1\cdot(4 H_1\cdot N_1^3 - 3y_1)}{64 H_1^2\cdot N_1^6 
     - 36 y_1\cdot H_1 \cdot N^3_1 + 9y_1^2}<1.
\end{align*}
%This system, together with \eqref{H1_N1}, leads to the 
These constraints can be written in a simple form:
\begin{subequations}
\label{restriction_7}
\begin{align}
    %H_1&>0, \quad N_1>0,\\
    y_1&>0, \label{y1}\\
    3y_1 &<4 H_1 \cdot N_1^3 \; ,
    \label{restriction_7a}
\end{align}
\end{subequations}
which also leads to 
\begin{equation}
\label{x1_v1_conditions}
    v_1>0, \quad x_1<0.
\end{equation}
It is worth noting that the value of $y$ at inflation has the opposite sign to
its value at contraction: $y_1 >0$ [see Eq.~\eqref{y1}] and $y_0<0$ (see Fig.~\ref{fig:region}),
respectively. On the contrary, the values of $x$ and $v$ have the same signs
at these two stages, see Eqs.~\eqref{restriction_6} and
\eqref{x1_v1_conditions}.

Now, let us comment on the possible behavior of the functions $x(t)$,
$v(t)$, $y(t)$ and $f(t)$ near
the bounce. Since $x(t)$ and $v(t)$ do not change signs during
the transition from
contraction to inflation, it is natural to take them monotonously
changing from $x_0$ to $x_1$ and from $v_0$ to $v_1$, respectively.
The function $f(t)$ flattens out, with $\dot{f} < 0$ both at
contraction and bounce. Near the bounce, at $t \approx t_b$, we have
$H(t)\approx 0$. Assuming that
the
functions $x(t)<0$ and $v(t)>0$  vary
slowly in comparison with $f(t)$ in the 
vicinity of the bounce,
%self-consistent, since this function tends to $+\infty$ at $t\to-\infty$ \eqref%{f_past},
%and tends monotonically to constant value \eqref{f_future} in future. 
%Next, we rewrite EoM \eqref{eoms_all_substitute} near the 
%bounce point $t_b$, where $H(t_b)\approx0$.
we obtain from the first 
equation of motion, Eq.~\eqref{eom_1} with $H(t_b)=0$, that
\begin{equation}
    N(t_b) \approx \sqrt{-\frac{3 v(t_b)}{x(t_b)} }.
\label{mar13-21-2}
\end{equation}
We note in passing
that with our choice of signs of $x$ and $v$ at contraction,
inflation and bounce  ($x(t_b) <0$, $v(t_b)>0$),
the argument of square root is positive.
Then, we take the time derivative of  Eq.~\eqref{eom_1} and solve it
together with
Eq.~\eqref{eom_2} for
$\dot{H}$ and $\dot{N}$ 
to find 
\begin{align*}
    \dot{H}(t_b)\approx\frac{2 v(t_b) \cdot x(t_b)\cdot \Big(-2 v(t_b) + \dot{f}(t_b)\cdot y(t_b)\cdot\big(2\mu+1 \big)\Big)}{3\sqrt{3}\sqrt{-\frac{v(t_b)}{x(t_b)}}\cdot f^2(t_b)\cdot\big(8v^2(t_b) - 3x(t_b)\cdot y^2(t_b)\big)},
\end{align*}
where we again neglect $\dot{x}(t)$, $\dot{v}(t)$, and $\dot{y}(t)$ in 
comparison with $\dot{f}(t)$. Since  $\dot{H}(t_b)>0$,
we see that $y(t_b)$ may be negative, but only
slightly: 
\begin{equation*}
    y(t_b)> \frac{2 v(t_b)}{\dot{f}(t_b)\cdot (2\mu+1)} \; .
\end{equation*}
With this qualification, most of the smooth functions interpolating
between \eqref{x_y_v_early} and \eqref{lagr_func_bounce_to_infl} indeed
give rise to the bouncing solution. We present a numerical example in
Sec.~\ref{sec:numerical_ex}.

\subsubsection{Kination epoch after transit from inflation}
\label{sec:kination}
To describe the final  kination epoch with GR and
free massless scalar field,
we make use of the
covariant formalism with the Lagrangian \eqref{Hor_L}.
It is convenient to use the freedom of field redefinition
and choose the background field $\phi$ as follows:
    \begin{equation}
        \text{e}^{\phi} = t, \;\;\;\;\;\; t \to + \infty \; .
        \label{phi}
    \end{equation}  
This choice corresponds to the Lagrangian
    \begin{equation*}
       \mathcal{L} = \frac{2}{3} X \; .
    \end{equation*}
    Indeed, it is straightforward to check that the scalar field
    equation and Friedmann equation have the solution \eqref{phi}
    with  $a =\mbox{const}\cdot t^{1/3}$,
    $N=\mbox{const}$, and
      $H=(3tN)^{-1}$.
        Note 
    that during the transition from inflation
    to kination, the coefficient of $X$ in the Lagrangian changes sign
    (it changes from $x_1 < 0$ at inflation to $2/3$ at kination). This is in
    complete accordance with
\cite{ArmendarizPicon:1999rj,BazrafshanMoghaddam:2016tdk}. 

Towards the kination epoch, 
other terms in the scalar field Lagrangian  should
tend sufficiently rapidly to zero.
This can be achieved, e.g.,
by requiring
the following asymptotics of functions in the covariant
Lagrangian \eqref{Hor_L} at large $\phi$ (distant future;
recall that GR is restored already at inflation, $G_4=1/2$): 
\begin{subequations}
\label{covar_lagr_distant_times}
\begin{align}
 G_2(\phi, X) &=   \frac{2}{3}  X  + \omega_2(\phi) \cdot X^2,
\end{align}
\begin{align}
   G_3(\phi, X) = \omega_3(\phi)\cdot X , 
\end{align}
\end{subequations}
where  $\omega_2(\phi)$ and $\omega_3(\phi)$ are
 damping factors,
 which  suppress higher order terms.
 They  can
 be chosen rather arbitrarily. We choose them
as follows:
\begin{equation*}
   \omega_2(\phi) = 4(v_2\cdot \text{e}^{-\phi} - y_2\cdot \text{e}^{-2\phi}),
  \quad \quad
%\end{equation*}
%and
%\begin{equation*}
    \omega_3(\phi) = 3y_2\cdot \text{e}^{-2\phi} \; .
\end{equation*}
 The reason for this choice
    is that
  we obtain simple ADM functions $A_2$ and $A_3$
  using conversion formulas
\eqref{ADM-trans2}--\eqref{F}: 
%we immediately obtain}
	\begin{align*}
	&A_2 = \frac{1}{3t^2\cdot N^2} + \frac{v_2}{t^5\cdot N^4}\text{,} \\ 
	&A_3 = \frac{y_2}{t^5\cdot N^3} \text{.}
        %\\
	%&A_4 = -\frac{1}{2} \text{.}
	\end{align*}
In fact, we have to generalize 
these expressions by introducing time shift, $t \to t-t_*$, where $t_*$
is a new parameter. The point is that once contraction
is described literally by formulas of Sec.~\ref{sec:early_times},
inflation
begins soon after $t=0$ and ends at some much
later time $t_{e}$. The time shift
(with $t_*$ of order of $t_e$ but
  somewhat smaller than $t_{e}$)
 has to be introduced to account for the fact that kination begins around
 $t_e \neq 0$.
% accomodate the
%latter property. 
Thus, the 
asymptotic behavior of $x(t)$, $v(t)$, and $y(t)$ 
at large $t$ is (recall that $f(t)=1$ at inflation and later)
\begin{subequations}
  \label{mar13-21-1}
\begin{align}
    x(t)&\to \frac{2}{3(t-t_*)^2},\\
    v(t)&\to \frac{v_{2}}{(t-t_*)^5},\\
    y(t)&\to \frac{y_{2}}{(t-t_*)^5}.
\end{align}
\end{subequations}
%\begin{subequations}
%\label{x_v_y_kination}
%\begin{align}
%v_2 &= \frac{1}{2}\cdot\left(\omega_2 + \frac{2}{3}\omega_3\right), \label{v_3}%\\
%y_{2} &=\frac{2}{3}\cdot\omega_3 \label{y_3},
%\end{align}
%\end{subequations}
%We will give concrete values 
%of $\omega_2$ and $\omega_3$ later when we turn to numerical example.
By choosing the functions $x(t)$, $v(t)$, and $y(t)$ in such a way that
they interpolate between constant values $x_1$, $v_1$, and $y_1$
at inflation and functions \eqref{mar13-21-1} at late times, we obtain
a smooth transition from inflation to kination. The issues of
stability and subluminality are, however, tricky at transition epochs;
designing a completely stable and subluminal model requires considerable
trial and error effort.

\subsubsection{Numerical example}
\label{sec:numerical_ex}
Here we present a concrete model which proves by example
that there exists stable and subluminal cosmology
with the desired properties listed in the beginning of this Section.
We emphasize again
that by rescaling the functions in the Lagrangian
in accordance with Eq.~\eqref{apr2-21-10}, one can make all time scales like
$c^{-1}$, inverse inflationary Hubble parameter $H_1^{-1}$, etc.,
arbitrarily long, much longer than the Planck time. This observation
applies also to
other models considered in this paper.

We choose the parameter
  $\mu$ near the center of allowed interval
$1/2 < \mu < 1$:
\[
\mu = 0.8 \; .
\]
As we already mentioned, at the contracting stage we choose, 
quite
arbitrarily,
\begin{equation}
c = 4\cdot 10^{-3} \; .
\label{mar16-21-1}
\end{equation}
The parameters $x_0$ and $y_0$ are then chosen from the
allowed region shown in   Fig.~\ref{fig:1b}; by trial and error we
find convenient values, consistent with the sign choice \eqref{restriction_6}:
\begin{equation}
%\begin{subequations}
\label{set_1}
    %\begin{align}
        x_0 = - 1.6 \cdot 10^{-5},
        %\frac{100}{(2500)^2}, 
        \quad
        y_0 = -1.2 \cdot 10^{-3} \; .
\end{equation}
The value of $v_0$ and the Hubble coefficient $\chi$ at the contraction stage
are found from \eqref{mar3-21-2} and  \eqref{mar3-21-3}:
\begin{equation}
   v_0 = 5.19\cdot 10^{-6} \; , \quad \chi = 0.01 \; .
  \label{mar16-21-2}
  \end{equation}
  
This completes the  description
of the contraction stage.

We would like to have bounce at some time before $t=0$;  we
  request [although
we do not have to do so in view of scaling \eqref{apr2-21-10}] that the
characteristic time scales are large compared to 1 (i.e., Planck time).
We begin with the function $f(t)$ which should interpolate between
$f = -ct$ at contraction and $f=1$ at inflation. A simple choice is
\begin{equation}
  f(t)
  = \frac{c}{2}\Big[-t+\frac{\text{ln}(2\text{cosh}(st))}{s}\Big] + 1 \; .
  \label{march13-21-5}
\end{equation}
The parameter $s$ is the inverse time scale
of the transition, and we set
\begin{equation}
  s = 2\cdot 10^{-3} \; .
  \label{march13-21-11}
\end{equation}
We wish the bounce to occur roughly at $t \sim - s^{-1}$, so the maximum
value of $|H|$ at contraction is estimated as
\[
|H|_{max} \sim \chi\cdot s \sim 2 \cdot 10^{-5} \; .
\]

Let us now turn to the  inflationary stage and transition to it.
A simple \textit{Ansatz} for the inflationary Hubble parameter $H_1$ is that
it is comparable to  the maximum
value  $|H|_{max}$ at contraction. There is no reason to
think that the lapse function at inflation is considerably different
from 1.
%\st{(note also that with $v_b/x_b \sim v_0/x_0$, the estimate
%\eqref{mar13-21-2}
%gives $N(t_b) \sim 1$)}.
We choose
\begin{equation*}
H_1 = 3.7\cdot 10^{-5} \; , \quad N_1 = 0.82\; .
\end{equation*}
We also have to specify the value of $y_1$ at inflation.
To this end, we  introduce
a simple \textit{Ansatz} of proportionality between
%for the behavior of
$x(t)$ and $v(t)$:
\begin{equation}
    v(t) = x(t)\frac{v_0}{x_0}\; ,
\label{mar13-21-10}
\end{equation}
so that $v(t)/x(t)$ is time independent at the transition from
contraction to inflation.  Note that
with the numerical values
\eqref{set_1} and \eqref{mar16-21-2}, the estimate \eqref{mar13-21-2}
is consistent with our choice
$N\sim 1$. Equation \eqref{mar13-21-10}
implies $v_1/x_1 = v_0/x_0$,
then \eqref{x1_v1}
gives
 \begin{equation*}
    y_1 = \frac{4 x_0 H_1^2 N_1^5 + 8v_0H_1^2N_1^3}{3x_0H_1 N_1^2 + 3 v_0 H_1};
 \end{equation*}
and numerically
\begin{equation*}
        y_1 = 1.2\cdot 10^{-6}\; .
\end{equation*}
This set of parameters is consistent with the stability and
subluminality constraints \eqref{restriction_7}.
The values of $x_1$ and $v_1$ are found from  \eqref{x1_v1}:
\begin{equation*}
  x_1  =  -1.07\cdot 10^{-8}
        %\frac{x_0}{1500}=  -\frac{1}{15\cdot 2500^2}
        , \quad
        v_1  =   3.47\cdot 10^{-9} \; . %0.022
\end{equation*}
Note that $|x_1|$, $v_1$, and $y_1$ are much smaller than
$|x_0|$, $v_0$, and $|y_0|$, respectively.
%\marginpar{\bf correct}
%\textcolor{red}{
  This has to do with two properties of the contracting
  stage which distinguish it from inflation. First, the constraints shown in 
  Fig.~\ref{fig:1b} are consistent with fairly large values of $|y_0|$,
  and it is indeed quite large in our example; on the contrary,
  inflationary
  $y_1$ is bounded by $H_1$, see \eqref{restriction_7a}. Second,
  equation of motion
  %there is a term in the second equation of motion,
  \eqref{eom_2} contains a term
  proportional to $\dot{f} y_0=-cy_0$, which is not so
  small at contraction
  and drives $|x_0|$ and $v_0$ to fairly large values; this term vanishes at
  inflation. We note in passing that Eq.~\eqref{eom_1} is satisfied
  at the contraction stage
  due to the partial cancellation between $x_0$ and $3v_0$.

The transition from contraction through bounce to inflation is described
by $x(t)$, $v(t)$, and $y(t)$ smoothly interpolating between
$x_0$, $v_0$, $y_0$ and $x_1$, $v_1$, $y_1$ [and with
$f(t)$ given by \eqref{march13-21-5}]. A nontrivial part of the construction
is to make sure that stability and subluminality conditions
\eqref{stability_conditions} and \eqref{velocities}
are satisfied. By trial and error, we find  appropriate forms
\begin{subequations}
  \label{mar15-21-1}
\begin{align}
  x(t) &= x_0 (1 - U_x (t)) + x_1 U_x (t) \; ,
  \\
  y(t) &=  y_0 (1 - U_y (t)) + y_1 U_y (t) \; ,
\end{align}
\end{subequations}
where the functions
\begin{subequations}
  \label{mar16-21-5}
\begin{align}
  U_x(t) & =
  \text{ln}\Big(\frac{\text{e}^{-1.5\cdot s\cdot(t-80)}+
    \text{e}^2}{\text{e}^{-1.5\cdot s\cdot(t-80)}+\text{e}}\Big)
  \\
  U_y(t) &= \text{ln}\Big(\frac{\text{e}^{-3.8\cdot s\cdot(t+180)}
    +\text{e}^2}{\text{e}^{-3.8\cdot s\cdot(t+180)}+\text{e}}\Big) 
\end{align}
\end{subequations}
interpolate between 0 and 1,
while $v(t)$ is given by \eqref{mar13-21-10}, and the parameter $s$ is the
same as in \eqref{march13-21-11}.

\begin{figure}[htb!]
\centering
\begin{minipage}{0.5\textwidth}
  \centering
\includegraphics[width=1.0\textwidth]{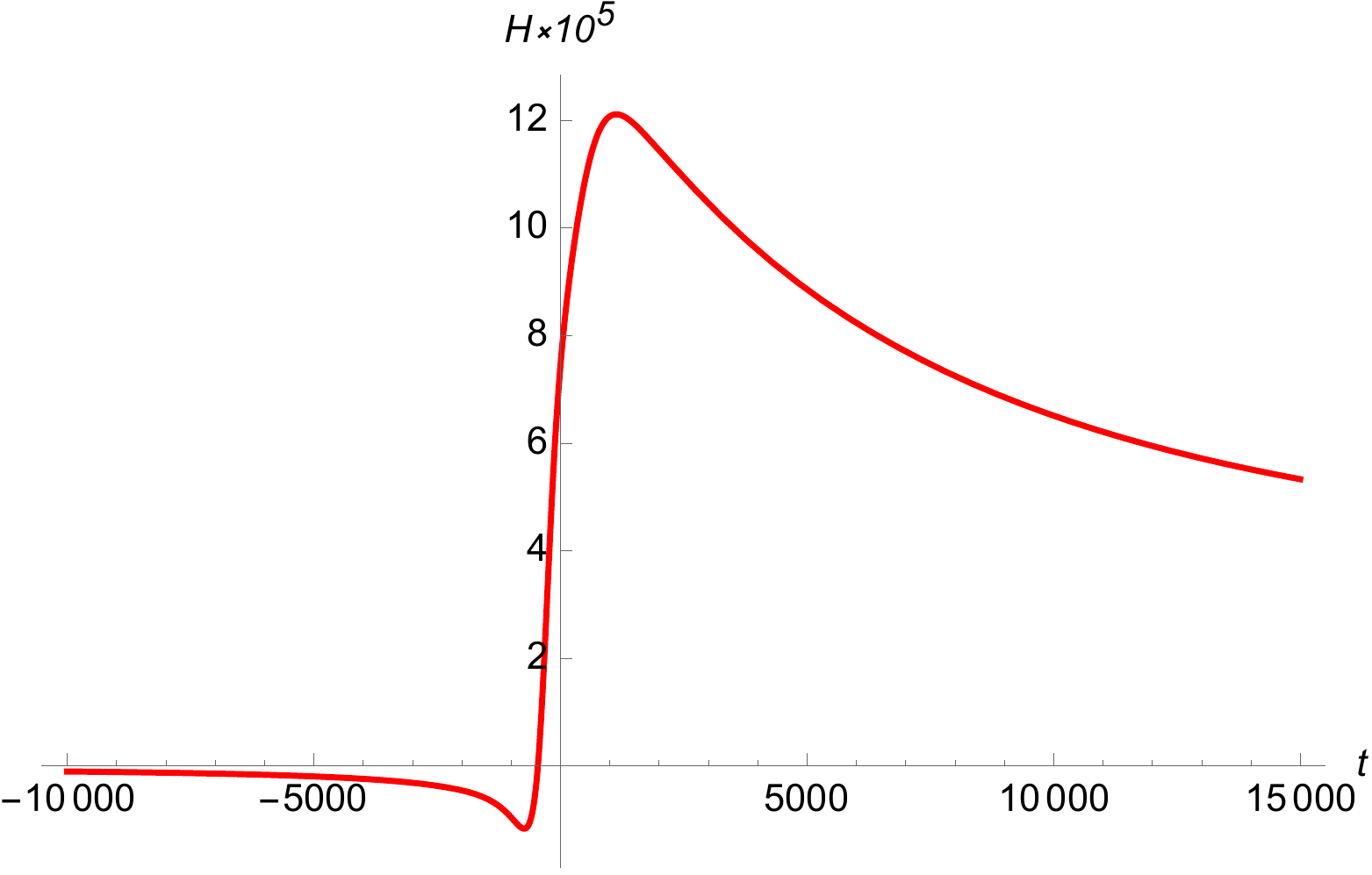}
\subcaption[first caption.]{}\label{fig:bounce_hubble1_1a}
\end{minipage}%
\begin{minipage}{0.5\textwidth}
  \centering
\includegraphics[width=1.0\textwidth]{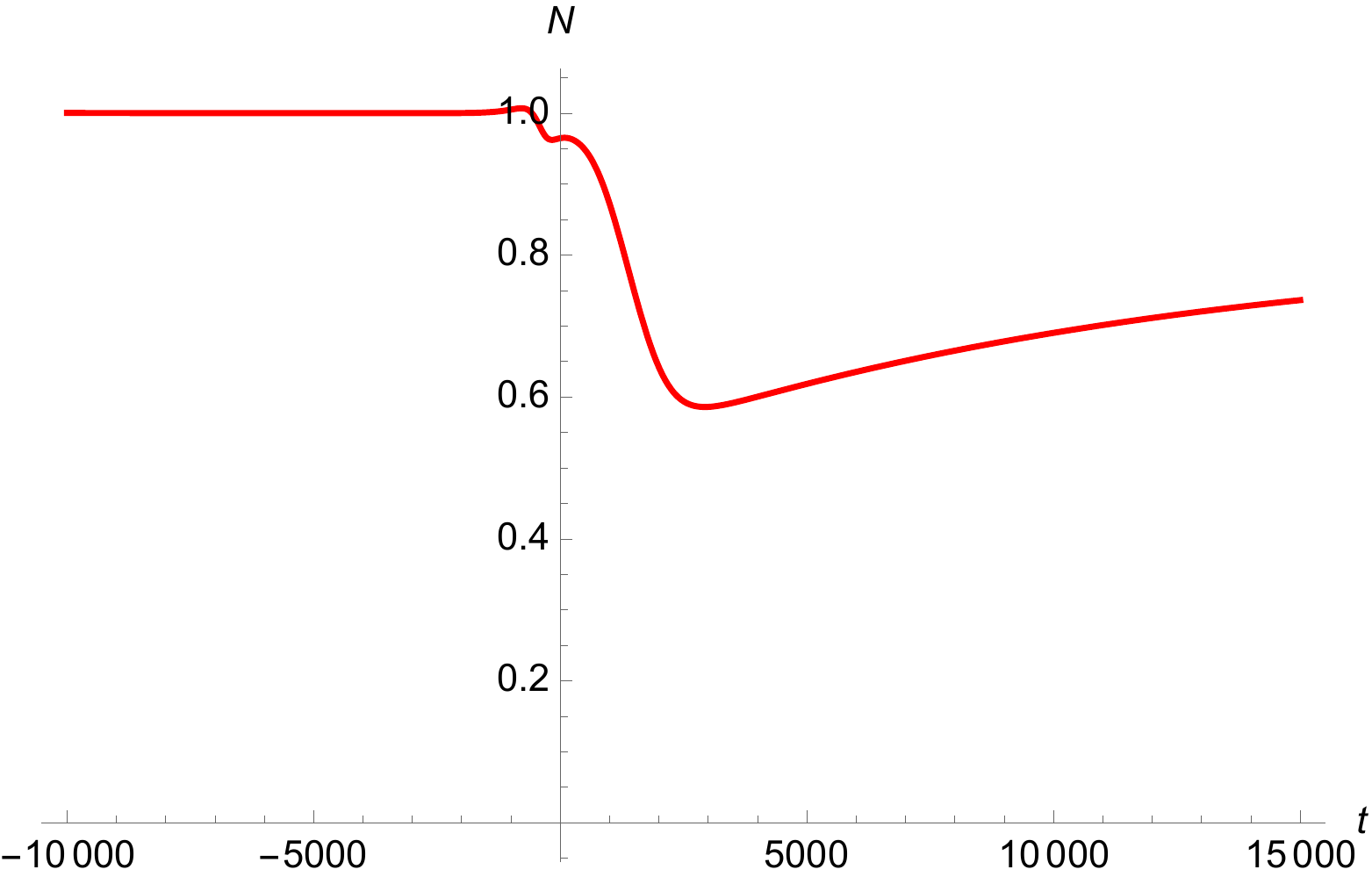}
\subcaption[second caption.]{}\label{fig:bounce_lapse1_1b}
\end{minipage}%

\caption{ Hubble parameter (left panel) 
and lapse function (right panel) 
for the model of Sec.~\ref{sec:numerical_ex}  at contraction,
bounce, and beginning of inflation.} 
\label{fig:bounce_H_N_1}
\end{figure}

\begin{figure}[htb!]
\centering
\begin{minipage}{0.5\textwidth}
  \centering
\includegraphics[width=0.95\textwidth]{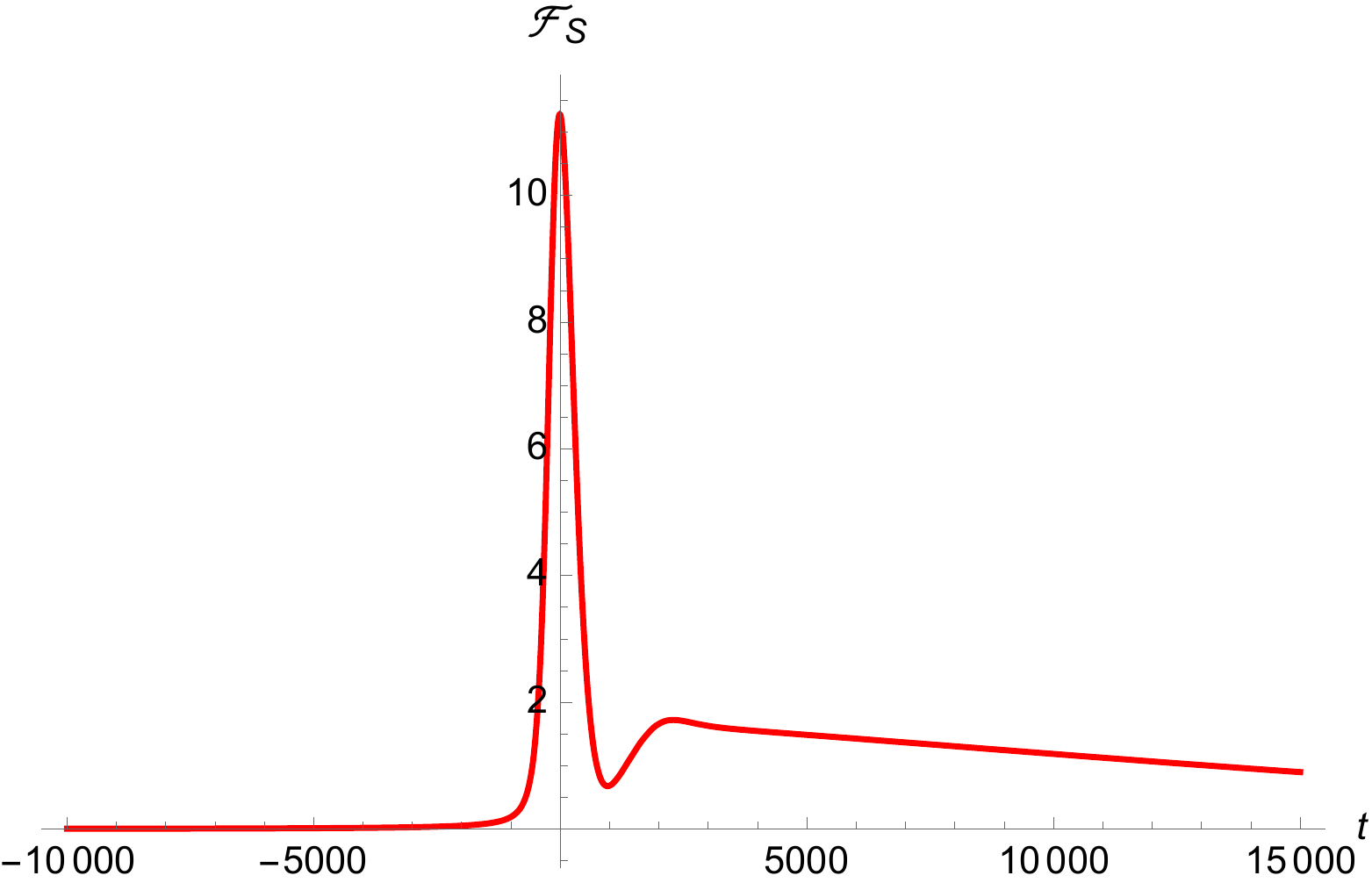}
\subcaption[first caption.]{}\label{fig:bounce_fs1b}
\end{minipage}%
\begin{minipage}{0.5\textwidth}
  \centering
\includegraphics[width=0.95\textwidth]{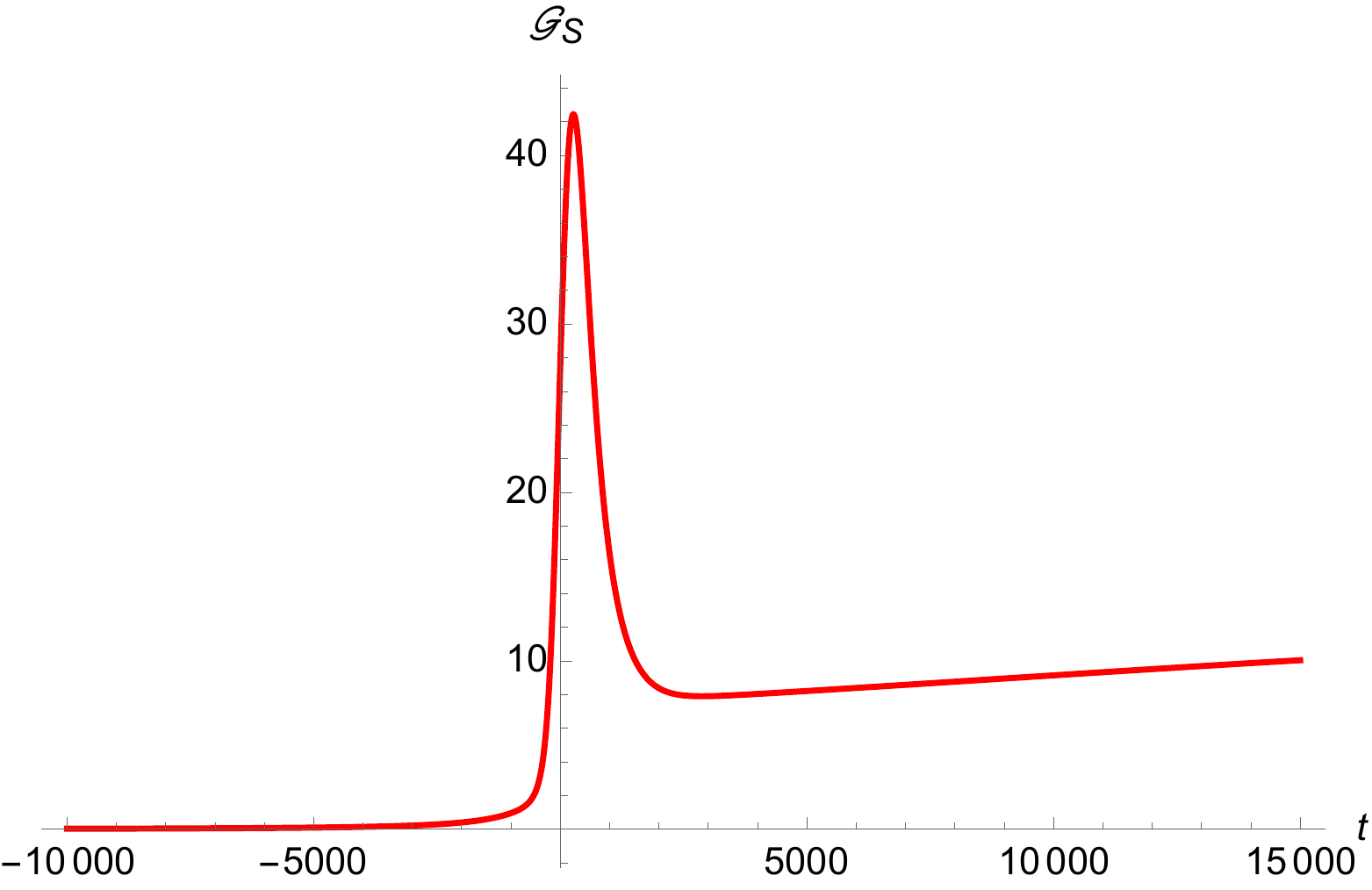}
\subcaption[second caption.]{}\label{fig:bounce_gs1c}
\end{minipage}%

\caption{ Coefficients $\mathcal{F}_S$ (left 
panel) and $\mathcal{G}_S$ (right
panel)  for the model of
Sec.~\ref{sec:numerical_ex}  at
contraction, bounce, and beginning of inflation. } 
\label{fig:bounce_FS_GS}
\end{figure}

\begin{figure}[htb!]
\centering
\begin{minipage}{0.5\textwidth}
  \centering
\includegraphics[width=0.95\textwidth]{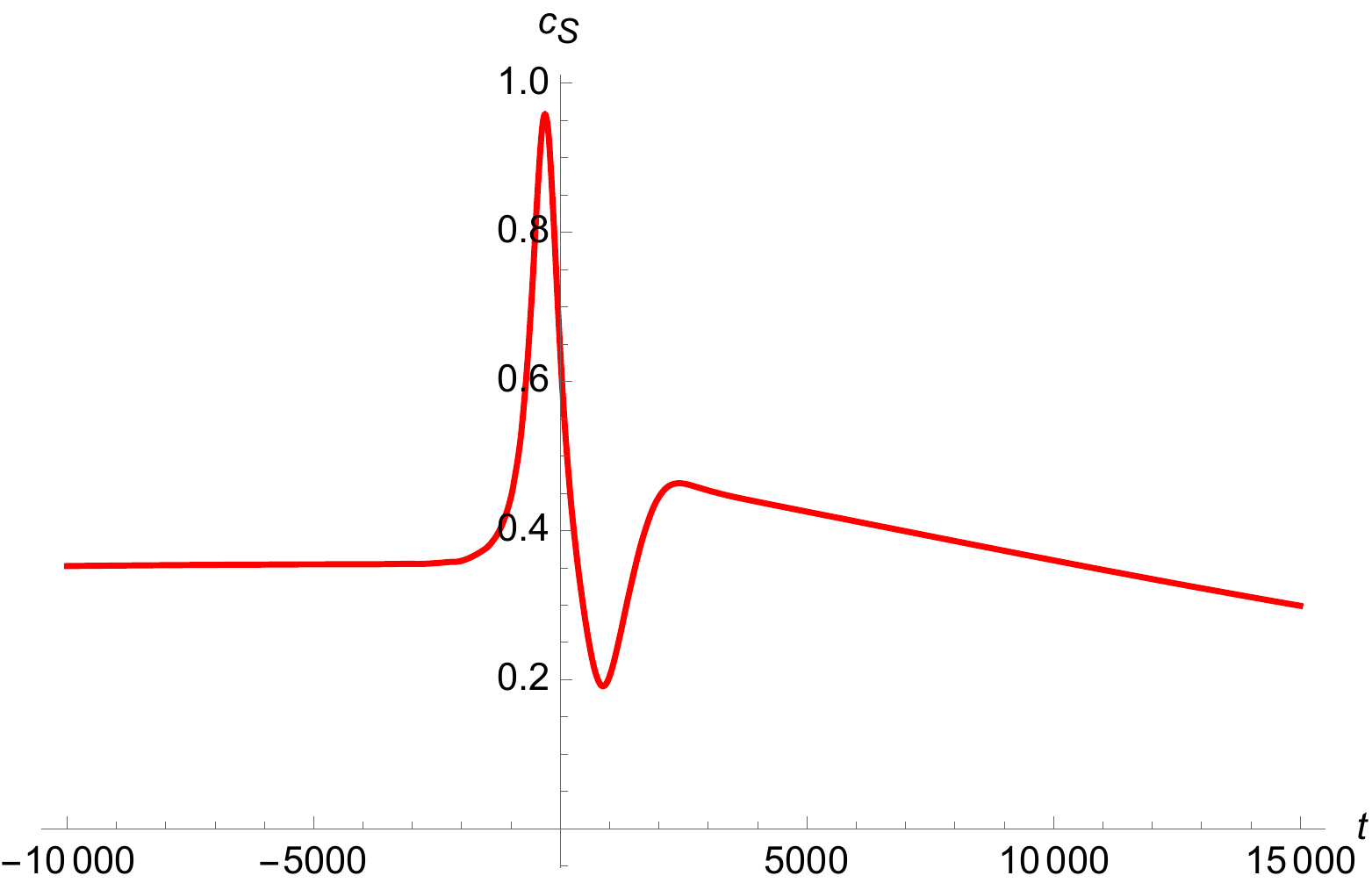}
\subcaption[first caption.]{}\label{fig:bounce_cs1d}
\end{minipage}%
\begin{minipage}{0.5\textwidth}
  \centering
\includegraphics[width=0.95\textwidth]{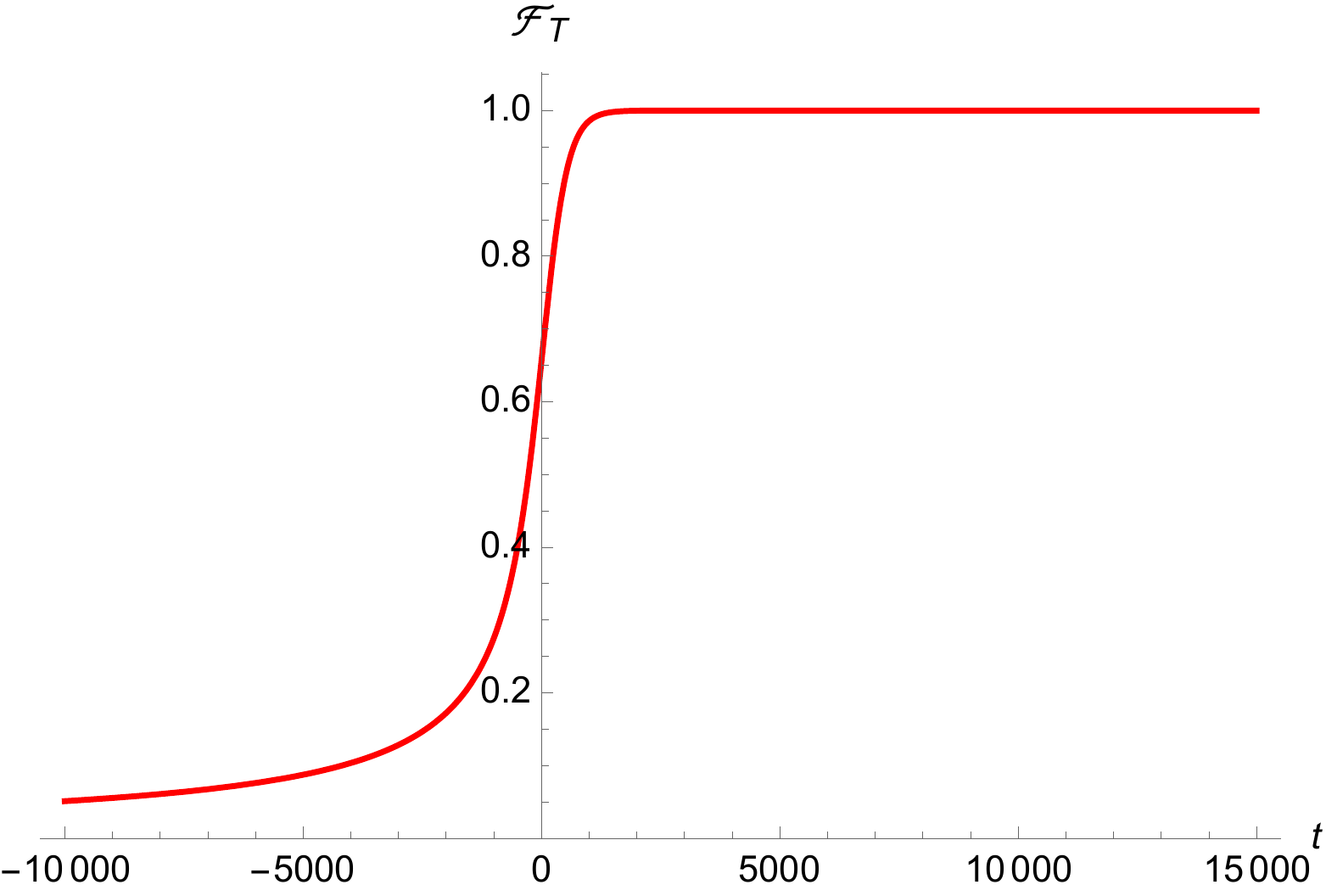}
\subcaption[second caption.]{}\label{fig:bounce_ft1a}
\end{minipage}%

\caption{ Sound 
speed of  scalar perturbations $c_S = \sqrt{\mathcal{F}_S/\mathcal{G}_S}$
(left panel) and  coefficient
$\mathcal{F}_T$ (right
panel)
for the model of
Sec.~\ref{sec:numerical_ex}  at
contraction, bounce, and beginning of inflation. } 
\label{fig:bounce_FT_cS_1}
\end{figure}

We show the behavior of the Hubble parameter and lapse function at
contraction, bounce, and beginning of inflation in Fig.~\ref{fig:bounce_H_N_1}. 
%\st{ Fig. 4
%  upper left,  Fig. 5
%  upper left}
The scalar coefficients  ${\cal F}_S$
  and ${\cal G}_S$ and scalar sound speed
$c_S$ are shown in Figs.~\ref{fig:bounce_fs1b}, \ref{fig:bounce_gs1c}
and \ref{fig:bounce_cs1d}, respectively;
%\ref{fig:bounce_FT_FS_GS_cS_1}:
%\st{\bf Fig. 7
%  upper left,  Fig. 8
%  upper left,  Fig. 9
%  upper left}:
the  stability and
subluminality are explicit. We show tensor coefficient ${\cal F}_T$
for completeness in Fig.~\ref{fig:bounce_ft1a}
%\st{\bf  Fig. 6
%  upper left}
(recall that ${\cal G}_T = {\cal F}_T$ and $c_T=1$ at all
times).

%\vspace{2cm}

%{\bf END OF REWRITTEN PART}

%\newpage

So, after a rather short transition period, the  inflationary stage sets in.
Depending on the parameters of the model, it can last for
a longer or shorter time. Note that this property may be of interest from a
phenomenological viewpoint~\cite{Tahara:2020fmn}.
We take, quite arbitrarily, the duration of inflation approximately
equal to $\Delta t_{inf} \approx
1.55 \cdot 10^6$ (in Planck units), which corresponds to the
number of e-foldings at inflation
$N_e = N_1 H_1 \Delta t_{inf} \approx 46$.
\begin{figure}[htb!]
  \centering
\includegraphics[width=0.6\textwidth]{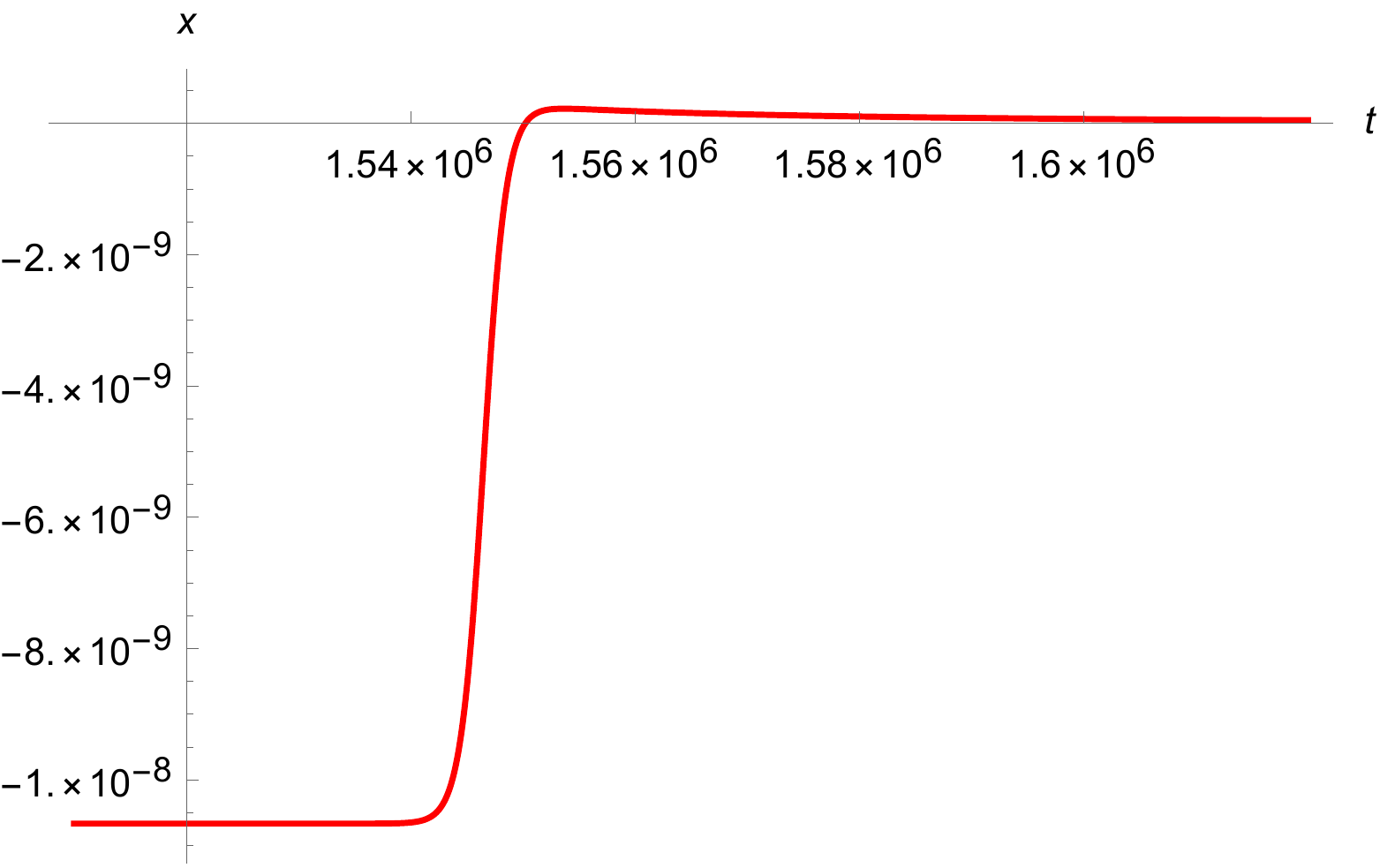}
\caption{Function $x(t)$ at transition epoch
from inflation to kination.} 
\label{fig:bounce_x2}
\end{figure}

\begin{figure}[htb!]
\centering
\begin{minipage}{0.5\textwidth}
  \centering
\includegraphics[width=0.95\textwidth]{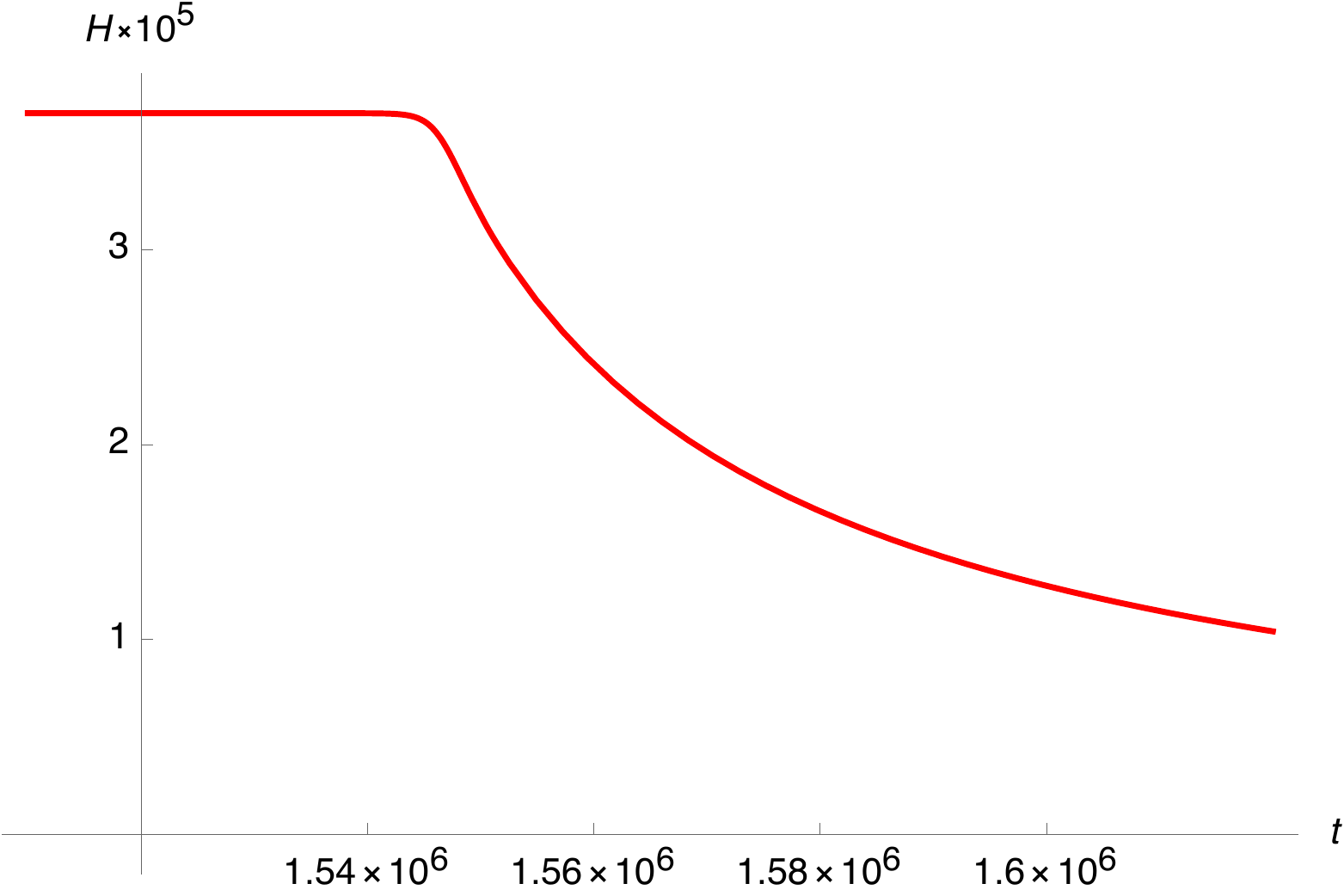}
\subcaption[first caption.]{}\label{fig:bounce_hubble2_1a}
\end{minipage}%
\begin{minipage}{0.5\textwidth}
  \centering
\includegraphics[width=0.95\textwidth]{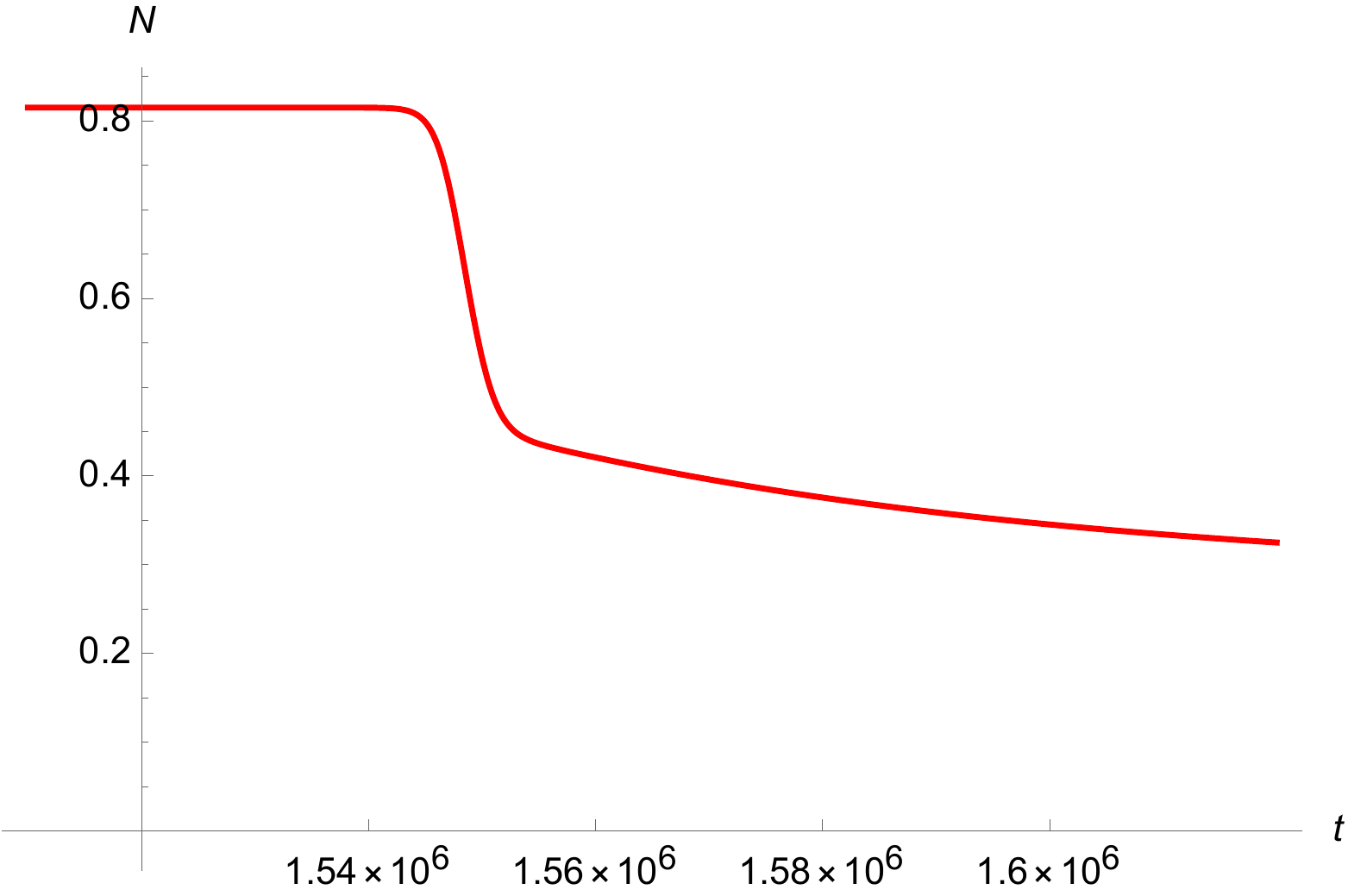}
\subcaption[second caption.]{}\label{fig:bounce_lapse2_1b}
\end{minipage}%

\caption{ Hubble
    parameter (left panel) 
and lapse function (right panel) for 
the model of Sec.~\ref{sec:numerical_ex}  at
  the end of  inflation
and beginning of kination.} 
\label{fig:bounce_H_N_2}
\end{figure}

\begin{figure}[htb!]
\centering
\begin{minipage}{0.5\textwidth}
  \centering
\includegraphics[width=0.95\textwidth]{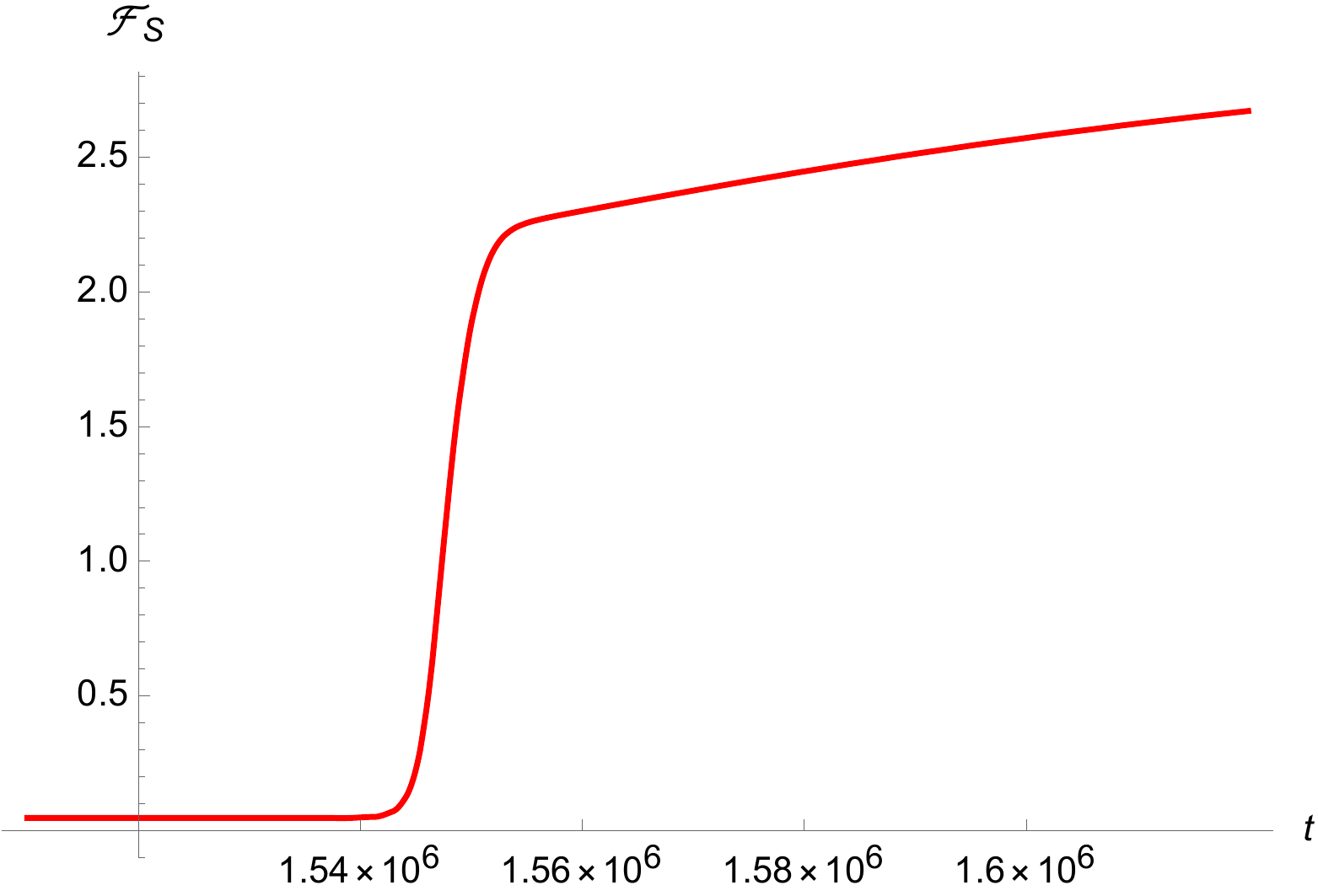}
\subcaption[second caption.]{}\label{fig:bounce_fs2a}
\end{minipage}%
\begin{minipage}{0.5\textwidth}
  \centering
\includegraphics[width=0.95\textwidth]{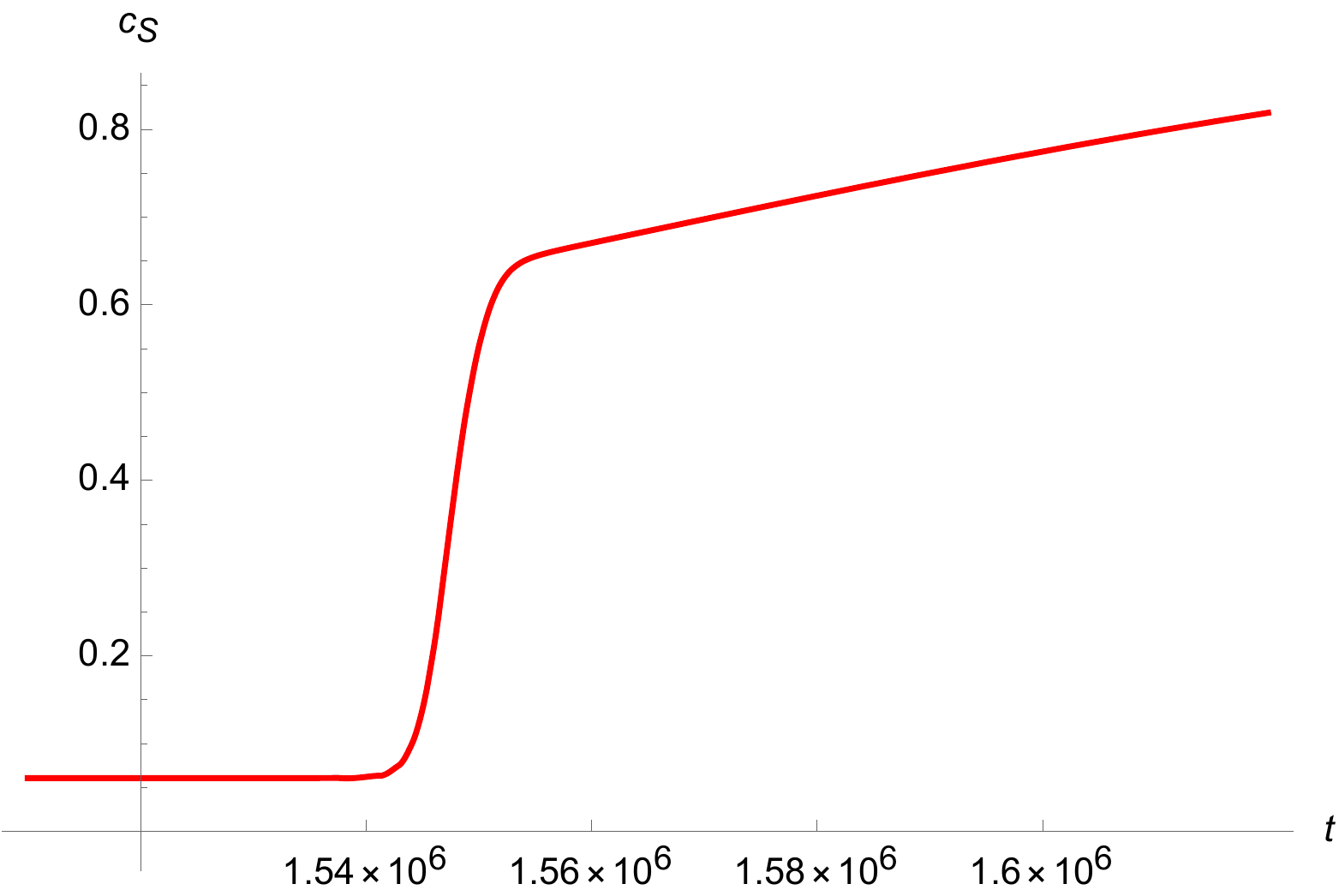}
\subcaption[second caption.]{}\label{fig:bounce_cs2b}
\end{minipage}%

\caption{ Coefficient  $\mathcal{F}_S$ (left panel)
    and  sound speed of  scalar 
perturbations $c_S = \sqrt{\mathcal{F}_S/\mathcal{G}_S}$ (right panel) 
for the model of Sec.~\ref{sec:numerical_ex}
at
  the end of  inflation
and beginning of kination.} 
\label{fig:bounce_FS_cS_2}
\end{figure}

To have the
transition from inflation to kination, we take at late times
\begin{subequations}
  \label{mar15-21-2}
\begin{align}
    x(t) &= x_1\big(1-V(t-t_*)\big) + x_2 \frac{V(t-t_*)}{(t-t_*)^2},\\
    v(t) &= v_1\big(1-V(t-t_*)\big) + v_2 \frac{V(t-t_*)}{(t-t_*)^5},\\
    y(t) &= y_1\big(1-V(t-t_*)\big) + y_2 \frac{V(t-t_*)}{(t-t_*)^5},
\end{align}
\end{subequations}
with the parameter which regulates
the duration of inflation equal to $t_* = 1.5\cdot 10^6$, and
%\begin{subequations}
%  \label{mar15-21-2}
\begin{align*}
    x_2 &=\frac{2}{3},\\ 
    y_2 &= -T^3y_1\cdot \frac{x_2}{x_1} = 6.83\cdot 10^{15},\\
    v_2 &= -T^3 v_1\cdot \frac{x_2}{ x_1}= 1.97\cdot 10^{13},
    \label{v_2}
\end{align*}
%\end{subequations}
where $T = 4.5\cdot 10^4$,
%\begin{align*}
 %   x(t) &\approx \frac{x_0}{1500}\cdot \Big[1 + V(t-t_0)\Big(\frac{r}{(t-t_0)^2} - 1\Big)\Big],\\
   % v(t) &\approx \frac{v_0}{1500}\cdot\Big[1 + V(t-t_0)\Big(-\frac{45000\cdot r}{(t-t_0)^3} - 1\Big)\Big],\\
  %  y(t) &\approx y_0\cdot (-0.001) \cdot\Big[1 + V(t-t_0)\Big(\frac{r}{(t-t_0)^2} - 1\Big)\Big].
%\end{align*}
 and the function
\begin{equation*}
  V(t) = 1+ \text{ln}\Big(\frac{\text{e}^{0.5\cdot s \cdot (t-T)}
    +\text{e}}{\text{e}^{0.5\cdot s \cdot (t-T)}+\text{e}^2}\Big), %\\
    %&V(-\infty) = 0, \nonumber\\
    %&V(+\infty) = 1.
\end{equation*}
again interpolates between 0 and 1 [the value of parameter $s$
is still given by \eqref{march13-21-11}].
  The parameters $t_*$ and $T$
are chosen in such a way that
the functions $x(t)$, $v(t)$, and $y(t)$ are reasonably smooth in
the transition
period, and
inflation smoothly
ends somewhat later than  $t_*$.
This is illustrated in Fig.~\ref{fig:bounce_x2}.
%\st{Function $x(t)$
%given by   
%\eqref{mar15-21-2},
%Fig. 27 of new-new file.}
At late times we obtain the correct kination
behavior 
given by \eqref{mar13-21-1}, and the Hubble parameter
asymptotes to $H= [3 (t-t_*)N]^{-1}$.
%\marginpar{\bf{add 
%N in magenta formula}}.
The Hubble parameter and lapse function
are shown in Fig.~\ref{fig:bounce_H_N_2},
%\st{ Figs. 16 and 18 from new file ``Long inflation''},
while the scalar
coefficient ${\cal F}_S$ and scalar sound speed $c_S$ are shown in
Fig.~\ref{fig:bounce_FS_cS_2}.
%\st{ Figs. 19 and 20 from new file ``Long inflation''}.
Clearly, the model is stable and subluminal at the transition from
inflation to kination.
The sound speed tends to 1 rather slowly, since the ratio $v(t)/x(t)$
exhibits slow
decay $(t-t_*)^{-3}$  (and $y(t)$ decays as $(t-t_*)^{-5}$). 

To end up this Section, we note that since the duration of inflation
is fairly long, the complete expressions for $x(t)$, $v(t)$, and $y(t)$, 
valid at all times, are obtained by simple
 superpositions of \eqref{mar15-21-1}
and  \eqref{mar15-21-2}, e.g.,
\begin{equation}
x(t) = x_0 (1 - U_x (t))  + x_1 U_x (t)\big(1-V(t-t_*)\big) +
x_2 \frac{V(t-t_*)}{(t-t_*)^2},
\label{step_x(t)}
\end{equation}
etc. This completes our discussion of the model with bounce, inflation,
and kination.

\subsection{Bounce directly to kination}
\label{sec:short_bounce_kin}

Contraction and bounce need not necessarily  proceed into
the inflationary stage: a short transition epoch after bounce
may end up directly at kination. In this scenario, the initial
stage is described in the same way as in Sec.~\ref{sec:early_times},
whereas the evolution after bounce proceeds as in Sec.~\ref{sec:kination}.
Let us give a numerical example which shows that stable and subluminal
cosmology of this sort is indeed possible.

\begin{figure}[htb!]
\centering
\begin{minipage}{0.5\textwidth}
  \centering
\includegraphics[width=0.95\textwidth]{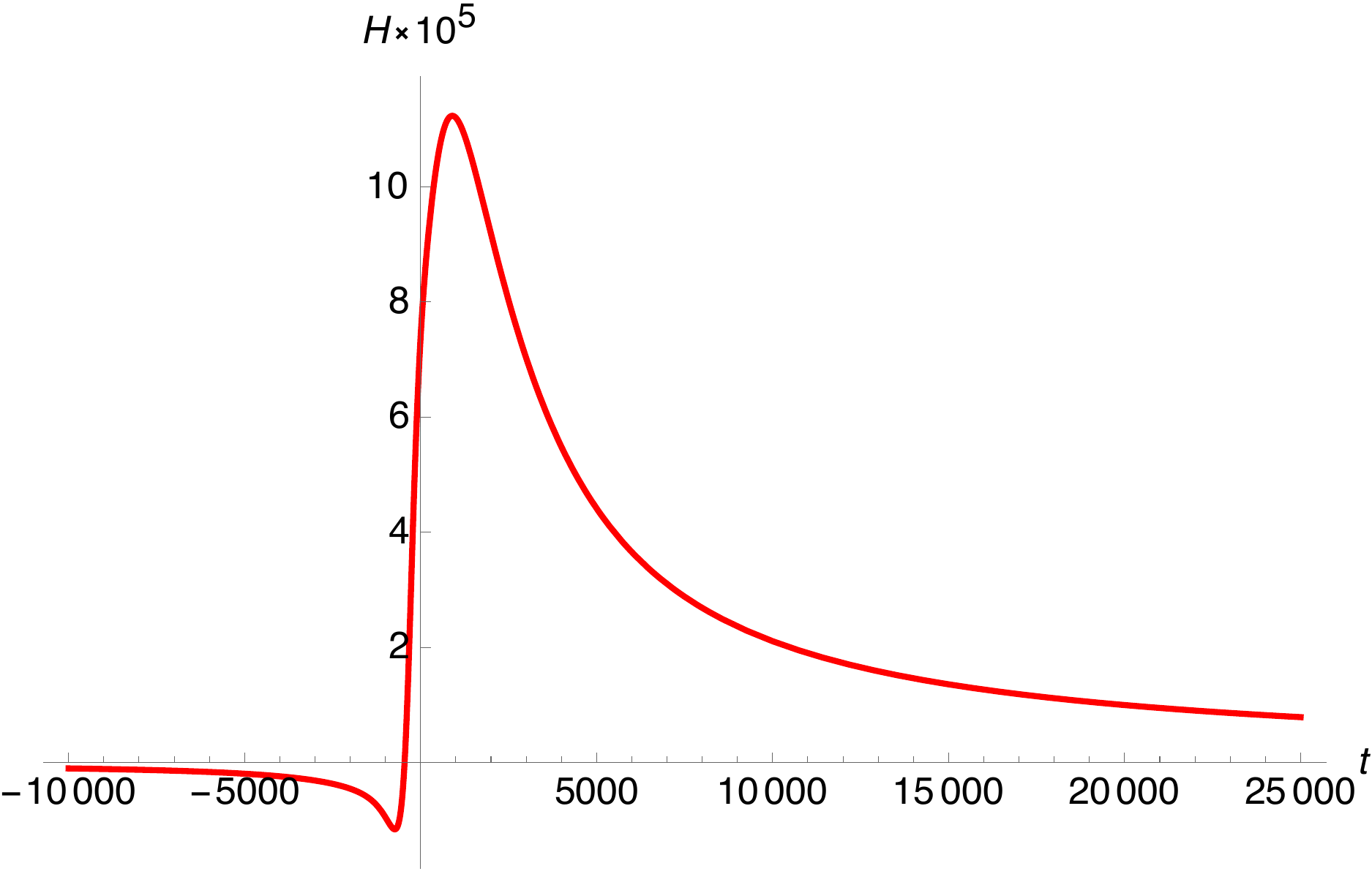}
\subcaption[first caption.]{}\label{fig:short_b_hubble_1a}
\end{minipage}%
\begin{minipage}{0.5\textwidth}
  \centering
\includegraphics[width=0.95\textwidth]{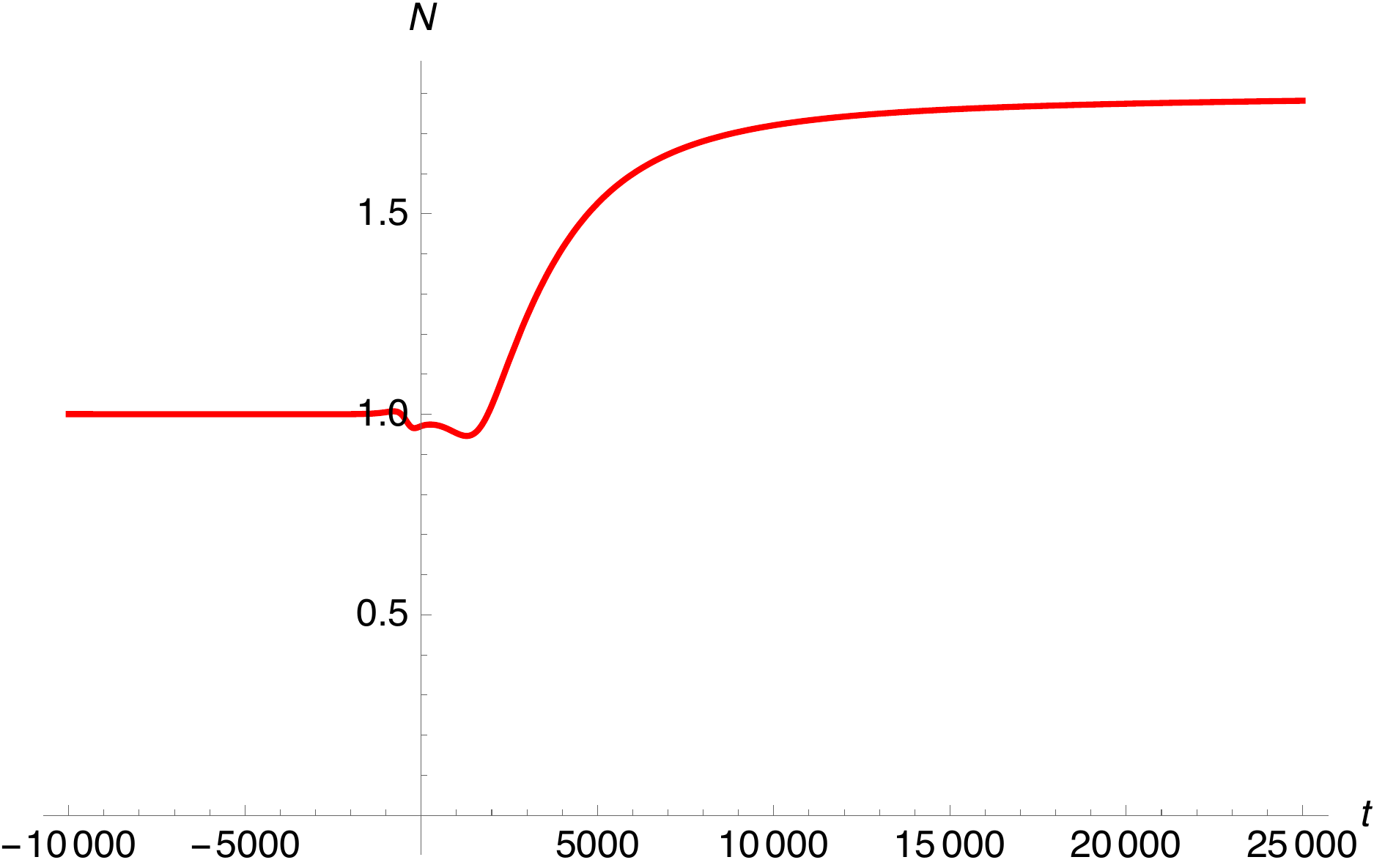}
\subcaption[second caption.]{}\label{fig:short_b_lapse_1b}
\end{minipage}%

\caption{Hubble parameter (left panel) 
and lapse function (right panel) 
for the model of Sec.~\ref{sec:short_bounce_kin}: bounce directly to kination.} 
\label{fig:short_b_H_N}
\end{figure}

\begin{figure}[htb!]
\centering
\begin{minipage}{0.5\textwidth}
  \centering
\includegraphics[width=0.95\textwidth]{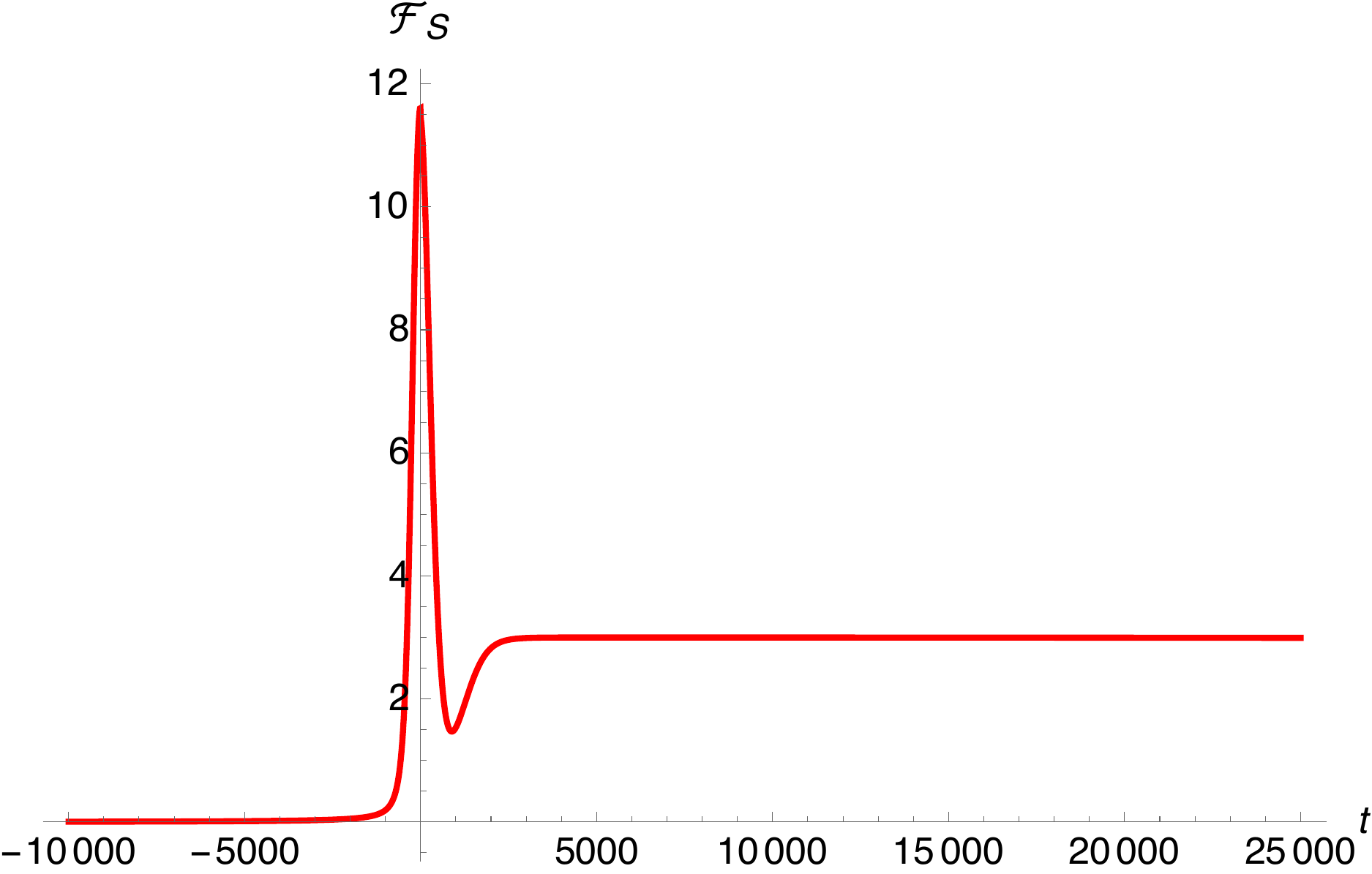}
\subcaption[second caption.]{}\label{fig:short_b_fsa}
\end{minipage}%
\begin{minipage}{0.5\textwidth}
  \centering
\includegraphics[width=0.95\textwidth]{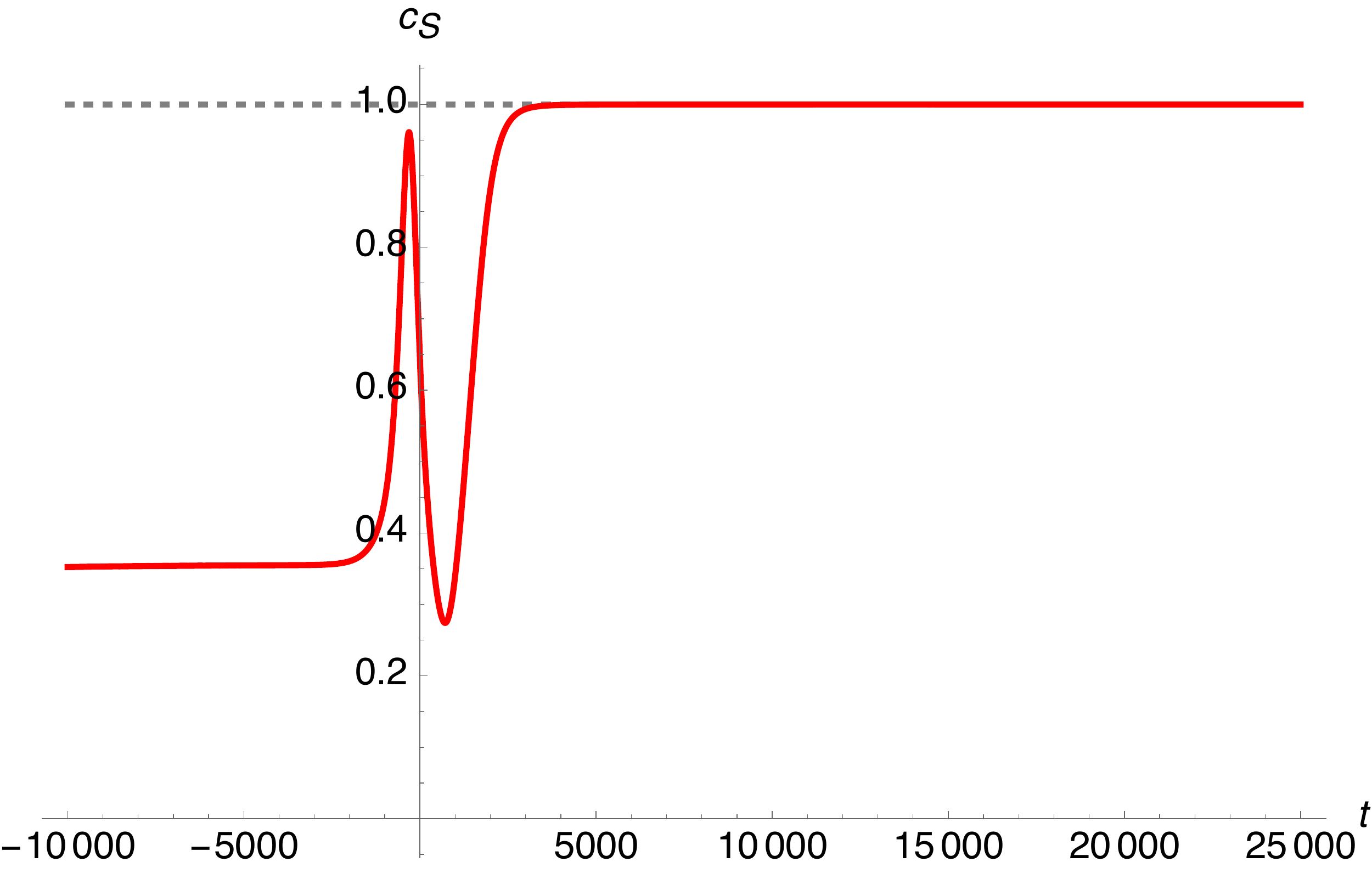}
\subcaption[second caption.]{}\label{fig:short_b_csb}
\end{minipage}%

\caption{Coefficient  $\mathcal{F}_S$ (left panel)
  and  sound speed of  scalar 
perturbations $c_S = \sqrt{\mathcal{F}_S/\mathcal{G}_S}$ (right 
panel) for the model of Sec.~\ref{sec:short_bounce_kin}.} 
\label{fig:short_b_FS_cS}
\end{figure}

We again consider a model with $\mu=0.8$ and choose the parameters
of the contraction stage as in \eqref{mar16-21-1},
\eqref{set_1}, and  \eqref{mar16-21-2}. The function $f(t)$ is again
given by \eqref{march13-21-5}, so that we restore GR
at later times. We would like to approach the behavior
\eqref{mar13-21-1} soon after bounce and, by trial and error,
end up with the following example:
    \begin{align*}    
      x(t) &= x_0\big(1- U_x(t)\big) + \frac{4}{3((t+2000)^2+(t-5000)^2)}
      \cdot U_x(t),\\
      v(t) &= v_0\big(1 - U_x(t)\big) +
      \frac{v_2}{(|t|+2000)^5} \cdot U_x(t),\\
      y(t) &= y_0\big(1- U_y(t)\big) +
      \frac{y_2}{(|t|+2000)^5} \cdot U_y(t),
    \end{align*}
where $U_x(t)$ and $U_y (t)$ are still given by \eqref{mar16-21-5}, and
 now 
\begin{equation*}
    v_2 = 1.04\cdot 10^8, \quad y_2 = 9.6\cdot 10^{10}. 
\end{equation*}
We show the Hubble parameter and lapse function for this model in
Fig.~\ref{fig:short_b_H_N}
%\st{ Figs. 10 and 12 below},
and the scalar coefficient
${\cal F}_S$ and scalar sound speed in Fig.~\ref{fig:short_b_FS_cS}.
%\st{ Figs. 13 and 14 below}.
The latter figure illustrates that the model
is stable and subluminal at all times.

%\begin{figure}[H]
%\centering
%\begin{minipage}{0.5\textwidth}
%  \centering
%\includegraphics[width=0.95\textwidth]{hubble_long_inf3.eps}
%\subcaption[first caption.]{}\label{fig:bounce_hubble3_1a}
%\end{minipage}%
%\begin{minipage}{0.5\textwidth}
%  \centering
%\includegraphics[width=0.95\textwidth]{lapse_long_inf3.eps}
%\subcaption[second caption.]{}\label{fig:bounce_lapse3_1b}
%\end{minipage}%
%\caption{Hubble parameter (left panel, Fig.~\ref{fig:bounce_hubble3_1a}) and lapse function (right panel, Fig.~\ref{fig:bounce_lapse3_1b}) for the model of Sec.~\ref{sec:numerical_ex}: kination epoch after transit from inflation.} 
%\label{fig:bounce_H_N_3}
%\end{figure}
%\begin{figure}[H]
%\centering
%\begin{minipage}{0.5\textwidth}
%  \centering
%\includegraphics[width=0.95\textwidth]{fs3.eps}
%\subcaption[second caption.]{}\label{fig:bounce_fs3a}
%\end{minipage}%
%\begin{minipage}{0.5\textwidth}
%  \centering
%\includegraphics[width=0.95\textwidth]{cs3.eps}
%\subcaption[second caption.]{}\label{fig:bounce_cs3b}
%\end{minipage}%
%\caption{Square root of the coefficient  $\mathcal{F}_S$ (left panel, Fig.~\ref{fig:bounce_fs3a}) and the sound speed for the scalar perturbations $c_S = \sqrt{\mathcal{F}_S/\mathcal{G}_S}$ (right panel, Fig.~\ref{fig:bounce_cs3b}) for the model of Sec.~\ref{sec:numerical_ex}:  kination epoch after transit from inflation. }
%\label{fig:bounce_FS_cS_3}
%\end{figure}
\section{Models with genesis}
\label{sec:genesismodels}

\subsection{\textit{Ansatz}}

%\marginpar{\bf Begin with Ansatz (sign and $1/2$ in $A_4$??)}

To illustrate that interesting cosmologies can be
obtained within  various \textit{Ansätze}, %\textcolor{red}{s},
in this Section
we construct genesis  models by choosing
the functions in the Lagrangian
\eqref{adm_lagr} in the following form:
\begin{subequations}
  \label{mar16-21-10}
	\begin{align}
	&A_2 =  \frac{1}{2}f^{-2\mu -2 -\delta} a_2 (N) \text{,} \\ 
	&A_3 =  \frac{1}{2}f^{-2\mu -1 -\delta} a_3 (t,N) \text{,} \\
	&B_4 = \frac{1}{2}f^{-2\mu}b_4(t,N),\\
	%\label{B_4}
	&A_4  =\frac{1}{2} f^{-2\mu}a_4(t,N) \text{.}
	\end{align}
\end{subequations}
The parameter $\mu>0$ is similar to that in the  previous Section,
and the parameter
$\delta>0$ is new.
The functions $A_4$ 
and $B_4$ depend on $N$ now; functions $a_2(N)$, $a_3(t,N)$,
$a_4(t,N)$, and $b_4(t,N)$ are chosen as follows:
%\marginpar{\bf \textcolor{blue}{correct?}}
\begin{subequations}
  \label{mar16-21-11}
\begin{align}
  &a_2(N) = x \Big(\frac{1}{N^2} - \frac{1}{3N^4}\Big), \quad x=\mbox{const},
  \label{apr23-21-1}\\
&a_3(t,N) = \frac{y(t)}{N^3},\\
  &a_4(t,N) = -\Big(1+\frac{z(t)}{N^2}\Big),
  \label{apr23-21-2}\\
  &b_4(t,N) = \Big(1-\frac{z(t)}{N^2}\Big).
  \label{apr23-21-3}
\end{align}
\end{subequations}
Note that the parameter $x$ is now time independent, unlike in the
models of Sec.~\ref{sec:bounce} where functions $x(t)$ and $v(t)$
entering $a_2$ exhibited step-function behavior.

Given the \textit{Ansatz} \eqref{mar16-21-10}, the background equations of motion \eqref{eoms} are
\begin{subequations}
\label{eoms_all_substitute_genesis}
\begin{align}
    &x f^{-2-\delta} \left(\frac{1}{N^4} - \frac{1}{N^2}\right) -9 y(t)\cdot f^{-1-\delta}\cdot \frac{ H}{N^3} 
    + 6 H^2\cdot \left(1+\frac{3z(t)}{N^2}\right)  = 0, \label{eom_1_gen} \\
    &xf^{-2-\delta}\left(\frac{1}{N^2} - \frac{1}{3N^4}\right) + 6 \cdot H^2 \cdot \left(1+\frac{z(t)}{N^2}\right) - \frac{1  }{N}\frac{d}{dt}\left[ \frac{y(t)}{N^3}\cdot f^{-1-\delta} - 4  \cdot H \cdot\left(1+\frac{z(t)}{N^2}\right)\right] \nonumber \\
    &+ 
    2\mu\cdot\frac{\dot{f}}{f}\frac{1}{N}\left[ \frac{y(t)}{N^3}\cdot f^{-1-\delta} - 4  \cdot H \cdot\left(1+\frac{z(t)}{N^2}\right)\right]  = 0.
    \label{eom_2_gen}
\end{align}
\end{subequations}
%\marginpar{\bf Write equations of motion and expressions for ${\cal G}_T$,
%  ${\cal F}_T$  ${\cal G}_S$  ${\cal F}_s$, just like in 
%  Sec.~\ref{sec:ansatz-bounce}}
The functions entering 
\eqref{eq:Fs_Gs_form} are given by
\begin{subequations}
\label{stability_all_subtitute_genesis}
\begin{align}
    \mathcal{F}_T &= f^{-2\mu}\cdot\left(1-\frac{z(t)}{N^2(t)}\right),\\ \mathcal{G}_T &=  f^{-2\mu}\cdot\left(1+\frac{z(t)}{N^2(t)}\right),
    %\mathcal{G}_T &=   f^{-2 \mu},\\
\end{align}
\begin{align}
    \mathcal{ F}_S=
     f^{-2\mu}&\cdot \frac{3y(t)\cdot z(t) - 3y(t)\cdot N^2 - 16f^{1+\delta}\cdot H\cdot N \cdot z^2(t)}{N^2\left[3y(t)-4f^{1+\delta}\cdot H\cdot N\cdot \big(N^2+3z(t)\big)\right]} \nonumber\\
     &- \frac{1}{N}\frac{d}{dt}\left(\frac{4 f^{-2\mu+\delta+1}(N^2+z(t))^2}{N\cdot\left[3y(t)-4f^{1+\delta}\cdot H\cdot N\cdot \big(N^2+3z(t)\big)\right]}\right),
\end{align}
\begin{align}
    &\mathcal{ G}_S=  f^{-2 \mu}\cdot \left(1+\frac{z(t)}{N^2}\right) \nonumber\\
    &\times\left(8f^{\delta}\cdot\big(N^2+z(t)\big)\cdot\frac{x\cdot(-2+N^2) + 18 f\cdot H\cdot N\cdot y(t)-6 f^{2+\delta}\cdot H^2 \cdot N^2 \cdot \big(N^2+6z(t)\big) }{\left[3y(t) - 4f^{1+\delta}\cdot H \cdot N \cdot \big(N^2+3z(t)\big)\right]^2}  + 3 \right). %\nonumber
    \label{G_s_full_contr}
\end{align}
\end{subequations}
We will make sure that the functions $f(t)$, $y(t)$
and $z(t)$ are such  that
inequalities \eqref{stability_conditions} and \eqref{velocities} are satisfied.
%\begin{subequations}
%\label{stability_all_subtitute_genesis}
%\begin{align}
%    \mathcal{F}_T &= f(t)^{-2\mu}\left(1+\frac{z_4(t)}{N^(t)}\right),\\ \mathcal{G}_T &=  f(t)^{-2\mu}\left(1-\frac{z_4(t)}{N^(t)}\right), \\ 
    %\mathcal{G}_T &=   f^{-2 \mu},\\
%    \mathcal{ F}_S&=
%     f^{-2 \mu} \cdot\left( \frac{f\cdot H}{f\cdot H 
%     - \frac{3  y}{4 N^3}} - 1 \right) 
%     + \frac{1}{N}\frac{d}{d t}\left(\frac{ f^{-2 \mu+1}}{f\cdot H 
%     - \frac{3 y }{4 N^3}}\right),\\
%    \mathcal{ G}_S&=  f^{-2 \mu} \left(\frac{z_2f^{-2\mu-\delta-2}\Big(-\frac{2}{N^4} + \frac{1}{N^2}\Big) + 18z_3f^{-2\mu-\delta-1}\frac{H}{N^3} - 6 f^{-2\mu}H^2(1+\frac{6z_4}{N^2})}{\Big(-\frac{3z_3 f^{-\delta-1}}{2N^4} + 2H\Big(1+\frac{3z_4}{N^2}\Big)\Big)^2}  - 3 \right). \label{G_s_full}
%\end{align}
%\end{subequations}

\subsection{Contracting genesis followed by bounce}
\label{sec:contr_gen}
%\marginpar{\bf Model of Appendix A.2. 
%Explain, why do you need $z_2 \neq 1$, $z_4 (t)$ and $b_4 \neq a_4$.
% Then the same way of presentation as in the previous subsection.} 
 
Here we construct  a model with genesis of contracting Universe.
This cosmology begins with the flat space-time,
then the Universe starts to contract, and the rate of contraction increases.
At some moment of time the bounce occurs: the contraction terminates
and expansion begins.
We consider for definiteness the case in which bounce
is followed by inflationary expansion;
inflation is assumed to end up as in Sec.~\ref{sec:bounce_to_inflation}.
Alternatively, bounce may lead directly to kination epoch like in
Sec.~\ref{sec:short_bounce_kin}; we do not elaborate on this possibility.

%This cosmology begins with the flat space-time, then the 
%Universe starts to contract. At some moment of time the bounce occurs: 
%the contraction terminates and expansion begins.

%In Section \ref{sec:bounce} we use specific step-functions $x(t)$ 
%and $v(t)$ (see \eqref{A_ansatz}, for example, with \eqref{step_x(t)}) 
%in $a_2$ in order to build stable bounce solution at all times. 
%%Also, the functions $a_4$, $b_4$ independent of $N$ in order to build stable s%olution at all times. 
%The stable contracting genesis and bounce could be built in the same way. 
%However, here, for the completeness, one can go the other way and 
%consider new possibility: set $a_4$ and $b_4$ as functions of both $t$
%and $N$, but get rid of time step-functions in $a_2$.      

%To this end, we are going to use the Ansatz \eqref{mar16-21-10}. 
We begin with early times  and consider the following asymptotics
\begin{subequations}
\label{x_y_z_early_contr_gen}
\begin{equation}
\label{f_past_contr_gen}
    f =  -ct, \quad c>0,
\end{equation}
\begin{equation}
    y(t) = y_0, \quad
    z(t) = z_0.
\end{equation}
\end{subequations}
This choice  leads to power-law behavior of the Hubble parameter.
%similarly 
%as for bounce \eqref{hubble_bounce}, \eqref{x_y_v_early}.
Indeed, by substituting \eqref{x_y_z_early_contr_gen} into equations of motion
\eqref{eoms_all_substitute_genesis} we arrive at
\begin{equation*}
    H = -\frac{\chi}{(-t)^{1+\delta}}, \quad N = 1, \quad t\to - \infty,
\end{equation*}
with %$\chi$ equals 
\begin{equation}
\label{chi_genesis_eq}
    \chi = \frac{\frac{2}{3}x-(2\mu+\delta+1)\cdot y_0\cdot c}{4(2\mu+\delta +1)\cdot(1+z_0)\cdot c^{\delta+2}} 
\end{equation}
[the fact that $N=1$ for this solution is due to
  the particular
choice of the coefficient $(-1/3)$ of $N^{-4}$ in \eqref{apr23-21-1};
this choice replaces in this model the constraint \eqref{mar3-21-2}].
For $\delta >0$, the scale factor tends to a constant as $t \to -\infty$:
\[
a = \mbox{const}\cdot \left( 1 - \frac{\chi}{\delta (-t)^\delta}\right) \; ,
\]
as required for genesis. Contracting genesis occurs for $\chi > 0$.
%in the same way as it was done in  \cite{Kobayashi:2016xpl}. That is why, we 
%choose the Lagrangian functions asymptotic \textcolor{magenta}{behavior} as follows
%where $c$, $x$, $y_0$ and $z_0$ are constant parameters.

Let us note that for small $\delta$, early time asymptotics is approached
slowly (backwards in time), since  
the expansion parameter is 
$(-t)^{-\delta}$:
\begin{subequations}
\begin{align}
    H(t) &= -(-t)^{-1-\delta}\cdot\left(\chi + \chi_1\cdot (-t)^{-\delta}+ \ldots\right), \\
    N(t) &= 1 + N_1\cdot (-t)^{-\delta}+ \ldots,
\end{align}
\label{H_N_series}
\end{subequations}
where coefficients $\chi_1$ and $N_1$ are not particularly small.
This complicates the numerical analysis;
%
%$\ldots$ means higher order terms by $(-t)^{-\delta}$,
% $\chi_1$ and $N_1$ is given by
%\begin{align*}
%    \chi_1 &= \frac{c^{-4-2\delta}\big(2x - 3 y_0 \cdot c\cdot (2\mu+\delta +1)%\big)}{576 x\cdot(2\mu+\delta+1)^3\cdot(2\mu+2\delta+1)\cdot (1+z_0)^4} \cdot\B%ig\{9 y_0^2 \cdot c^2\cdot(2\mu+\delta+1)^2\cdot(2\mu+2\delta+1)\cdot (3+z_0)\c%dot(5+3z_0)  \\
%    &+ 4 x^2\cdot\big[\delta\cdot(3+5z_0\cdot(2+3z_0)) + (2\mu +1)\cdot (3+z_0\%cdot(8+9z_0))\big]\\
%    &+6 x\cdot y_0\cdot c\cdot(2\mu+\delta+1)\cdot \big[2z_0\cdot(2\mu+1)\cdot %(7+3z_0)+\delta\cdot(3+z_0\cdot(34+15z_0))\big]\Big\},
%\end{align*}
%\begin{align*}
%    N_1 &= \frac{2x - 3 c\cdot y_0 - 3c\cdot \delta \cdot y_0 - 6 c \cdot \mu \%cdot y_0}{48c^{2+\delta} \cdot x \cdot (2\mu +\delta +1)^2\cdot (1+z_0)^2 }\cdo%t \big\{2 x + 15 c\cdot y_0 + 15 c \cdot \delta\cdot y_0 + 30 c \cdot \mu \cdot% y_0 \\
%    &+6 x \cdot z_0 + 9 c\cdot y_0 \cdot z_0 + 9 c \cdot \delta \cdot y_0 \cdot% z_0 + 18 c \cdot \mu \cdot y_0 \cdot z_0\big\}.
%\end{align*}
%This series \eqref{H_series} for Hubble has the dumping 
%factor $(-t)^{-\delta}$, so for sufficiently small $\delta$ 
%it converges only at fairly large negative times. Since we obtain 
%all solutions numerically, these large values of time lead to some  
%difficulties with numerical calculations.
we explain how we get around 
this obstacle in Appendix \ref{app:variable_u}. The same comment applies to
the genesis model of Sec.~\ref{sec:genesis_kob}.

The asymptotic behavior of coefficients
\eqref{stability_all_subtitute_genesis}
entering the quadratic action for perturbations
 is  
\begin{align*}
    \mathcal{F}_T &\propto (-ct)^{-2\mu}, \quad
    \mathcal{G}_T \propto (-ct)^{-2\mu},\\
    \mathcal{F}_S &\propto (-ct)^{-2\mu+\delta}, \quad
    \mathcal{G}_S \propto (-ct)^{-2\mu+\delta},
\end{align*}
The constraints on  parameters arise from the same 
requirements as in Sec.~\ref{sec:early_times}. Let us  list them:
%\begin{itemize}
%\item

  (i) Contraction at early times:
    \begin{equation}
        \chi>0;
        \label{chi_contr_genesis}
    \end{equation}
  %\item

    (ii) Stability of background and subluminality of perturbations:
\begin{subequations}
\label{stability_contr_gen}
\begin{align}
&\mathcal{ F}_T,  \mathcal{ G}_T,\mathcal{ F}_S,  \mathcal{ G}_S > 0\, ,\\
&c_T^2 \leq 1, \quad c_S^2 \leq 1;
\end{align}
\end{subequations}
%\item

(iii) Evading the no-go argument of Ref.~\cite{Kobayashi:2016xpl}:
    \begin{equation}
        2\mu>1+\delta;
\label{apr2-21-1}
    \end{equation}
  %\item

    (iv) Absence of strong coupling in the past. This issue has been
    studied in detail in Ref.~\cite{Ageeva:2020buc} with the result
    \begin{equation}
\label{old_no_SC}
    \mu + \frac{3}{2}\delta < 1;
\end{equation}
  %\item

    (v)  Belinsky--Khalatnikov--Lifshitz phenomenon.
    In the same way as  in Sec.~\ref{sec:bounce}, we obtain for
    time-dependent superhorizon
perturbations 
  \[
  h_{ij} \propto \int~dt~ \frac{1}{a^3  \mathcal{G}_T} \propto
  (-t)^{2\mu+1} \, .
  \]
  and
    \[
  \zeta \propto \int~dt~ \frac{1}{a^3  \mathcal{G}_S} \propto
  (-t)^{2\mu-\delta+1} \, .
  \]
They decay as $t$ increases towards zero, provided that
  %\begin{equation*}
   %$ 2\mu +1 > 0 $
  %\end{equation*}
  %and
  %\begin{equation*}
     $2\mu +1 > \delta$.
  %\end{equation*}
  %This is our last constraint
  %on the parameters characterizing  the early epoch. 
  This constraint is weaker than (iii).

  All these constraints can be satisfied without much of fine-tuning.
In view of \eqref{apr2-21-1}, the constraints give
\begin{align*}
  {\cal F}_T >0 ~:& ~~~~~~~~~ z_0 < 1\; ;
  \\
  c_T^2 \leq 1~:& ~~~~~~~~~ z_0 \geq 0\; ;
  \\
    {\cal G}_S >0 ~:& ~~~~~~~~~ x_0 < 0\; ;
    \\
    \chi > 0 ~:&~~~~~~~~~
    %y_0 <0 \; ;
     %\\
    %\chi > 0 ~:&
    %~~~~~
     3(2\mu+\delta +1) c |y_0| > 2 |x_0| \; ; \quad \quad  y_0 <0 \; ;
    %\\
    %{\cal F}_S >0 ~:& ~~~~~~~~~ |x_0|(1+3z_0)+3c|y_0|(2\mu+\delta+1)>0, \quad \text{identically satisfied} ;
\end{align*}
and $c_S^2 \leq 1$ gives
\begin{equation*}
    3 (2\mu - \delta -1)
        (2\mu+\delta +1)  c |y_0|
      \leq \left[ 4(\mu+\delta+1)+6z_0(\delta+1) \right]
      |x_0|.
\end{equation*}
The constraint ${\cal F}_S >0$ is satisfied
   automatically.
%Note that non-trivial inequalities involve $y_0$ in combination
   %$(2\mu+\delta +1) c |y_0|$.
   This
  set of inequalities can be
satisfied for all $\mu$ and $\delta$
from their allowed range.

Let us turn to inflationary epoch. 
Like in Sec.~\ref{sec:bounce_to_inflation}, inflation occurs
for time-independent coefficients
in the Lagrangian:
\begin{subequations}
\label{lagr_func_gen_bounce_inf}
\begin{align}
    & ~~~~f = 1 \; , \label{f_future_gen_bounce} \\
    y &= y_1, \quad
    z = z_1.
\end{align}
\end{subequations}
In this case, equations of motion
\eqref{eoms_all_substitute_genesis} read
%One can substitute \eqref{lagr_func_bounce_to_infl} into \eqref{eoms_all_substitute} 
%and it leads to the following equations of motion:
\begin{subequations}
\label{eoms_all_substitute_contr_bounce_to_infl}
\begin{align}
  &x\left( \frac{1}{N^4}-\frac{1}{N^2}\right) -
  \frac{9 y_1\cdot H}{N^3} 
    + 6 H^2\cdot\left(1+\frac{3z_1}{N^2}\right)  = 0, \\
    &x\left(\frac{1}{N^2} - \frac{1}{3N^4}\right) + 6 H^2\cdot\left(1+\frac{z_1}{N^2}\right) = 0\; , 
    %- 
    %\frac{1}{N}\frac{d}{dt}\left( \frac{y_1}{N^3} - 4 H\right) = 0,
\end{align}
\end{subequations}
and we denote the (time-independent) solution to these equations by
$H = H_1$ and $N=N_1$. We require $H_1 > 0$, $N_1 >0$. 
%are constants and the solution of the above EoM. 
%It means that at these times the inflation takes place.  However, it is 
%not guaranteed at all, that the requirement \eqref{lagr_func_bounce_to_infl} le%ads to 
%the bounce at some moment of time, 
%since there may be not exist the solution, which connects the demanded asymptot%ics  
%\eqref{hubble_bounce} and ($H_1,N_1$). So, for each set of our parameters 
%one should check the existence of such a solution numerically.

\begin{figure}[t]
\centering
%\begin{minipage}{0.5\textwidth}
%  \centering
\includegraphics[width=0.40\textwidth]{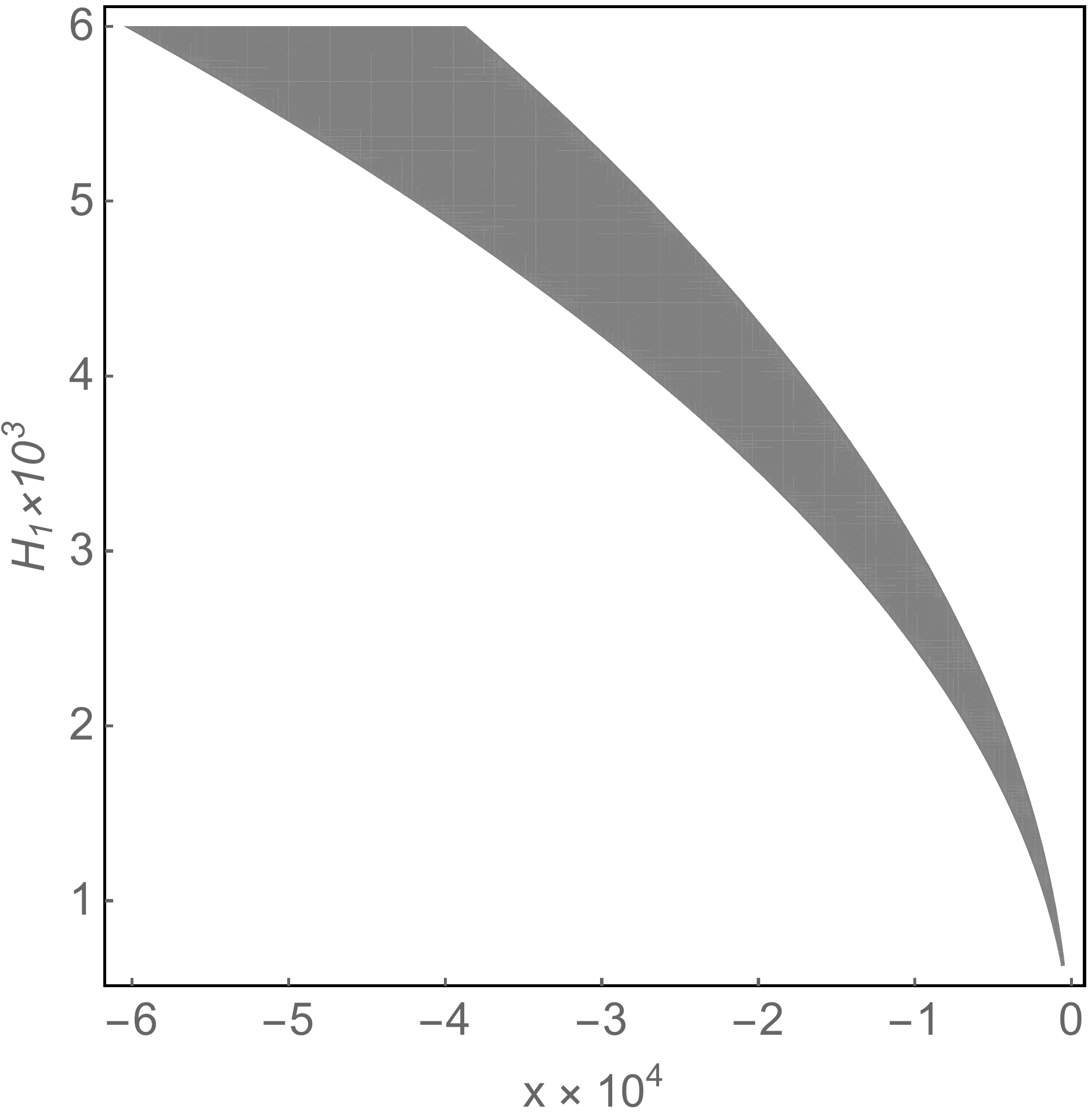}
%\subcaption[first caption.]{}\label{fig:contr_bounce_3D_c2_1a}
%\end{minipage}%
%\begin{minipage}{0.5\textwidth}
%  \centering
%\includegraphics[width=0.6\textwidth]{cakeOURback.eps}
%\subcaption[second caption.]{}\label{fig:contr_bounce_3D_c2_1b}
%\end{minipage}%
\caption{Space of parameters $x$ and $H_1$ characteristic of
%determining the Lagrangian \eqref{mar16-21-10} with
%  asymptotics \eqref{lagr_func_gen_bounce_inf} during 
 inflation in Sec.~\ref{sec:contr_gen}. 
%Grey region 
%corresponds to
 Parameters $x$ and $H_1$ in the gray region
 %which 
 satisfy constraints \eqref{inflation_after_gen}. Other parameters are
   $\mu = 0.8$, $\delta = 0.1$, and $N_1 = 0.74$.} 
\label{fig:contr_bounce_infl}
\end{figure}

In analogy to  Sec.~\ref{sec:bounce_to_inflation}, it
is convenient to treat $H_1$ and $N_1$
as independent parameters and express  $y_1$ and $z_1$ through
these parameters  using
\eqref{eoms_all_substitute_contr_bounce_to_infl}:
\begin{subequations}
\label{y1_z1}
\begin{align}
    y_1 &= -\frac{12 H^2_1 \cdot N_1^4 -2x + 4x\cdot N_1^2}{9 H_1 \cdot N_1}, 
\label{y1_z1a} \\
z_1 &=
\frac{-18 H_1^2\cdot N_1^4 + x - 3 x\cdot N_1^2}{18 H_1^2 \cdot N_1^2}.
\label{y1_z1b}
\end{align}
\end{subequations}
We emphasize that unlike in
  the bounce scenario of Sec.~\ref{sec:bounce}, 
  the inflationary epoch in our current model
  is not described by GR since $z_1\neq0$.
The conditions for the
%Our first constraint comes from the fact that we want to obtain the 
%inflation in the future, i.e.
%\begin{equation}
%\label{H1_N1}
%    H_1 > 0, \quad N_1 > 0.
%\end{equation}
%The second requirement is the necessity of
background stability 
and subluminal propagation of perturbations,
Eqs.~\eqref{stability_all_subtitute_genesis},
read
\begin{subequations}
\label{inflation_after_gen}
\begin{align}
     \mathcal{F}_T &= 2+\frac{x\cdot(3N_1^2-1)}{18 H_1^2\cdot N_1^4}  >0,
%\end{align}
%\begin{align}
 \quad \quad    \mathcal{G}_T = \frac{x\cdot(1-3N_1^2)}{18 H_1^2\cdot N_1^4}  >0,
\end{align}
\begin{align}
     c_T^2 &= -1 + \frac{36H_1^2\cdot N^4_1}{x\cdot(1-3N_1^2)}  \leq 1,
\label{apr2-21-2}
\end{align}
\begin{align}
  \mathcal{F}_S &=
  \frac{-648 H_1^4 \cdot N_1^8 + 18 x\cdot H_1^2\cdot N_1^4 \cdot (1-9 N_1^2)+x^2 \cdot(-1 + 9 N_1^2 - 18 N_1^4)}{54 H_1^2 \cdot N_1^6 \cdot (6 H_1^2\cdot N_1^2 + x)}>0,
\end{align}
\begin{align}
  \mathcal{G}_S = -\frac{x\cdot (-1 + 3N_1^2)\cdot \big(108 H_1^4 \cdot N_1^6 + 6 x\cdot H_1^2 \cdot N_1^2 (1 + 3 N_1^2) + x^2\cdot (-1+6 N_1^2)\big)}{18 H_1^2 \cdot N_1^6 \cdot(6H_1^2\cdot N_1^2 + x)^2}>0,
\end{align}
\begin{align}
     c^2_S =\frac{1}{3} - \frac{12 H_1^2\cdot N_1^4}{x\cdot(1-3N_1^2)} + \frac{72 H_1^4 \cdot N_1^6 - 4 x\cdot H_1^2\cdot N_1^2}{108 H_1^4 \cdot N_1^6 + 6 x\cdot H_1^2\cdot N_1^2\cdot(1+3N_1^2)+x^2\cdot (-1+6N_1^2)}\leq 1.
\end{align}
\end{subequations}
%This system, together with \eqref{H1_N1}, leads to the 
%It is possible to satisfy all  constraints both at early times 
%\eqref{chi_contr_genesis}-\eqref{old_no_SC} and at the inflation stage \eqref{i%nflation_after_gen}.
These can also be satisfied without much of fine-tuning.
As an example, the allowed
range of
$x$ and $H_1$ is  shown in Fig.~\ref{fig:contr_bounce_infl}
% for fixed values of  $\mu = 0.8$, $\delta = 0.1$
for rather arbitrarily chosen $\mu = 0.8$, $\delta = 0.1$,
and $N_1 = 0.74$. %is  shown in Fig.~\ref{fig:contr_bounce_infl}.
Plots for other values of $N_1$ are similar, provided that
$N_1 \lesssim 1$. The fact that $x$ is negative and $|x|$ is small for
small $H_1$ (inflationary expansion rate much lower than the Planck
scale) is clear from, e.g.,  Eq.~\eqref{apr2-21-2}.

%The allowed
%regions of
%$x$, $H_1$ and $N_1$ are  shown in Fig.~\ref{fig:contr_bounce_infl_3D}
%for $\mu = 0.8$ and $\delta = 0.1$
%(same set as in Fig.~\ref{fig:contr_bounce_c2_3D}); we display the range of
%$x$ in which healthy initial contracting stage is possible, see
%Fig.~\ref{fig:contr_bounce_c2_3D}.
%is  shown in Fig.~\ref{fig:contr_bounce_infl_3D}.
%\begin{figure}[htb!]
%\centering
%\begin{minipage}{0.5\textwidth}
%  \centering
%\includegraphics[width=0.8\textwidth]{inflFRONT.eps}
%\subcaption[first caption.]{}\label{fig:contr_bounce_3D_infl_1a}
%\end{minipage}%
%\begin{minipage}{0.5\textwidth}
%  \centering
%\includegraphics[width=0.8\textwidth]{inflBack.eps}
%\subcaption[second caption.]{}\label{fig:contr_bounce_3D_infl_1b}
%\end{minipage}%
%
%\caption{
%3-dimensional  space of parameters $x$, $H_1$ and $N_1$
%relevant to the inflationary stage.  Values of other parameters
%are   $\mu = 0.8$,  $\delta = 0.1$.
%The regions consistent wiht stability and subluminality constraints
%  is inside the
%  yellow envelopes.}
%  The right panel is
%  the same figure  rotated by 180\si{\degree}.
%\label{fig:contr_bounce_infl_3D}
%\end{figure}
%The end of treated evolution is inflation, which is similar 
%to what we construct in subsection~\ref{sec:infl-bounce}. 
%Next, one can pass to the kination stage in the same manner
%as in subsection~ \ref{sec:kination}.

To see that the contracting genesis stage can consistently
pass through bounce to inflationary expansion, we now
give an explicit numerical example.
As we alluded to above, we choose 
%\begin{equation}
$\mu = 0.8$, $\delta =0.1$. The parameter relevant
to the contracting genesis stage is chosen as
%\label{mu_d_contr_gen}
%\end{equation}
%Next parameter is $c$ and it is again rather
%arbitrarily,
%\begin{equation}
$c = 1.7545\cdot10^{-2}$.
%\label{mar22-21-1}
%\end{equation}
%Since the latter is roughly the inverse characteristic 
%time scale in Planck units, it should be small.
%The parameters $x$ and $y_0$ are then chosen from the
%allowed region shown in Fig.~\ref{fig:1b}; 
By trial and error we
find convenient values for other Lagrangian parameters at early times, 
consistent with the system of constraints
\eqref{chi_contr_genesis}-\eqref{old_no_SC} and
\eqref{inflation_after_gen}:
\begin{equation}
%\begin{subequations}
\label{set_genesis_1}
    %\begin{align}
        x = - 2.097 \cdot 10^{-4},
        %\frac{100}{(2500)^2}, 
        \quad
        y_0 = -2.481\cdot 10^{-2} , \quad
        z_0 = 0.905.
\end{equation}
The value of the Hubble coefficient $\chi$ at the contraction stage
is found from \eqref{chi_genesis_eq},
%\marginpar{\bf V!}
%\begin{equation*}
  $\chi = 0.25$.
  %\label{mar16-21-2}
  %\end{equation*}
Next, we turn to the function $f(t)$ which should interpolate 
between
$f = -ct$ at contraction and $f=1$ at inflation.
To ensure stability and subluminality at all times,
we choose this function in somewhat more complicated form
than in \eqref{march13-21-5}: 
\begin{equation*}
  f(t)
  = \frac{c}{2}\Big[-t+\frac{\text{ln}(2\text{cosh}(st))}{s}\Big] +
  0.89\cdot U_f(t) + 1 \; ,
  %\label{march13-21-5}
\end{equation*}
where
\begin{equation*}
    U_f(t) = \text{ln}\Big(\frac{\text{e}^{4\cdot s\cdot(t-600)}
    +\text{e}}{\text{e}^{4\cdot s\cdot(t-600)}+\text{e}^2}\Big)
\end{equation*}
interpolates between $-1$ and $0$. The parameter $s$ 
is the same as in  \eqref{march13-21-11}, $s=2\cdot 10^{-3}$.

We now define the parameters of the inflationary stage and describe
transition to it through bounce. We
choose the (time-independent)
Hubble parameter and lapse function at inflation as follows:
\begin{equation*}
H_1 = 3.71\cdot 10^{-3} \; , \quad N_1 = 0.74\; .
\end{equation*}
This set of parameters is consistent with the stability and
subluminality constraints \eqref{inflation_after_gen}.
Then Eq.~\eqref{y1_z1} with $x$ given by
\eqref{set_genesis_1} 
leads to
\begin{equation}
        y_1 = -4.01\cdot 10^{-4}, \quad z_1 = 0.445.
\label{apr23-21-4}
\end{equation}
The transition from contraction  to inflation is described
by $y(t)$ and $z(t)$  smoothly interpolating between
$y_0$, $z_0$ and $y_1$, $z_1$. By trial and error, we find  appropriate functions
\begin{subequations}
  \label{general_y_z_genesis}
\begin{align}
  y(t) &= y_0(1-U_y(t)) + y_1 U_y (t),
  \\
  z(t) &=  z_0(1-U_z(t)) + z_1 U_z (t),
\end{align}
\end{subequations}
where the functions
  %\label{mar16-21-5}
\begin{align*}
  U_y(t) & =
  1+\text{ln}\Big(\frac{\text{e}^{3.8\cdot s\cdot(t+150)}+
    \text{e}}{\text{e}^{3.8\cdot s\cdot(t+150)}+\text{e}^2}\Big),
  \\
  U_z(t) &= \text{ln}\Big(\frac{\text{e}^{-5.8\cdot s\cdot(t-605)}
    +\text{e}^2}{\text{e}^{-5.8\cdot s\cdot(t-605)}+\text{e}}\Big)
\end{align*}
interpolate between 0 and 1. A nontrivial requirement
%part of the construction
leading to \eqref{general_y_z_genesis}
is again
%to guarantee that
stability
and subluminality of perturbations
%conditions are satisfied
at all times.

\begin{figure}[htb!]
\centering
\begin{minipage}{0.5\textwidth}
  \centering
\includegraphics[width=0.95\textwidth]{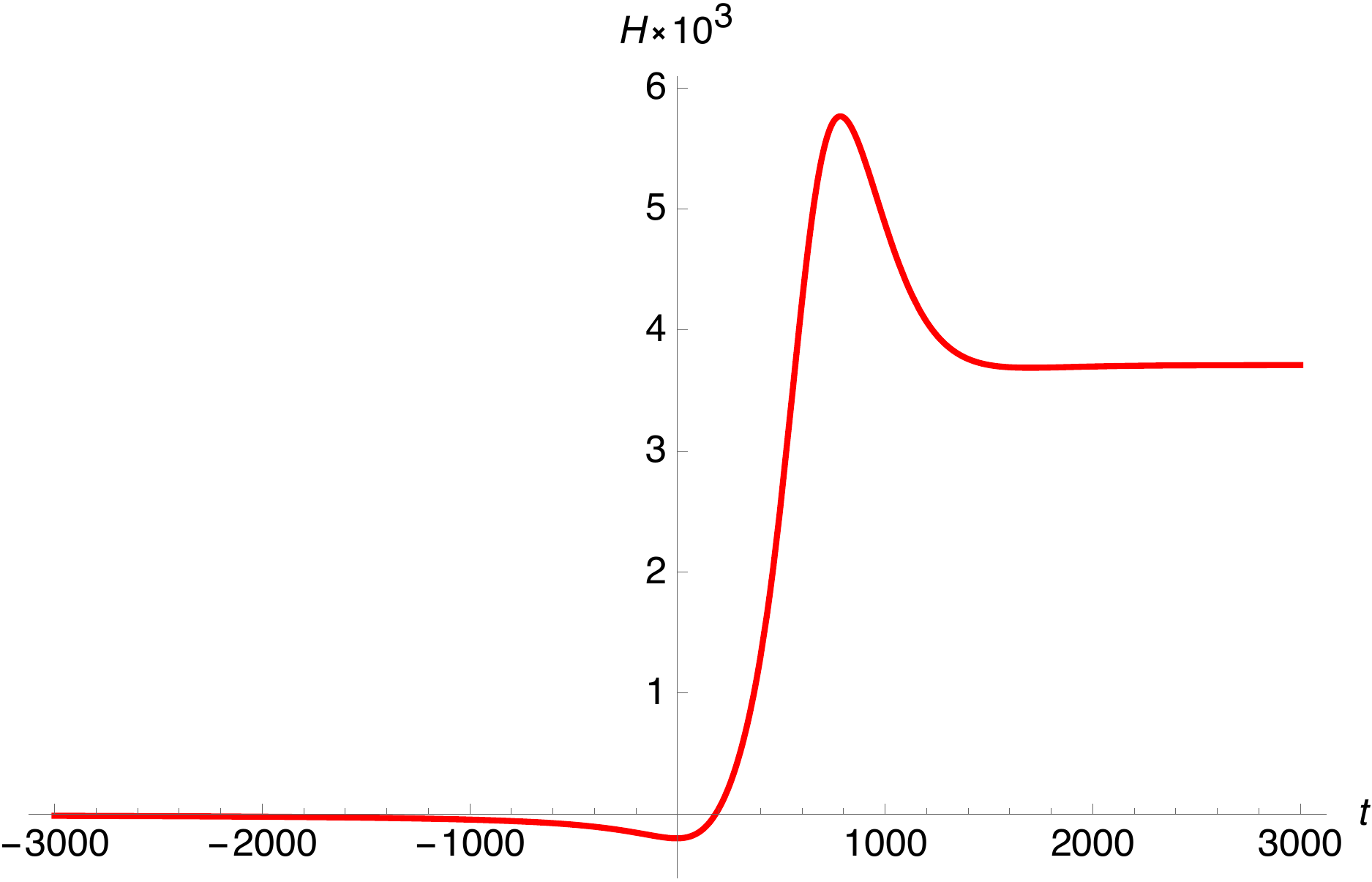}
\subcaption[first caption.]{}\label{fig:hubble_contr_gen_1a}
\end{minipage}%
\begin{minipage}{0.5\textwidth}
  \centering
\includegraphics[width=0.95\textwidth]{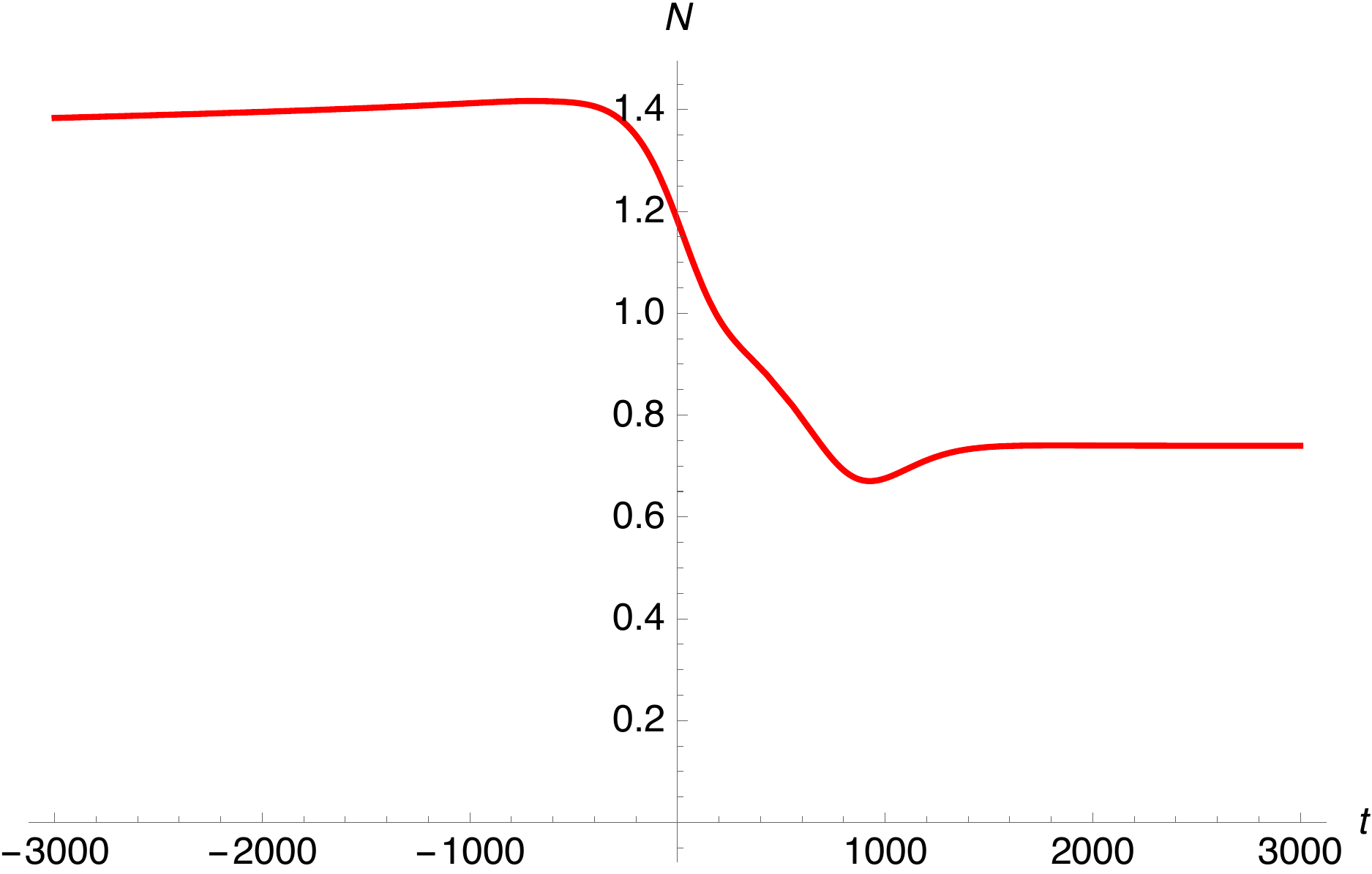}
\subcaption[second caption.]{}\label{fig:lapse_contr_gen_1b}
\end{minipage}%

\caption{Hubble parameter (left panel) 
and lapse function (right panel) for 
the model of Sec.~\ref{sec:genesis_kob}: contracting genesis and bounce.} 
\label{fig:contr_gen_H_N}
\end{figure}

\begin{figure}[H]
\centering
\begin{minipage}{0.5\textwidth}
  \centering
\includegraphics[width=0.95\textwidth]{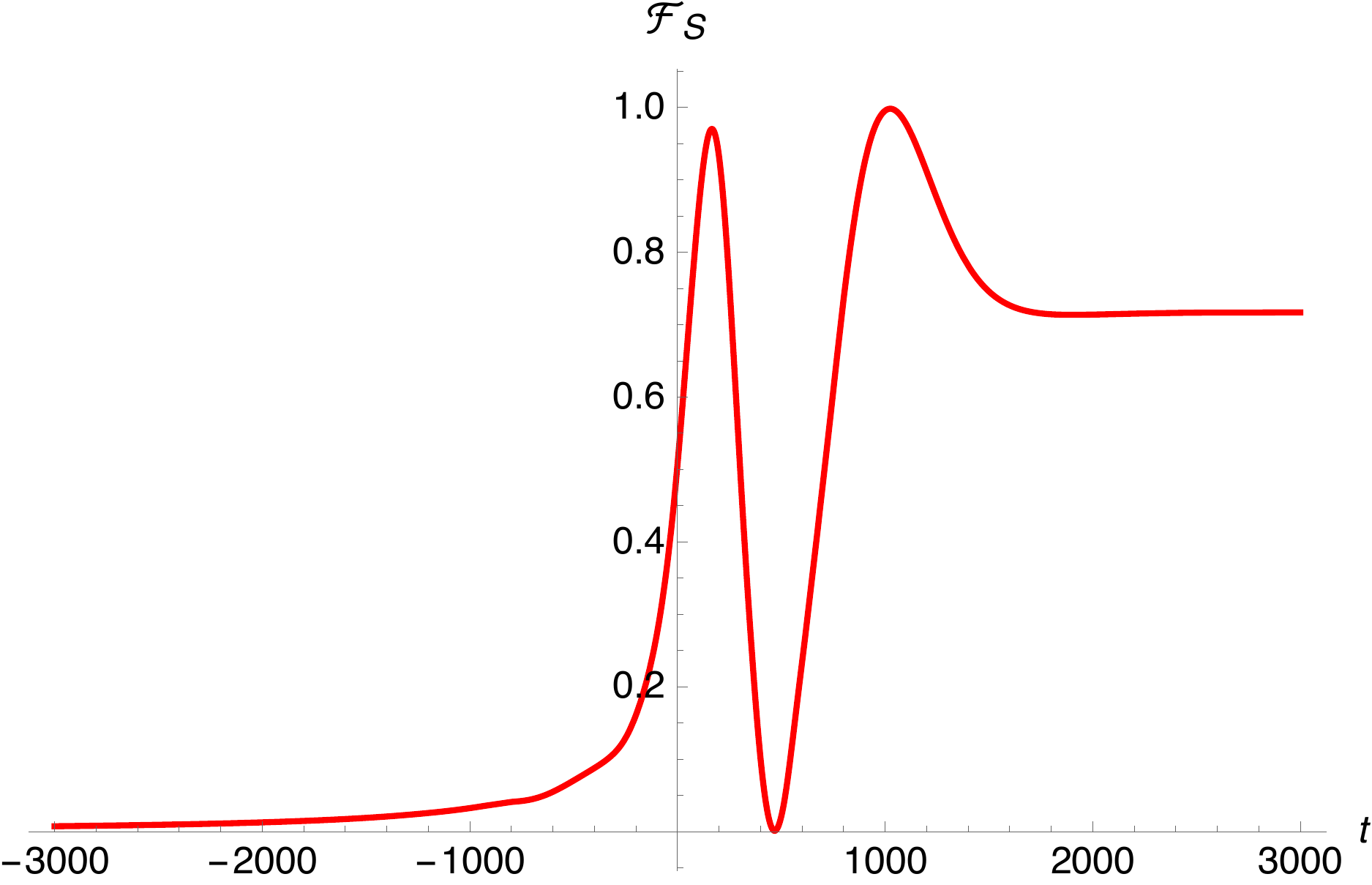}
\subcaption[second caption.]{}\label{fig:contr_genesis_fsa}
\end{minipage}%
\begin{minipage}{0.5\textwidth}
  \centering
\includegraphics[width=0.95\textwidth]{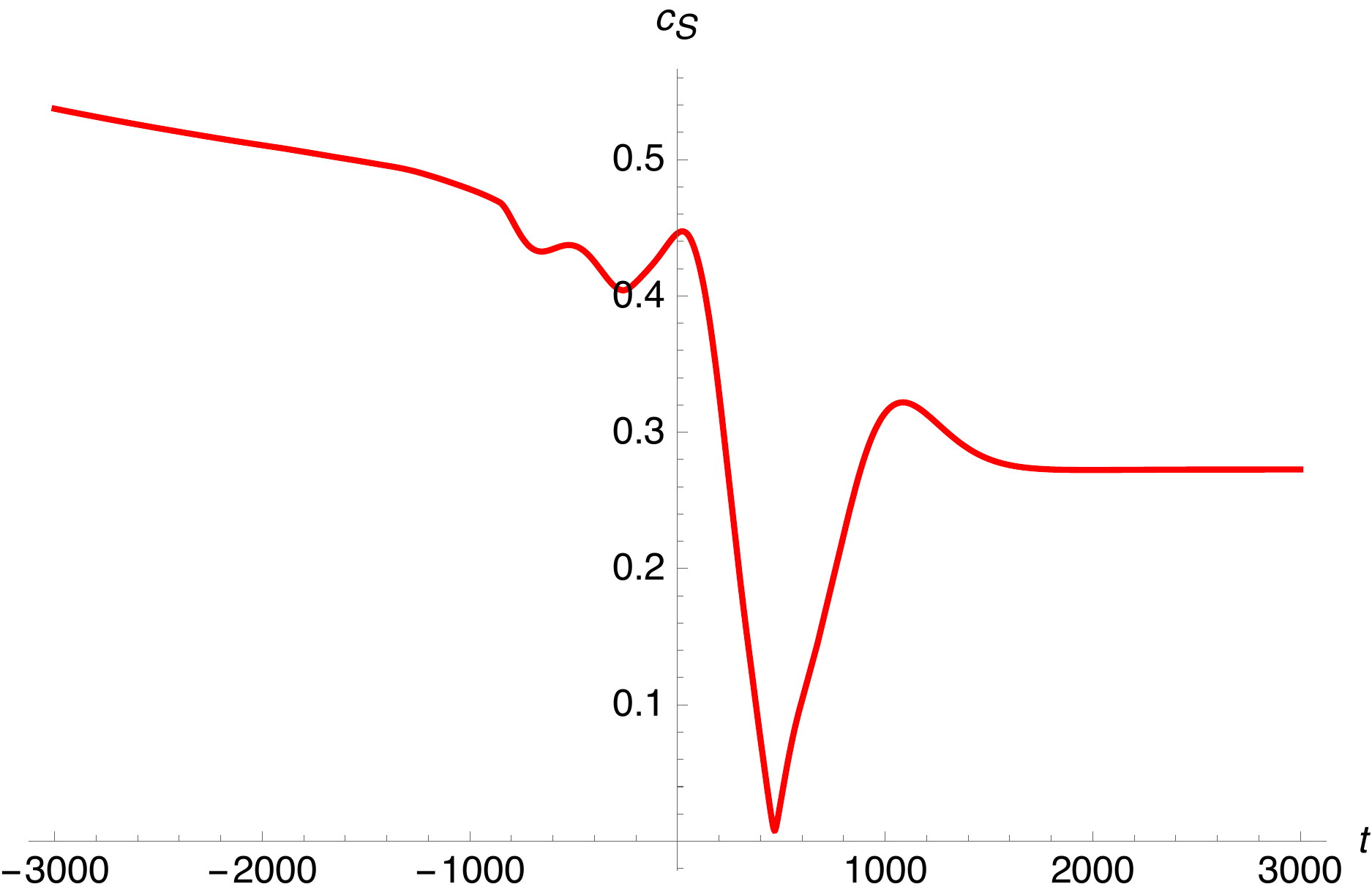}
\subcaption[second caption.]{}\label{fig:contr_genesis_csb}
\end{minipage}%

\caption{The coefficient  $\mathcal{F}_S$ (left panel)
  and the sound speed for the scalar perturbations $c_S = \sqrt{\mathcal{F}_S/\mathcal{G}_S}$ (right panel)
for the model of Sec.~\ref{sec:contr_gen}.} 
\label{fig:contr_gen_FS_cS}
\end{figure}

We show the behavior of the Hubble parameter and lapse function at
contraction, bounce and beginning of inflation in Fig.~\ref{fig:contr_gen_H_N}.
%\marginpar{\bf Vert. axis should start from 0 for $N$ and everywhere)}
%\st{ Fig. 4
%  upper left,  Fig. 5
%  upper left}.
The scalar  coefficient  ${\cal F}_S$
and scalar sound speed
$c_S$ are shown in Fig.~\ref{fig:contr_gen_FS_cS}: 
%\marginpar{\bf ${\cal F}_S$, not $\sqrt{{\cal F}_S}$ in Figs.}
%\st{ Fig. 7
%  upper left,  Fig. 8
%  upper left,  Fig. 9
%  upper left}
 the  stability and
 subluminality are explicit
(although not obvious in Fig.~\ref{fig:contr_gen_FS_cS}a,
  the
   coefficient  ${\cal F}_S$ is, in fact, strictly positive at all times). 


We end up
  this Section by the following remark. As we pointed out above,
  gravity at the inflationary epoch is not yet
  described by GR, since $z_1 \neq 0$
and, therefore $A_4 \neq -B_4$, see Eqs.~\eqref{apr23-21-2}, \eqref{apr23-21-3}.
To ensure that inflation at its last stage proceeds
within GR, one chooses $z(t)$ which,
at  the intermediate inflationary stage,
smoothly evolves from $z=z_1$ to $z=0$.
The function $y(t)$ should also be nontrivial, as it should interpolate
between the negative value $y_1$ [see~\eqref{apr23-21-4}] and some
positive value $y_2$.
  The latter property follows from the expression
  for $\mathcal{F}_S$, which for  $z=0$ has the following form:
\begin{align*}
    \mathcal{F}_S = \frac{3 y_2}{4 H_2\cdot N_2^3- 3 y_2}>0,
\end{align*}
where $N_2>0$ and $H_2>0$ are lapse function and Hubble parameter
at GR inflation. 
%eqs.~\eqref{y1_z1a},
%\eqref{y1_z1b} with the left hand side of \eqref{y1_z1b} equal to zero).
It is straightforward to design appropriate functions $z(t)$ and $y(t)$
without spoiling the stability and subluminality properties. An example is
\begin{align*}
  z(t) &= z_1 \cdot (1 - U_{z_1} (t- t_{**})) \; ,
  \\
  y(t) &= y_1 \cdot (1-U_{y_1} (t-t_{**})) + y_2 \cdot U_{y_1} (t-t_{**}) \; ,
  \end{align*}
where $y_2 = 2 \cdot 10^{-4}$, $t_{**} = 1.8\cdot 10^4$ and
\begin{align*}
  U_{z_1} (t) &= \text{ln}\Big(\frac{\text{e}^{-5.8\cdot s\cdot(t-500)}
    +\text{e}^2}{\text{e}^{-5.8\cdot s\cdot(t-500)}+\text{e}}\Big), 
  \\
  U_{y_1} (t) &= 1+\text{ln}\Big(\frac{\text{e}^{3.8\cdot s\cdot t}
    +\text{e}}{\text{e}^{3.8\cdot s\cdot t}+\text{e}^2}\Big).
\end{align*}
 The transition from late GR inflation
  to kination proceeds in the same way
  as in Sec.~\ref{sec:bounce_to_inflation}.

\subsection{Genesis without strong coupling}
\label{sec:genesis_kob}
As we pointed out in the Introduction, the genesis model of
Ref.~\cite{Kobayashi:2016xpl} suffers from the strong coupling problem
at early times. For completeness, we present here a version of this model
which is free of the strong coupling  problem.
%Here we are going to present the healthy genesis with subsequent inflation mode%l which is 
%based on the ansatz from \cite{Kobayashi:2016xpl}.
The Lagrangian 
is the same as in \cite{Kobayashi:2016xpl} but with different
parameters. Namely, the Lagrangian functions 
are defined by \textit{Ansatz} \eqref{mar16-21-10} with $x=const$, $y=const$, and 
%\marginpar{\bf{Model of Appendix A.1 (sign in $A_4$??)}}
%$x =-10^{-4}$, $y=2.5\cdot 10^{-3}$, 
$z=0$, i.e.,
\begin{subequations}
  \label{apr3-21-1}
\begin{align}
	&A_2 =  \frac{1}{2}f^{-2\mu -2 -\delta} a_2 (N) \text{,} \\ 
	&A_3 =  \frac{1}{2}f^{-2\mu -1 -\delta} a_3 (N) \text{,} \\
	&A_4 = -B_4  = -\frac{1}{2}f^{-2\mu} \text{,}
\end{align}
\end{subequations}
where $\mu >0$, $\delta >0$ and 
\begin{align*}
&a_2(N) = x\cdot\left(\frac{1}{N^2} - \frac{1}{3N^4}\right),\\
&a_3(N) =  \frac{y}{N^3}.
%\label{app:a2_a3}
\end{align*}
As before, we choose the asymptotic behavior %of $f(t)$ as follows
% \begin{equation*}
%\label{f_past_contr_gen}
%
$f =  -ct$  ($c>0$) as $t\to - \infty$,
%\end{equation*}
and using equation of motion \eqref{eoms_all_substitute_genesis} obtain
the genesis behavior
\begin{equation}
    H = \frac{\xi}{(-t)^{1+\delta}}, \quad N = 1, \quad t\to - \infty,
\label{apr2-21-3}
\end{equation}
where $\xi$ is given by
\begin{equation*}
%\label{chi_genesis_eq}
   \xi = \frac{3(2\mu+\delta+1)\cdot
    c\cdot y -2x}{12(2\mu+\delta +1)\cdot c^{\delta+2}}.
\end{equation*}
The asymptotics of the scalar coefficients and
scalar sound speed squared  are
%\marginpar{\bf write expressions}
\[
 {\cal F}_S = -(-c\cdot t)^{-2\mu+\delta}
  \cdot\frac{6 c^2\cdot (2\mu - \delta - 1 )
    \cdot(2\mu + \delta + 1)}{x+3c\cdot y\cdot(2\mu+\delta+1)} \; ,
  \quad \quad {\cal G}_S = -(-c\cdot t)^{-2\mu+\delta}\cdot \frac{18 \cdot c^2
    \cdot x\cdot(2\mu + \delta +1)^2}{\big(x+3c \cdot y
    \cdot(2\mu + \delta +1)\big)^2}\; ,
  \]
  \[
  c_S^2
    = \frac{(2\mu -\delta-1)(x+3c \cdot y \cdot (2\mu+\delta+
      1))}{3x\cdot (2\mu+\delta+1)}.
  \]
%Like in the model of Sec.~\ref{sec:contr_gen}, corrections to \eqref{apr2-21-3}
%are of order $(-t)^{-\delta}$, which makes the numerical treatment
%somewhat involved, see Appendix~\ref{app:variable_u}.
%where $c$ is a constant parameter.

%We want to mark, that this background solution for Hubble parameter has the fol%lowing expansion by $(-t)^{-\delta}$:
%\begin{equation}
%    H(t) = -(-t)^{-1-\delta}\cdot\left(\chi + \chi_1\cdot (-t)^{-\delta}+ \ldot%s\right),
%    \label{H_series}
%\end{equation}
%where $\ldots$ means terms higher order by $(-t)^{-\delta}$,
%and $\chi_1$ is given by
%\begin{align*}
%    \chi_1 = -\frac{c^{-4-2\delta} \big(8+ 3c\cdot(2\mu+\delta + 1)\big)}{12288%(2\mu+\delta + 1)^2\cdot(2\mu+2\delta+1)}\cdot\big\{64 - 24 c\cdot\delta + 45 c%^2 \cdot(2\mu+\delta+1)\cdot(2\mu+2\delta+1)\big\},
%\end{align*}
%Similarly as in previous subsection, 
%this series \eqref{H_series} for Hubble has the dumping factor
%$(-t)^{-\delta}$, so for sufficiently small $\delta$ it converges 
%only at fairly large negative times. These large values of time again 
%lead to   difficulties with numerical calculations. We solve this problem
%in  completely the same way as for contraction genesis and bounce model. 
Parameters of this model should obey several constraints. The first
one comes from the
requirement of evading the
no-go argument of Ref.~\cite{Kobayashi:2016xpl}:
\begin{equation}
    2\mu>1+\delta \; .
\label{apr14-21-1}
\end{equation}
The second constraint ensures that 
the classical treatment of early time evolution is
legitimate \cite{Ageeva:2020buc}:
\begin{equation}
    \mu + \frac{3}{2}\delta<1.
\label{apr14-21-2}
\end{equation}
The third one is the requirement that the Universe expands
at early times:
    \begin{equation}
        \xi>0.
     \label{xi_genesis}
    \end{equation}
Finally, one requires the
background stability 
and subluminal propagation of perturbations. For $y>0$ (as needed for
healthy inflation, see below), all these constraints
  are satisfied, provided that $x<0$ and
  %, ${\mathcal F}_S >0$, ${\mathcal G}_S >0$ and $c_S<1$, respectively. 
  %The explicit forms of these  constraints are ~~~~~~~~~~{\bf $y>0$ or not?!!!}
  %together with \eqref{xi_genesis} can be satisfied.
  %In view of no-go argument, we have
\begin{align}
\label{constr_gen_minus_infty}
    %{\cal G}_S >0 ~:& ~~~~~~~~~ x<0 ;
    %\\
%{\cal F}_S >0 ~:& ~~~~~~~~~
x<-3c\cdot y(2\mu+\delta+1) \; .
%      \\
%      c_S^2 \leq 1  ~:& ~~~~~~~~~ x < \frac{3c\cdot y\cdot (2\mu+\delta+1)\cdot%(2\mu-\delta-1)}{4\cdot(\mu+\delta+1)};\\
%      \xi>0~:& ~~~~~~~~~ x<\frac{3}{2}c\cdot y(2\mu+\delta+1) \; .
\end{align}
The
transition from the genesis stage to inflation is achieved simply by
flattening out the function $f(t)$ to $f=1$.
For $f=1$, equations of motion
\eqref{eoms_all_substitute_genesis} read
%One can substitute \eqref{lagr_func_bounce_to_infl} into \eqref{eoms_all_substitute} 
%and it leads to the following equations of motion:
\begin{subequations}
\label{eoms_all_substitute_gen}
\begin{align}
  &x\left( \frac{1}{N^4}-\frac{1}{N^2}\right) -
  \frac{9 y\cdot H}{N^3} 
    + 6 H^2  = 0, \\
    &x\left(\frac{1}{N^2} - \frac{1}{3N^4}\right) + 6 H^2 = 0\; , 
    %- 
    %\frac{1}{N}\frac{d}{dt}\left( \frac{y_1}{N^3} - 4 H\right) = 0,
\end{align}
\end{subequations}
and we denote the (time-independent) solution to these equations by
$H = H_1$ and $N=N_1$. We require $H_1 > 0$, $N_1 >0$. 
%are constants and the solution of the above EoM. 
%It means that at these times the inflation takes place.  However, it is 
%not guaranteed at all, that the requirement \eqref{lagr_func_bounce_to_infl} le%ads to 
%the bounce at some moment of time, 
%since there may be not exist the solution, which connects the demanded asymptot%ics  
%\eqref{hubble_bounce} and ($H_1,N_1$). So, for each set of our parameters 
%one should check the existence of such a solution numerically.
The requirements of
%Our first constraint comes from the fact that we want to obtain the 
%inflation in the future, i.e.
%\begin{equation}
%\label{H1_N1}
%    H_1 > 0, \quad N_1 > 0.
%\end{equation}
%The second requirement is the necessity of
background stability 
and subluminal propagation of perturbations,
Eqs.~\eqref{stability_all_subtitute_genesis},
read in this case
%~~~~~~~~~~~{\bf  \textcolor{magenta}{Exact expressions?!!!}}
\begin{subequations}
\label{stabil_gen_kob_infl}
\begin{align}
     \mathcal{F}_S &= \frac{3y}{4 H_1\cdot N_1^3 - 3 y}>0,
%\end{align}
%\begin{align}
     \quad \quad     \mathcal{G}_S
       = \frac{8 N_1^2\cdot x\cdot  (N_1^2-2)
         + 72 H_1\cdot N_1^3 \cdot y + 27 y^2}{(4H_1\cdot N_1^3 - 3y)^2}  >0,
\end{align}
\begin{align}
  c_S^2 &= \frac{3y\cdot (4H_1\cdot N_1^3 - 3y)}{8 N_1^2\cdot x\cdot
    (N_1^2-2)+ 72 H_1\cdot N_1^3 \cdot y + 27 y^2}  \leq 1,
%\label{apr2-21-2}
\end{align}
\end{subequations}
%In analogy to  previous subsection,
 As before, it
is convenient to treat $H_1$ and $N_1$
as independent parameters and  express
$x$ and $y$ through
these parameters  using
\eqref{eoms_all_substitute_gen}:
\begin{subequations}
\label{x_y}
\begin{align}
    x &= \frac{18 H_1^2\cdot N_1^4}{1-3N_1^2}, \\
    y &=  \frac{4 H_1\cdot N_1^3\cdot(3N_1^2-2)}{(9N_1^2-3)}.
\end{align}
\end{subequations}
%Thus the requirements of
%background stability 
%and subluminal propagation of perturbations \eqref{stabil_gen_kob_infl}
%read~~~~~~~~~~~{\bf  \textcolor{magenta}{Exact expressions?!!!}}
%\begin{align*}
%    {\cal G}_S = 6-27N_1^2 + 54 N_1^4 &>0;
%    \\
%    {\cal F}_S = -2+3N_1^2 &>0;
%      \\
%      c_S^2 = \frac{-2+3N_1^2}{6-27 N_1^2 + 54 N_1^4} &\leq 1.
%\end{align*}
Then the
constraints \eqref{constr_gen_minus_infty} and 
\eqref{stabil_gen_kob_infl} reduce to
%
%together with \eqref{constr_gen_minus_infty} and \eqref{x_y} give:
\begin{subequations}
\label{H_1_N_1_gen}
\begin{align}
    N_1 &> \frac{\sqrt{6}}{3},\\
    H_1 &> \frac{2 c \cdot (2\mu+\delta +1)
      \cdot (3N_1^2 -2)}{9 N_1}.
\end{align}
\end{subequations}
In accordance
  with \eqref{x_y}, these constraints translate into constraints on
  $x$ and $y$. It is straightforward to
  see that the latter are satisfied, provided that $y>0$ and $x$ obeys
  \eqref{constr_gen_minus_infty}.
%where we also show the bound \eqref{constr_gen_minus_infty}.}
%\marginpar{\bf \textcolor{magenta}{What about
%    bound \eqref{constr_gen_minus_infty}?}}

\begin{figure}[htb!]
\centering
\begin{minipage}{0.5\textwidth}
  \centering
\includegraphics[width=0.95\textwidth]{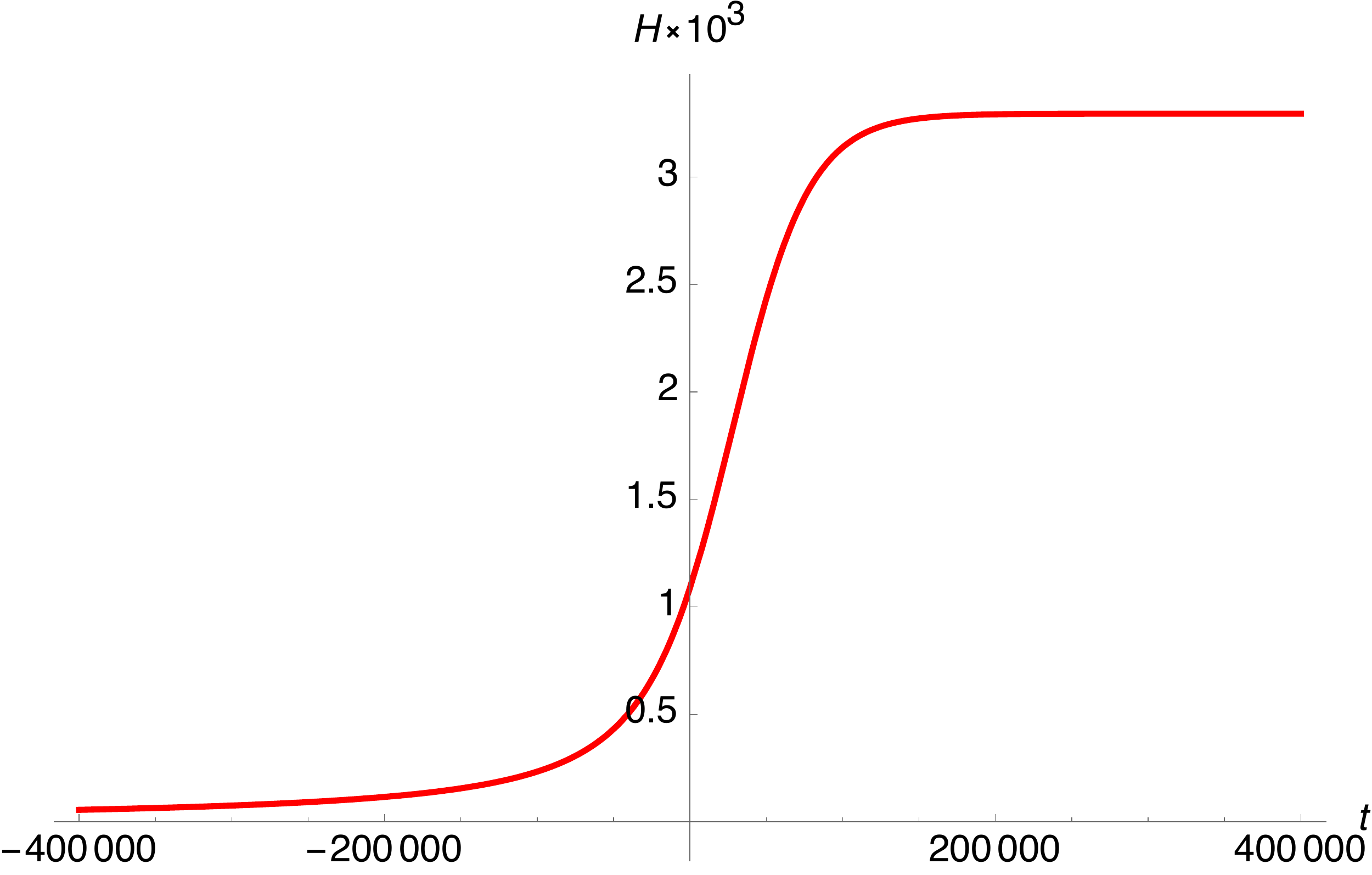}
\subcaption[first caption.]{}\label{fig:hubble_gen_1a}
\end{minipage}%
\begin{minipage}{0.5\textwidth}
  \centering
\includegraphics[width=0.95\textwidth]{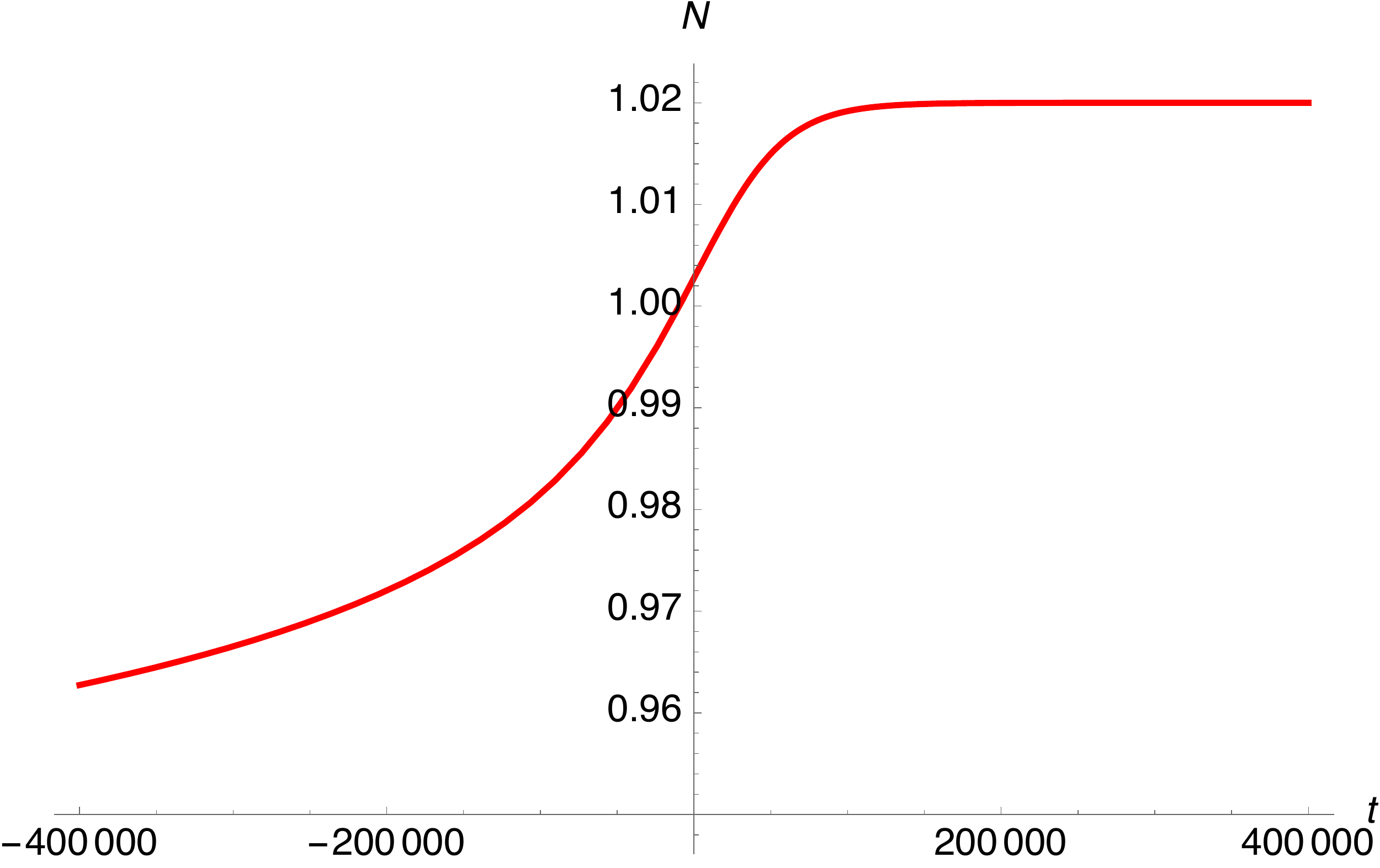}
\subcaption[second caption.]{}\label{fig:lapse_gen_1b}
\end{minipage}%

\caption{Hubble parameter (left panel) 
and lapse function (right panel) for the
model of Sec.~\ref{sec:genesis_kob}: genesis without strong coupling.} 
\label{fig:gen_H_N}
\end{figure}

\begin{figure}[htb!]
\centering
\begin{minipage}{0.5\textwidth}
  \centering
\includegraphics[width=0.95\textwidth]{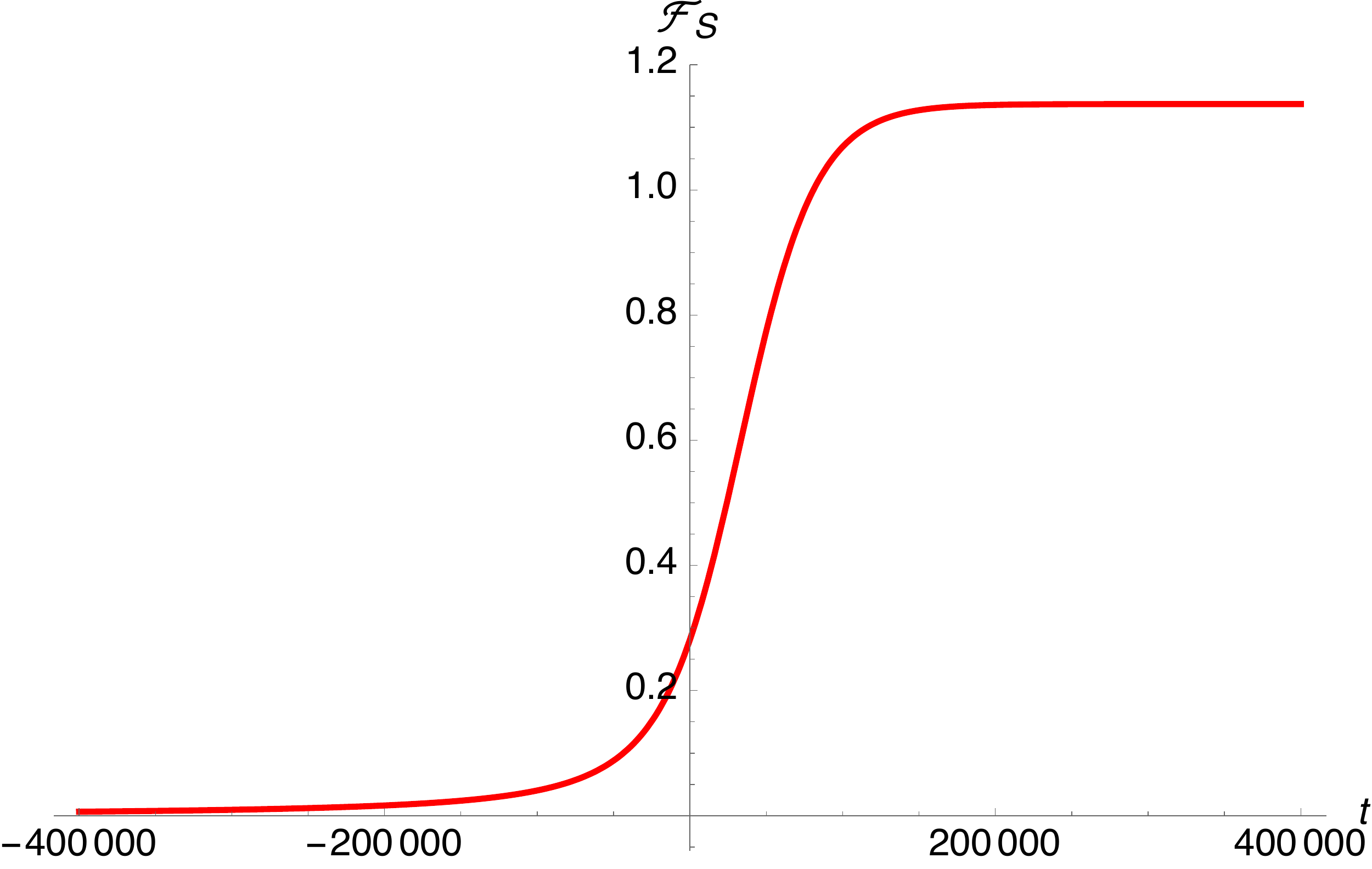}
\subcaption[second caption.]{}\label{fig:genesis_fsa}
\end{minipage}%
\begin{minipage}{0.5\textwidth}
  \centering
\includegraphics[width=0.95\textwidth]{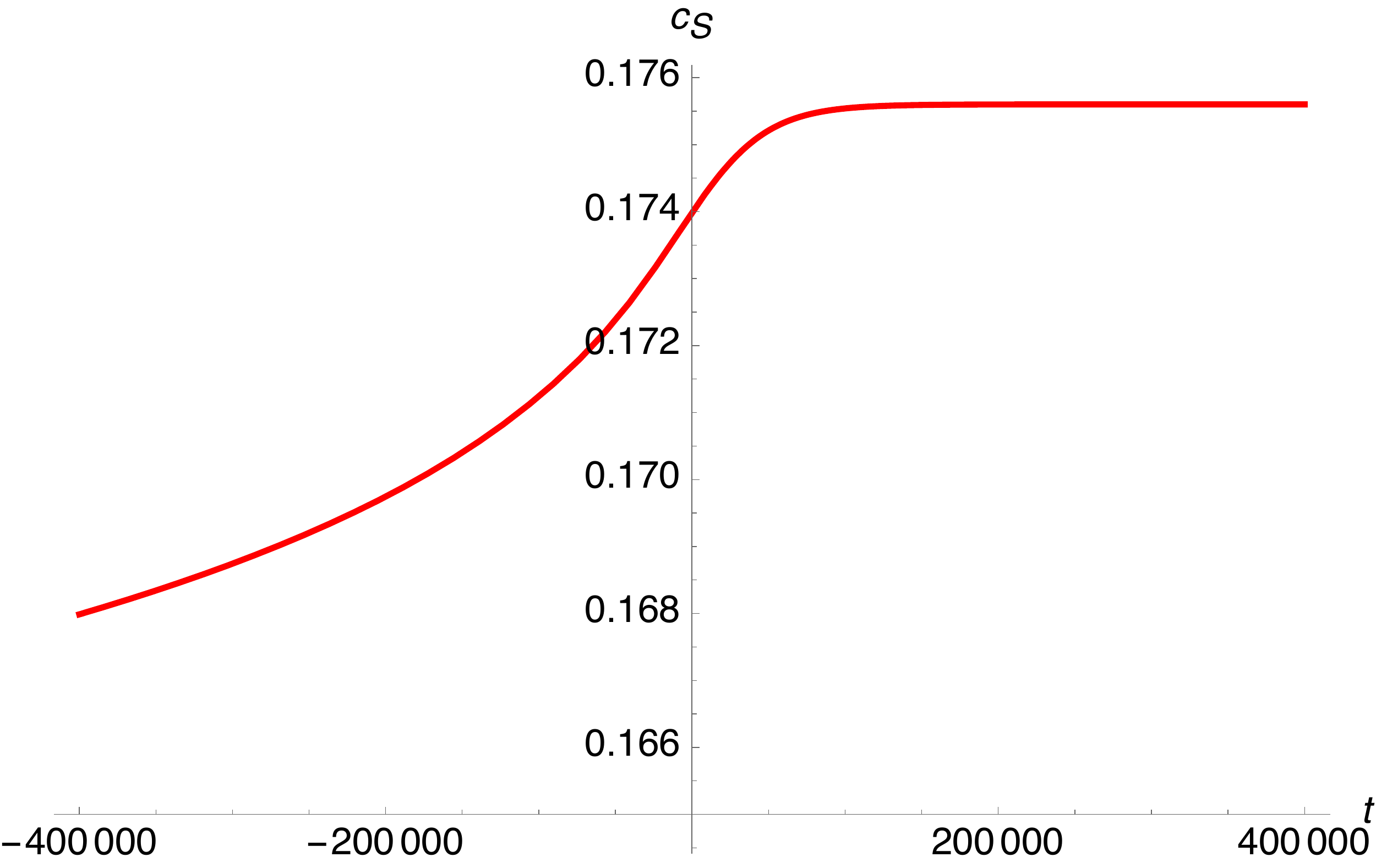}
\subcaption[second caption.]{}\label{fig:genesis_csb}
\end{minipage}%

\caption{The coefficient  $\mathcal{F}_S$ (left panel)
  and scalar sound speed
  $c_S = \sqrt{\mathcal{F}_S/\mathcal{G}_S}$ (right panel)
  for the model of Sec.~\ref{sec:genesis_kob}.} 
\label{fig:gen_FS_cS}
\end{figure}

Now, let us turn to our numerical example.  We choose 
\begin{equation*}
    \mu = 0.65, \quad \delta=0.2.
\end{equation*}
This choice is consistent  with the constraints
  \eqref{apr14-21-1} and
\eqref{apr14-21-2}.
%
%with no-go argument and absence of strong coupling regime.
We choose
    \begin{align*}
        &f(t) = \frac{c}{2}\Big[-t+\frac{\text{ln}(2\text{cosh}(st))}{s}\Big] + 1,\nonumber \\
        &c = 10^{-4}, \quad s = 2\cdot10^{-5}.
    \end{align*}
    We obtain the values of $x$ and $y$ by considering the
    %We now define the parameters of
    inflationary stage. We
choose  $H_1$ and $N_1$ at inflation as follows:
\begin{equation*}
H_1 = 3.3\cdot 10^{-3} \; , \quad N_1 = 1.02\; .
\end{equation*}
This set of parameters is consistent with constraints \eqref{H_1_N_1_gen}.
Then Eq.~\eqref{x_y}  
leads to
\begin{equation*}
        x = - 10^{-4}, \quad y = 2.5 \cdot 10^{-3} \; ,
\end{equation*}
which is consistent with \eqref{constr_gen_minus_infty}.
We show 
    the evolution of the Hubble parameter, lapse function, scalar
    coefficient ${\cal F}_S$,
    and scalar sound speed in Figs.~\ref{fig:gen_H_N}
    and \ref{fig:gen_FS_cS}. In the tensor sector we have
    ${\cal F}_T = {\cal G}_T = f^{-2\mu} > 0$, $c_T=1$.
    Thus, our background solution is fully stable and
free of the strong coupling problem at early times;  
perturbations about it are not superluminal. We conclude that
our model gives
an example of healthy genesis with strong gravity in the past.

%  1) Solution for early genesis,
 % with corrections of order
  %\[
  %\frac{1}{(-t)^\delta}
  %\]
 %Still NOT numerics!  Do NOT introduce $u$.

  %2) Say that the end of treated evolution is inflation, similar to
  %Sec.~\ref{sec:infl-bounce}.

  %3) Give numerical example:

  %beginning;

  %give values of new $x_1$, $v_1$, $y_1$ at inflation,
  %explain them;

  %give numerical expressions for  
  %all functions at and around transition epoch. In particular, why do you
 % need $f_0$ in $f(t)$?

  %4) Figures for Hubble, lapse, ${\cal F}_T$,  ${\cal F}_S$
  %and two sound speeds.

  %In Appendix: explain what is the problem and how it is cured by
%introducing variable $u$.

%\begin{figure}[htb!]
%  \centering
%  \begin{subfigure}[b]{0.45\linewidth}
%    \includegraphics[width=\linewidth]{c(t)_1_S.jpeg}
%     \caption{Coffee.}
%  \end{subfigure}
%  \begin{subfigure}[b]{0.45\linewidth}
%    \includegraphics[width=\linewidth]{c(t)_2_S.jpeg}
%    \caption{More coffee.}
%  \end{subfigure}
%  \begin{subfigure}[b]{0.8\linewidth}
%    \includegraphics[width=\linewidth]{c(t)_3_S.jpeg}
%    \caption{Too much coffee.}
%  \end{subfigure}
%  \caption{The same cup of coffee. Multiple times.}
% \end{figure}

\section{Conclusion}
\label{sec:summary}

This
  paper demonstrates that it is relatively straightforward
to construct, within the Horndeski class of
scalar-tensor theories, nonsingular
cosmological models which are healthy, i.e., free of
instabilities and superluminal propagation of
perturbations. The price to pay is strong gravity in the past,
the property that effective Planck masses tend to zero
as $t\to -\infty$. We have made sure, however, that the latter property
does not spoil the description of the background within classical
field theory. In this way we have constructed bouncing
  models,
genesis cosmology, and a combination thereof. These may or may not pass through the
inflationary stage, as we explicitly demonstrated in Sec.~\ref{sec:bounce}.

In
  our constructions, we heavily exploited the functional freedom
that exists in the Horndeski class of theories. On the one hand,
this freedom is instrumental for designing models with prescribed properties
(in other words, for employing
the ``inverse method'' \cite{Ijjas:2016tpn}); on the other,
it makes the whole approach not so appealing. It is certainly desirable to
have a better idea of which Horndeski theories, if any, have a chance to
be realistic as low energy effective theories.

\section*{Acknowledgments} 

The authors are grateful to Victoria Volkova and Sergei Mironov 
for useful comments and fruitful discussions. This work has been 
supported by Russian Science Foundation Grant No. 19-12-00393.

\appendix

\section{ABSENCE OF STRONG COUPLING AT EARLY 
TIMES}
\label{app:strong_coupl}
\numberwithin{equation}{section}

\setcounter{equation}{0}
The purpose of this Appendix is to study
the issue of
strong coupling at early times in the model with the Lagrangian
\eqref{adm_lagr} and \eqref{A_ansatz} and contracting background solution
\eqref{hubble_bounce} and \eqref{x_y_v_early}. As we outlined in
Sec.~\ref{sec:intro}, we are going to compare the energy scale of classical
evolution $E_{class} \sim |t|^{-1}$ with quantum strong coupling
scale $E_{strong}$ inferred from the analysis of the nonlinear terms in the
Lagrangian for perturbations about this background.
%The idea and many
%technicalities here are similar to the ones introduced for
%the genesis case~\cite{Ageeva:2020buc}.
%: obtain and use the \textit{unconstrained} action written 
We begin with the action written in terms of %variables $\zeta$ and
 metric variables,
%(we sometimes
%omit indices in the notation of tensor perturbations $h_{ij}$ for brevity): 
%Now let us turn to the analysis of the strong coupling in such a model.
%our action is 
\begin{equation*}
  \mathcal{S} = \int{d^4 x \sqrt{-g}\mathcal{L}} = \int dt d^3 x
  N a^{3} (1+\alpha)  \text{e}^{3\zeta}  \mathcal{L},
\end{equation*}
with the Lagrangian \eqref{adm_lagr} (recall that $N$ denotes the background
lapse function).
Different terms in the Lagrangian  \eqref{adm_lagr} contain different
powers of the scale factor, which in the contracting Universe
nontrivially depends on time, $a(t) \propto (-t)^{-\chi}$. It is therefore
convenient to work with
\textit{physical} 
momenta and frequencies. Assuming that they are higher than $E_{class}$,
i.e., assuming
that  $E_{strong} \gg E_{class}$, we can neglect slow dependence of
the scale factor
on
$t$ and at a given time treat $a$ as an (instantaneously)
time-independent parameter (with the exception of expressions that
involve the Hubble parameter explicitly). Of
course, this assumption must be justified {\it a posteriori}: the whole
analysis is valid if the classical treatment of the background is legitimate,
$E_{strong} \gg E_{class}$.
Having  this in mind, we introduce
physical spatial and temporal coordinates
\begin{align*}
    \tilde{x} &\equiv x a, \nonumber \\
    \tilde{t} &\equiv t N = t.
\end{align*}
Note that
$N=1$ in our case, but we keep the notation $\tilde{t}$ 
for concordance with spatial coordinates.
Then the derivatives are, respectively,
\begin{subequations}
\label{phys_der}
\begin{align}
    \tilde{\partial}_i &\equiv \frac{1}{a} \partial_i, \\
    \tilde{\partial}_t &\equiv \frac{1}{N} \partial_t = \partial_t.
\end{align}
\end{subequations}
We now rewrite our Lagrangian \eqref{adm_lagr} in terms 
of physical coordinates.
%We have
%$\sqrt{-g} d^4 x =  (1+\alpha) e^{3\zeta} d \tilde{t} d^3 \tilde{x} $.
We define 
\begin{equation}
E_{ij} = K_{ij}  N (1+\alpha)\; ,
\label{app:E_K}
\end{equation}
and find 
\begin{align}
    E^i_j &= \gamma^{ik} E_{kj} = \gamma^{ik} \Big[\frac{1}{2}\Big(\dot\gamma_{kj} -\;^{(3)}\nabla_{k}N_{j}-\;^{(3)}\nabla_{j}N_{k}\Big)\Big] \nonumber \\
    & = \frac{1}{2} \Big[\gamma^{ik}\dot{\gamma}_{kj} - 2 \gamma^{ik} \partial_j \partial_k \beta + 2 \Gamma^{k}_{lj}\partial_k \beta \gamma^{il} + 2 \Gamma^k_{lj} N^T_k \gamma^{il} - \gamma^{ik}\partial_k N^T_j - \gamma^{ik}\partial_j N^T_k \Big] \nonumber \\
    & = \frac{1}{2} \Big[\gamma^{ik}\dot{\gamma}_{kj} - 2 \text{e}^{-h_{ik}}\text{e}^{-2\zeta} \tilde{\partial}_j \tilde{\partial}_k \beta + 2 \tilde{\Gamma}^{k}_{lj}\tilde{\partial}_k \beta \text{e}^{-h_{il}}\text{e}^{-2\zeta} + 2 \tilde{\Gamma}^k_{lj} \tilde{N}^T_k \text{e}^{-h_{il}} \text{e}^{-2\zeta} \nonumber \\
    &- \text{e}^{-h_{ik}} \text{e}^{-2\zeta}\tilde{\partial}_k \tilde{N}^T_j - \text{e}^{-h_{ik}}\tilde{\partial}_j \tilde{N}^T_k \Big],
    \label{app:E}
\end{align}
where $\text{e}^{h_{ij}} \equiv (\text{e}^h)_{ij}$,
    \begin{align}
      \gamma^{ik}\dot{\gamma}_{kj} = \frac{1}{a^2}\text{e}^{-h_{ik}}\text{e}^{-2\zeta}\frac{\partial}{\partial t} \Big(a^2 \text{e}^{2\zeta} \text{e}^{h_{kj}}\Big) = 2
      H \delta^i_j+ \text{e}^{-h_{ik}}e^{-2\zeta}\frac{\partial}{\partial \tilde{t}}
      \Big( \text{e}^{2\zeta} \text{e}^{h_{kj}}\Big) \; ,
\label{mar18-21-1}
    \end{align}
 the new (physical) transverse shift vector is given by
\begin{align*}
    \tilde{N}^T_k \equiv \frac{N^T_k}{a}
\end{align*}
and $\tilde{\Gamma}^l_{ij}$ (and  $^{(3)}\tilde{R}$ below)
 are made of metric $\tilde{\gamma}_{ij}=
\text{e}^{h_{ij}} \text{e}^{2\zeta}$.
We rewrite $^{(3)}R$ in the same way:
\begin{align}
  ^{(3)}R &= \gamma^{ij} \, ^{(3)}R_{ij}
  = \gamma^{ij} \Big[ \partial_l \Gamma^l_{ij} - \partial_i \Gamma^l_{lj} + \Gamma^l_{ij}\Gamma^m_{ml} - \Gamma^m_{il}\Gamma^l_{jm} \Big] \nonumber \\
    &= \text{e}^{-h_{ij}}\text{e}^{-2\zeta} \Big[ \tilde{\partial}_l \tilde{\Gamma}^l_{ij} - \tilde{\partial}_i \tilde{\Gamma}^l_{lj} + \tilde{\Gamma}^l_{ij}\tilde{\Gamma}^m_{ml} - \tilde{\Gamma}^m_{il}\tilde{\Gamma}^l_{jm} \Big] =  ^{(3)}\tilde{R}.
\label{app:R}
\end{align}
%where  $\tilde{\Gamma}^l_{ij}$ is made of metric $\tilde{\gamma}_{ij}=
%e^{h_{ij}} e^\zeta$.
Finally, we have
$\sqrt{-g} d^4 x =  (1+\alpha) \text{e}^{3\zeta} d \tilde{t} d^3 \tilde{x} $.
Thus, the action written in terms of
physical variables does not contain the scale factor
anymore, but otherwise  the Lagrangian has the same structure as the
original  one \eqref{adm_lagr}, except for the first term
in the right-hand side of \eqref{mar18-21-1}.

The action for perturbations, written in terms of physical variables
$\tilde{x}^i$, $\tilde{t}$, is identically the same as the limiting case
of the action that we encountered in
Ref.~\cite{Ageeva:2020buc}, where we studied the strong coupling issue
in the model with genesis %, which is the same as the model
%. Namely, the model that admits genesis is
described
in Sec.~\ref{sec:genesis_kob}.  Namely, the model with genesis
has an additional
parameter $\delta > 0$ in \eqref{apr3-21-1}, while the bouncing model
we discuss does not.
By direct inspection we find that
the action for perturbations in the bouncing model,
written in terms of physical variables, is obtained
from the action for perturbations in the genesis model by sending
$\delta \to 0$ [this includes also the first term
in the right-hand side of \eqref{mar18-21-1}]. Thus, the sufficient condition
for the absence of the strong coupling problem is obtained from
the result of Ref.~\cite{Ageeva:2020buc} in the limit $\delta \to 0$
and reads
\[
\mu < 1 \; .
\]
 This result is quoted in Sec. \ref{sec:bounce}.

 Let us illustrate this constraint by considering 
 the quadratic and cubic action for tensor modes.  The
   complete expression is~\cite{Gao:2011vs}
\begin{align*}
  \mathcal{S}^{(2)}_{hh} +
  \mathcal{S}^{(3)}_{hhh} =  \int d\tilde{t}  d^3 \tilde{x} \mathcal{F}_T
  (\tilde{\partial} h_{ij})^2 +
  \int d\tilde{t}  d^3 \tilde{x} \frac{\mathcal{F}_T}{4} \big( h_{ik} h_{jl}  -  \frac{1}{2}h_{ij} h_{kl}\big) \tilde{\partial}_k \tilde{\partial}_l h_{ij}  \; .
\end{align*}
We recall that  $\mathcal{ F}_T \propto   (- \tilde{t})^{-2\mu}$,
see \eqref{asy_Ft}. To figure out the associated strong coupling scale,
we introduce the canonically normalized field (omitting indices)
\[
h_{c} = {\mathcal F}_T^{1/2} h \propto (-\tilde{t})^{-\mu} h \; ,
\]
and find that the interaction term is (modulo numerical factor)
\[
\int d\tilde{t}  d^3 \tilde{x} \mathcal{F}_T^{-1/2} h_c h_c
\tilde{\partial}^2 h_c \; .
\]
Thus, on dimensional grounds, the strong coupling scale
is
\begin{align*}
    E_{strong} \propto \mathcal{F}_T^{1/2} \propto (-\tilde{t})^{-\mu} \;.
\end{align*}
This scale is much higher than the classical scale
$E_{class} = \tilde{t}^{-1}$ for $\mu <1$, as promised.

\section{CONTRACTING GENESIS:  SUBTLETY
  OF NUMERICAL SOLUTION}
\label{app:variable_u}
\numberwithin{equation}{section}
\setcounter{equation}{0}

As we pointed out in  Sec.~\ref{sec:contr_gen}, 
corrections to the leading asymptotics of classical solutions at
early times behave as $(-t)^{-\delta}$. This makes  straightforward
numerical treatment problematic for small $\delta$. Here we sketch our way of
dealing with this problem. We consider for
definiteness the model of Sec.~\ref{sec:contr_gen}
with $\delta=0.1$.

We study the time interval $-\infty < t < t_1$, where $t_1$
is negative and $|t_1|$ is
large enough, so that it is a very good approximation to
use the asymptotics 
\begin{equation}
  \label{app:y_z}
    y(t) = y_0\; , \quad \quad
    z(t) = z_0, \; , \quad \quad f(t) =  -ct
\end{equation}
(recall that these asymptotics are approached exponentially fast
backwards in time). We introduce
a    new  variable $u$ instead of $t$:    
   \begin{align}
   \label{app:u}
   u &\equiv (-ct )^{-\delta} > 0\; ;
   \end{align}
   early-time asymptotics occur as $u \to 0$. Then
   corrections to  the leading asymptotics of classical solutions
   are  of order $u$. The key point is
to introduce, instead of $H$, a new unknown  
function $k(u)$ as follows:
   \begin{align*}
   %\label{app:k_def}
   k(u) \equiv u^{-1/\delta-1} \cdot H(u)\cdot N(u).
   \end{align*}
   Then coefficient $u^{2/\delta}$ factors out in equations of
   motion~\eqref{eoms_all_substitute_genesis}, and coefficients in these
   equations become linear polynomials in $u$. This form of
   equations of motion enables one to solve them in a straightforward way.

The initial condition 
in distant past is set by expanding $N(u)$ and $k(u)$ in $u$ at small $u$.
The first nontrivial terms in this expansion are straightforward to calculate.
One writes, in notations \eqref{H_N_series},
 \begin{equation*}
    %\label{k}
    N(u) = 1 + N_1 \cdot c^{\delta} \cdot u  +\ldots,
 \end{equation*}
 and
    \begin{equation*}
   % \label{app:k}
    k(u) = - \chi\cdot c^{1+\delta}  - (\chi \cdot N_1 + \chi_1 )\cdot c^{1+2\delta}\cdot u+\ldots \; .
\end{equation*}
For 
our choice  of parameters
$x$, $y_0$, and $z_0$ in  \eqref{set_genesis_1}, and $\mu = 0.8$, $\delta =0.1$, and 
$c = 1.75\cdot10^{-2}$,
the coefficients are
\begin{align*}
    \chi \cdot c^{1+\delta} = 0.0029,\quad
    (\chi \cdot N_1 + \chi_1 )\cdot c^{1+2\delta} = 0.0043, \quad
    N_1\cdot c^{\delta} = 1.09.
\end{align*}
Thus, corrections are small for small enough $u=u_0$, where initial
conditions are imposed. In practice we choose 
 $u_0 = 10^{-7}$, 
 which corresponds to 
 $t_0 = -6 \cdot 10^{71}$. This huge number is the reason why we
 have invented our procedure.

 We solve the equations of motion written in terms of $u$ until
$u$ becomes roughly of order 1; in practice we choose
 $u_1 = (-ct_1 )^{-\delta} = 0.67$, so that $t_1 = 3000$. At that time
 Eq.~\eqref{app:y_z} is still a good approximation.
 Then we continue solving equations of motion using  time $t$,
 with obvious matching at $t=t_1$.

\end{document}